\documentclass[aps,prd,twocolumn,showpacs,superscriptaddress,nofootinbib,10pt]{revtex4-2}

\usepackage{acronym}
\usepackage{amsfonts}
\usepackage{amsmath}

\usepackage{amssymb}
\usepackage{bm}
\usepackage{cases}
\usepackage{derivative}
\usepackage{color}
\usepackage{comment}
\usepackage{booktabs}
\usepackage[dvipsnames]{xcolor}
\usepackage{enumerate}
\usepackage{epsfig}
\usepackage{etoolbox}
\usepackage{fancyref}
\usepackage{fancyhdr}
\usepackage[T1]{fontenc} 
\usepackage{graphicx}
\usepackage{hyperref}
\usepackage{indentfirst}
\usepackage{latexsym}
\usepackage{mathrsfs}
\usepackage{mciteplus}
\usepackage{multirow}
\usepackage{rotating}
\usepackage{scrextend}
\usepackage{soul}
\usepackage{subfigure}
\usepackage{verbatim}
\usepackage{amsthm} 

\usepackage[normalem]{ulem}

\def\APJL{Astrophys. J. Lett.}

\def\ACS{ASP Conf. Ser.}

\def\BAAS{Bull. Am. Astron. Soc.}
\def\CQG{Classical Quantum Gravity}
\def\CMP{Commun. Math. Phys.}

\def\GRG{Gen. Rel. Grav.}

\def\JMP{J. Math. Phys.}

\def\JPCS{J. Phys. Conf. Ser.}

\def\LRR{Living Reviews in Relativity}
\def\MNRAS{Mon. Not. R. Astron. Soc.}
\def\NA{Nature Astron.}

\def\NCB{Nuovo Cim. B}

\def\PR{Phys. Rev.}
\def\PRD{Phys. Rev. D}                               
\def\PRL{Phys. Rev. Lett.}   
\def\PRX{Phys. Rev. X}

\def\RPP{Rept. Prog. Phys.}
\def\RMP{Rev. Mod. Phys.}

\begin{document}

\title[Probing Kerr symmetry breaking with LISA extreme-mass-ratio inspirals]{Probing Kerr symmetry breaking with LISA extreme-mass-ratio inspirals}

\author{Pablo F. Muguruza}
\email{pfernandez@ice.csic.es}
\affiliation{Institut de Ci\`encies de l'Espai (ICE, CSIC), Campus UAB, Carrer de Can Magrans s/n, 08193 Cerdanyola del Vall\`es, Spain}
\affiliation{Institut d'Estudis Espacials de Catalunya (IEEC), Edifici Nexus, Carrer del Gran Capit\`a 2-4, despatx 201, 08034 Barcelona, Spain}
\affiliation{Department of Physics, Universitat Autònoma de Barcelona, Facultat de Ci\`encies, Edifici C, 08193 Bellaterra (Cerdanyola del Vallès), Spain}

\author{Carlos F. Sopuerta}
\email{carlos.f.sopuerta@csic.es}
\affiliation{Institut de Ci\`encies de l'Espai (ICE, CSIC), Campus UAB, Carrer de Can Magrans s/n, 08193 Cerdanyola del Vall\`es, Spain}
\affiliation{Institut d'Estudis Espacials de Catalunya (IEEC), Edifici Nexus, Carrer del Gran Capit\`a 2-4, despatx 201, 08034 Barcelona, Spain}

\date{\today}

\begin{abstract}
Extreme-mass-ratio inspirals (EMRIs) are one of the main sources of gravitational waves expected in the low-frequency band, where space-based detectors like Laser Interferometer Space Antenna (LISA) will operate. 
The large number of gravitational-wave cycles accumulated in the EMRI signal in the strong-field regime makes them very precise probes of the local spacetime geometry that are highly sensitive to deviations from the Kerr black hole paradigm. 
In this work, we investigate EMRIs around generic, non-Kerr compact objects characterized by an arbitrary and rich multipolar structure. 
At leading post-Newtonian and linear mass-ratio orders, we incorporate in the waveform model both the axisymmetric and nonaxisymmetric components of the mass quadrupole and octupole moments, parametrizing in this way the breaking of two fundamental symmetries of the Kerr metric. 
We study the impact of these modifications on the waveform following the philosophy of the EMRI \emph{analytic kludge} models. In our case, the orbital evolution follows from a Lagrangian that contains of the multipolar structure of the primary, including for the first time the nonaxisymmetric multipole moments that break axial symmetry. 
Then, using Fisher-matrix analysis, we assess the capability of LISA to constrain deviations of the multipole moments from their Kerr values and, in particular, the possibility of detecting symmetry-breaking effects.
In this way, we analyze how effectively LISA will be able to probe models beyond General Relativity that predict horizon-scale modifications, such as the fuzzball model proposed in string theory.
Our results demonstrate that future LISA observations of EMRIs will provide powerful and unprecedented tests of the black hole structure and the underlying theory of gravity. 
In particular, with one year of LISA data from the inspiral of a $10\,M_{\odot}$ compact object into a rotating supermassive black hole of $10^{6}\,M_{\odot}$ and signal-to-noise ratio of 30, it will be possible to place tight bounds on deviations from the two fundamental symmetries of the Kerr metric, constraining equatorial symmetry breaking in the mass multipole sector to the $10^{-1}$ level and axial symmetry breaking to the $10^{-2}$ level.
\end{abstract}

\maketitle


\section{Introduction}\label{Sec: introduction}

On September 14, 2015, a new form of astronomy, Gravitational Wave (GW) astronomy, was inaugurated with the first direct detection of GWs by the ground-based Laser Interferometer GW Observatory (LIGO)~\cite{LIGOScientific:2016aoc}. Since then, the LIGO-Virgo-KAGRA (LVK) Collaboration has observed more than 200 GW signals from compact binary coalescences across all the observing runs~\cite{LIGOScientific:2018mvr,LIGOScientific:2021usb,LIGOScientific:2021djp,LIGOScientific:2025slb}, with the majority coming from binary black hole (BH) mergers.

The sensitivity of current laser-interferometer ground-based GW detectors, operating in the high-frequency band of the GW spectrum ($10-10^4\,$Hz), is limited at low frequencies by seismic and Newtonian gravity-gradient noises. The future planned third-generation ground-based detectors, such as the Einstein Telescope (ET)~\cite{Punturo:2010zz} and Cosmic Explorer~\cite{Reitze:2019iox}, are expected to extend the low-frequency sensitivity down to $\sim 1\,$Hz, while atom-interferometric GW detectors will probe the decihertz band~\cite{Badurina:2019hst,Canuel:2019abg}.  To access the low-frequency GW band ($10^{-4}-1\,$Hz) we need a space-based detector that avoids the terrestrial noise limitations. The Laser Interferometer Space Antenna (LISA)~\cite{LISA:2017pwj} will be the first GW detector in space, following the successful demonstration of key technologies by the LISA Pathfinder mission~\cite{Armano:2016bkm,Armano:2018kix,LISAPathfinder:2024ucp}. The science theme of LISA was selected by the European Space Agency (ESA) in 2013~\cite{eLISA:2013xep}, the mission concept was selected as an ESA Large-class mission in 2017~\cite{LISA:2017pwj}, and it was formally adopted by the ESA Senior Programme Committee in 2024~\cite{Colpi:2024xhw}, with a launch expected in 2035. 

By unlocking the millihertz frequency band, LISA will enable the observation of sources that are inaccessible to ground-based detectors~\cite{LISA:2017pwj,Colpi:2024xhw}, including: 
(i) The coalescence of massive BH (MBH) binaries (MBHBs) with masses in the range $10^5-10^8\,M_\odot$; 
(ii) the capture and inspiral of a stellar-mass compact object (typically a neutron star or a stellar-origin BH (SOBH) with mass $\mu\sim 1-10^2\,M_\odot$) into a MBH, the so-called \emph{extreme-mass-ratio inspirals} (EMRIs). Variations of these systems that LISA can observe include the \emph{intermediate-mass-ratio inspirals} (IMRIs), in which either an intermediate-mass BHs [(IMBHs) $10^2-10^5\,M_\odot$] inspirals into a MBH (\emph{heavy} IMRIs) or a SOBH inspirals into an IMBH (\emph{light} IMRIs); (iii) ultracompact Galactic binaries (GBs). Millions of them are expected in the LISA band, mainly white dwarf binaries; (iv) SOBH binaries with sufficient signal-to-noise (SNR) ratio during their early inspiral in the LISA band before the become LVK GW sources (such as GW150914~\cite{LIGOScientific:2016aoc} or GW250114~\cite{LIGOScientific:2025rid,LIGOScientific:2025wao}); (v) stochastic GW backgrounds of cosmological origin generated by high-energy processes (around the TeV) in the early Universe; and (vi) unforeseen GW sources that may appear in this unexplored frequency window and for which we should be prepared.

The LISA sources just described allow for a revolutionary science program~\cite{Colpi:2024xhw} with strong impact in astrophysics~\cite{LISA:2022yao}, cosmology~\cite{Laghi:2021pqk, LISACosmologyWorkingGroup:2022jok}, and fundamental physics~\cite{Gair:2012nm,Berti:2019xgr,Barausse:2020rsu, LISA:2022kgy}, the latter being the focus of this work. In particular, we consider one of the GW sources that LISA will observe for the first time, namely EMRIs, to explore further their potential as high-precision probes of the nature of BHs. Specifically, EMRIs provide a unique opportunity to test whether BHs are well described by the Kerr solution~\cite{Kerr:1963ud} of general relativity (GR) or instead correspond to an alternative (exotic) compact objects (see~\cite{Cardoso:2016oxy,Cardoso:2019rvt} for reviews of different scenarios).

EMRIs can form through a variety of astrophysical channels and dynamical mechanisms~\cite{Amaro-Seoane:2007osp,Amaro-Seoane:2012lgq,Amaro-Seoane:2014ela,2020arXiv201103059A}. The most studied formation channel occurs in gas-poor galactic nuclei, where two-body relaxation and related processes drive the stellar-mass compact object (hereafter referred to as the \emph{secondary}) onto an inspiraling orbit around the  putative central MBH (hereafter referred to as the \emph{primary}). Depending on the underlying astrophysical assumptions, the predicted EMRI event rate for LISA spans a wide range, between $1-10^3\,$yr$^{-1}$~\cite{Babak:2017tow,Gair:2017ynp}.  
Other EMRI formation channels include: tidal disruption of stellar-mass binaries by the primary~\cite{ColemanMiller:2005rm}, capture and migration of stellar cores in accretion disks~\cite{DiStefano:2001ci, Davies:2005tc}, formation and subsequent inspiral of SOBHs in active galactic nucleus (AGN) disks~\cite{Levin:2003ej, Levin:2006uc, Secunda:2020cdw, Pan:2021ksp, Pan:2021oob}, and supernova kicks that place the compact remnant onto low-angular-momentum orbits around the primary~\cite{Bortolas:2019sif}.

The remarkable potential of EMRIs for fundamental physics~\cite{Berry:2019wgg,Datta:2019euh,Cardenas-Avendano:2024} lies in the fact that their orbital motion is highly relativistic and, due to the extreme mass ratios involved, $q\equiv \mu/M\ll 1$ (typically $10^{-6} < q < 10^{-4}$), they are long-lived GW sources that can emit of the order of $N\sim q^{-1}\sim 10^{4-6}$ cycles of a highly phase-coherent GW signal within the LISA frequency band. As a result, the EMRI waveform encodes a detailed map of the spacetime geometry of the primary, effectively allowing a precise determination of its multipolar structure (see, e.g.~\cite{Sopuerta:2010fte,Sopuerta:2012hg}). Observations of EMRIs with LISA are therefore expected to provide unprecedented access to the near-horizon region of massive compact objects~\cite{Berry:2019wgg,Cardenas-Avendano:2024}, making them a natural laboratory for testing deviations from the classical BH description predicted by GR~\cite{Hawking:1973uf}.  

The program of testing the \emph{Kerrness} property of ultracompact astrophysical objects through their multipolar structure using EMRI GW signals was initiated by Ryan~\cite{Ryan:1995wh,Ryan:1995zm,Ryan:1995xi,Ryan:1997hg,Ryan:1997kh}. Assuming that the primary is stationary and axisymmetric, it is well-known that its external spacetime geometry can be uniquely characterized by its mass and current multipole moments defined at spatial infinity~\cite{Geroch:1970rg,Hansen:1974ro,Thorne:1980rm} (see also~\cite{Fodor:1989gf}), denoted $(\mathcal{M}^{}_\ell, \mathcal{S}^{}_\ell)$. An alternative characterization based on quasilocal definitions of multipole moments have also been developed, for instance using the framework of isolated horizons for BHs~\cite{Ashtekar:2004gp,Ashtekar:2021wld,Gourgoulhon:2026nes}. A description in terms of multipole moments provides a general and model-independent way to parametrize the geometry of the primary. Ryan demonstrated that the GW signal from a quasi-circular, adiabatic inspiral encodes the full multipolar structure of the background spacetime. These results established, at least in principle, that sufficiently precise GW observations of EMRIs could be used to test the Kerr hypothesis by verifying whether the measured multipoles satisfy the Kerr relations 
\begin{equation}
\mathcal{M}^{}_\ell + i \mathcal{S}^{}_\ell = M (i a )^\ell \quad(\ell=0,1,\ldots) \,.
\end{equation}
The fact that the Kerr multipole moments depend only on the mass $M$ and spin angular momentum $S$ of the BH, $(M,S)$,  or equivalently, on the BH mass $M$ and spin parameter $a = S/M$, $(M,a)$,  is a consequence of the BH uniqueness (\emph{no-hair}) theorems~\cite{Israel:1966rt,Carter:1971zc,Hawking:1972} (see also~\cite{Chruciel:2012}).

Ryan further quantified the accuracy with which the lowest multipole moments could be extracted from the waveform, thereby laying the theoretical foundation for using EMRIs as probes of the strong-field geometry of massive compact objects. Specifically, he showed that observations of EMRIs with the classic LISA configuration (see, e.g.~\cite{Danzmann:1996da,Cutler:1998rf,Prince:2002hp}) could allow for the measurement of three to five multipole moments, and hence providing one to three independent tests of the Kerr hypothesis~\cite{Ryan:1997hg}. More developments of this program can be found in~\cite{Sotiriou:2004bm, Brink:2008xx, Kastha:2018bcr, Datta:2019euh, Datta:2019epe}. In parallel, a complementary line of research has focused on constructing phenomenological \emph{bumpy} BH metrics, introducing controlled deformations of the Kerr geometry within the framework of metric perturbation theory~\cite{Manko:1992mn, AbhishekChowdhuri:2023gvu, Collins:2004na, Vigeland:2009pr, Vigeland:2010xe, Vigeland:2011ji, Moore:2017lxy, Xin:2018urr, Datta:2020axm,LaHaye:2025ley, Zou:2025fsg}. Using a bumpy BH metric, a connection between entropy and the no-hair theorem has been established recently~\cite{Faraji:2025duh}.

A fundamental ingredient for these tests is the development of waveform models that encode the multipole moments of the primary (or a subset of them). The simplest description is the Peters and Matthews approximation~\cite{Peters:1963pm,Peters:1964PhRv..136.1224P}, in which the orbital dynamics is Newtonian and the GW emission mechanism is described by the quadrupole formula.  A widely used extension of the Peters and Matthews approximation is the \emph{analytic kludge} (AK) model introduced by Barack and Cutler~\cite{Barack:2003fp}, which incorporates all the remaining ingredients in EMRI dynamics not present in the Peters and Matthews approximation using (low-order) post-Newtonian (PN) approximations: pericenter precession, Lense-Thirring precession, and inspiral driven by radiation reaction. The AK model enables fast waveform generation and has been extensively used in data-analysis and parameter estimation studies. An improved version of the AK model, the \emph{augmented} AK (AAK) model~\cite{Chua:2017ujo}, replaces the PN orbital frequencies with the relativistic Kerr frequencies for geodesic motion, which achieved a more faithful frequency evolution. The next development worth mentioning for the purposes of this work, is the introduction in the AK model of the mass quadrupole as an additional independent parameter~\cite{Barack:2006pq}:
\begin{equation}
\mathcal{M}^{}_2 \equiv Q = -Ma^2 = -\frac{S^2}{M} \,,
\end{equation}
where the last two equalities hold for a Kerr BH. It has been shown~\cite{Babak:2017tow} that LISA can constrain fractional deviations away from the Kerr mass quadrupole $Q$ at the level of $\sim 10^{-4}-10^{-2}$. 
To move beyond consistency tests of the Kerr quadrupole and begin probing more general departures from the Kerr geometry, probing more in depth the nature of the primary, measurements of higher-order multipole moments are required (see, e.g.~\cite{Cardoso_2016}). Fransen and Mayerson~\cite{Fransen:2022jtw} went beyond the Barack-Cutler work~\cite{Barack:2006pq} by including, for the first time, a current quadrupole and a mass octupole moment as independent parameters. The contribution from these parity-odd moments breaks equatorial symmetry, which is one of the two fundamental symmetries of the Kerr metric. They found that LISA could constrain such symmetry-breaking effects at the level of $10^{-2}$~\cite{Fransen:2022jtw}. As these results illustrate, extending the analysis beyond the quadrupole to the octupolar order already reduces the sensitivity by roughly two orders of magnitude, indicating the increasing challenge of probing higher multipole moments with EMRI observations.

In this paper we go beyond the work of Fransen and Mayerson~\cite{Fransen:2022jtw}, not only by adding more multipole moments as independent parameters to get deeper in the study of tests of the nature of BHs with EMRIs, but also by looking at nontrivial multipolar configurations that can provide not only quantitative but also more qualitative features to our understanding of BHs. 
Specifically, in our model we incorporate nonaxisymmetric components of the mass quadrupole and mass octupole, which characterize arbitrary nonspin-induced deviations from Kerr~\cite{Loutrel:2022ant, Loutrel_2024}. These nontrivial multipole moments, not considered up to date in EMRI models, explicitly break axial symmetry, which is the other underlying fundamental symmetry of the Kerr metric.  

This generalization allows us to constrain the geometry of compact (exotic) objects that violate the symmetries underlying the Kerr solution, such as \emph{fuzzball} configurations that arise in the framework of string theory~\cite{Mathur:2005zp,Skenderis:2006ah,Chowdhury:2007jx,Mathur:2008kg,Chowdhury:2008bd,Skenderis:2008qn,Bacchini:2021fig}. Fuzzballs are expected to possess a highly intricate multipolar structure, which provides a concrete avenue for distinguishing them observationally from the classical black holes predicted by GR~\cite{Bianchi:2020bxa}. At the same time, due to the complex structure of fuzzballs~\cite{Melis_2025}, explicit dynamical calculations of the GW emission from systems involving these objects remain out of reach. In this work, and in its particular application to fuzzballs~\cite{Muguruza:2026zcs},
we adopt a phenomenological approach in which we parametrize the symmetry-breaking features expected in such scenarios by introducing a nonaxisymmetric quadrupole to capture axial-symmetry breaking and an octupolar component to encode equatorial-symmetry breaking.  We incorporate them directly into the EMRI waveform model as potentially observable signatures in the GW signal.

Regarding the EMRI waveform model that we use, it shares with the AK model~\cite{Barack:2003fp}, and its extensions by Barack and Cutler~\cite{Barack:2006pq} and Fransen and Mayerson~\cite{Fransen:2022jtw}, the Peters and Matthews dynamics as the lowest-order approximation (the Newtonian approximation). In particular, for the structure of the waveform itself we use the Barack and Cutler construction~\cite{Barack:2003fp} (AK model). The inspiral trajectory generates a multiharmonic Peters and Matthews waveform, which is then mapped onto the two time-delay interferometry (TDI) data streams through the standard long-wavelength LISA response, as in~\cite{Barack:2003fp} and its extensions~\cite{Barack:2006pq,Fransen:2022jtw}. The main novely that we have introduced in our model has to do with the orbital evolution equations. Instead of importing PN corrections from different derivations in the literature, we rederive them from first principles, starting from a Lagrangian that contains the leading-order terms for each multipole moment of the primary, extending also the multipolar structure to the nonaxisymmetric sector. As is common in kludge models, our equations combine terms of mixed PN order. Our conservative precession equations for $\tilde{\gamma}$ and $\alpha$ are the same as those of Fransen and Mayerson~\cite{Fransen:2022jtw} when we take into account only the multipoles in common (mass quadrupole and octupole), and they also contain the standard general-relativistic periastron advance. The only differences with respect to their model appear in the definition of the orbital frequency $\nu$, which removes the $1/e$ terms from $d\nu/dt$, and in the dissipative sector $de/dt$ (see Sec.~\ref{Sec: equations}).

We start from the multipolar expansion of the gravitational Newtonian potential, which determines the orbital motion of the system, i.e. the conservative dynamics. From here, using the quadrupolar approximation, we   compute the corresponding energy and angular-momentum fluxes. 
The evolution of the eccentricity and frequency, $(e,\nu)$, which is dissipative in nature, includes in this way the contributions from both the axisymmetric and nonaxisymmetric multipolar deviations.
This includes, for the first time in this type of Newtonian-based frameworks, the dissipative corrections of the axisymmetric and nonaxisymmetric quadrupole and octupole moments on the secular evolution of $e$, which was not included in previous AK-based studies. 
It is not \emph{a priori} clear which of the two dominates the resulting constraint, and a full comparison is left for future work. 
While the apsidal, $\tilde{\gamma}$, and nodal, $\alpha$, precession angles include conservative corrections as in earlier work, the radiation-reaction sector of our model is fundamentally different. As we have mentioned above,  the apsidal ($\tilde{\gamma}$) and nodal ($\alpha$) precession equations are compatible with those in Fransen and Mayerson~\cite{Fransen:2022jtw}, including the general-relativistic periastron advance. Our radiation-reaction (dissipative) sector is instead different, since it includes the leading dissipative contributions of the axisymmetric and nonaxisymmetric quadrupole and octupole moments to the evolution of the eccentricity and the frequency. In this way we extend the original AK-based analyses.

While we use the standard AK waveform machinery~\cite{Barack:2003fp} to map the inspiral onto the LISA response, our model goes beyond it, and beyond the latter extensions~\cite{Barack:2006pq,Fransen:2022jtw}, in three ways. First, we promote to independent parameters not only the axisymmetric mass quadrupole and octupole but also, for the first time in an EMRI model for LISA, their nonaxisymmetric components, so that we can probe the breaking of axial symmetry as well as equatorial symmetry. Second, instead of importing PN corrections of different origin, as in the AK model, we derive the orbital evolution equations from a single multipolar Lagrangian. Third, we include, for the first time in this type of Newtonian-based framework, the dissipative contributions of these multipole moments to the secular inspiral. Taken together, these developments extend kludge-based EMRI models to a much more general non-Kerr multipolar structure.

Using our 20-dimensional EMRI model, we perform an extensive Fisher matrix analysis for one year of LISA data with a normalized SNR of 30. We find that LISA can constrain axial symmetry breaking, parametrized by the nonaxisymmetric quadrupole moment $Q_+$, at the level of $10^{-4}$--$10^{-2}$ across a broad range of source parameters. In contrast, equatorial symmetry breaking, associated with the mass octupole moment, can be measured at the level of $10^{-1}$--$10^{0}$, implying that axial symmetry tests are up to two orders of magnitude more sensitive in the mass multipole sector.
The constraints are robust under variations of the spin and eccentricity and improve with increasing secondary mass. Our results demonstrate that a single EMRI observation can place stringent bounds on deviations from the Kerr symmetries, establishing axial symmetry breaking as a new and precise probe of the Kerr geometry and providing a unique pathway to probe exotic compact objects such as fuzzballs (see~\cite{Muguruza:2026zcs} for details). A related study that uses the ringdown signal from comparable-mass binaries, relevant for ground-based GW detectors, can be found in~\cite{Ghosh:2024het}, where constraints to deviations from Kerr are placed using the event GW150914.

The paper is organized as follows. 
In Sec.~\ref{Sec: AKludge}, we introduce the EMRI waveform model constructed in this work and motivate its use for probing non-Kerr spacetimes through deviations in the multipolar structure of the EMRI primary. 
The 20-dimensional parameter space of the EMRI model is described in Sec.~\ref{Sec: parameterspace}.
The orbital dynamics is developed in Sec.~\ref{Sec:orbital-dynamics}, while the radiation-reaction effects are presented in Sec.~\ref{sec:inspiralviaRR}. 
These ingredients are then combined in Sec.~\ref{Sec: equations} to derive
the secular evolution equations for the orbital elements. 
In Sec.~\ref{Subsec: Numerical Validation}, we present a series of validation and consistency checks for our model, including comparisons with previous results in the literature. 
In Sec.~\ref{Sec: Framework Parameter Estimation}, we outline the parameter-estimation framework adopted in this work. In particular, Sec.~\ref{Sec:waveform-construction} describes the construction of the EMRI waveform in the source frame and its mapping to the LISA detector response, while Sec.~\ref{Sec:signal-dynamics} details the computation of the Fisher matrix, the covariance matrix, and the resulting forecasts for the statistical parameter uncertainties. 
In Section~\ref{Sec: results}, we present our main results on LISA’s capability to constrain the different EMRI parameters, including the extra multipolar structure of the EMRI primary, with particular emphasis on the nonaxisymmetric quadrupole moment responsible for the breaking of axial symmetry. We also discuss the implications of these constraints for detecting deviations from Kerr symmetries and for identifying potential signatures of exotic compact objects different from BHs. 
Finally, in Sec.~\ref{Sec: conclusion}, we summarize our main results and discuss future research directions to realize this program.

Throughout this paper, otherwise stated, we use geometric units in which $G = c = 1\,$.

\section{Our EMRI Model} \label{Sec: AKludge}
EMRI waveform models suitable for both detection and accurate parameter estimation in space-based detectors such as LISA are currently under development. The main approach is the \emph{self-force} program (see, e.g., \cite{Poisson:2011nh, Barack:2018yvs}), which is based on BH perturbation theory as numerical relativity is not computationally feasible due to the very different spatial and temporal scales involved in the problem. In the self-force framework, the EMRI spacetime is described as perturbations of the spacetime geometry of the primary induced by the secondary. The perturbative expansion parameter is the mass ratio. At leading order, the small compact object follows a geodesic of the background spacetime (the primary geometry), while at first order the self-force introduces both dissipative effects, that drive the inspiral through energy-angular momentum fluxes, and conservative corrections that modify the orbital frequencies and phase evolution. The self-force program can provide EMRI waveforms with the accuracy required for LISA~\cite{vandeMeent:2017bcc}, but they are only available for limited specific cases. 

To overcome these limitations, a common simplification used is the adiabatic approximation, which captures only the orbit-averaged dissipative part of the first-order self-force and models the inspiral as a sequence of geodesics evolving via balance laws~\cite{Drasco:2005kz}. This approach yields computationally efficient waveform models that are in general not sufficiently accurate for data analysis purposes but are good enough for parameter-estimation forecasts.
Postadiabatic schemes improve the phase accuracy by incorporating conservative and oscillatory corrections beyond the adiabatic limit~\cite{Hinderer:2008dm}.

In this work, we adopt a more pragmatic and flexible strategy aimed at probing non-Kerr spacetimes, prioritizing adaptability to beyond-Kerr geometries over waveform accuracy, being the main goal to capture all the relevant physical effects required for producing reliable parameter-estimation forecasts.  
To this end, we develop a semianalytic EMRI waveform model in which the orbital dynamics is described at leading order by a Newtonian framework, with relativistic effects incorporated through PN and post-Minkowskian corrections derived from first principles. This approach provides a computationally efficient description of the inspiral that remains sufficiently general to accommodate deviations from the Kerr geometry through modifications of the multipolar structure of the EMRI primary. The relatively low computational cost of our framework is particularly advantageous for exploring the high-dimensional parameter space of the model (with dimension $d=20$ in our implementation), enabling efficient phenomenological studies while retaining the key features of EMRI dynamics.

\subsection{Description of the parameter space} \label{Sec: parameterspace}

A generic Kerr BH binary system can be fully described by 17 parameters ($d=17$), excluding the parameters characterizing the final compact remnant, which can be considered to be functions of the parameters of the initial configuration. In the standard EMRI scenario, the secondary is typically assumed to be nonspinning. The dimensionality of the parameter space in this case is then reduced to $d=14$. This approximation is well-justified, as the spin of the smaller body has only a subdominant effect on the waveform and does not significantly affect the overall parameter estimation at the level of accuracy required for our study. 

Since our model shares with the standard AK model~\cite{Barack:2002mha} a common underlying Newtonian description of the orbital dynamics, we follow a very similar parametrization. Nevertheless, as it has been discussed above, our model departs from the standard AK model in that we rederive the orbital evolution from first principles using a multipolar expansion of the gravitational field of the primary, together with consistent energy and angular-momentum fluxes.

To probe deviations from the Kerr geometry, we introduce six additional parameters that encode non-Kerr departures in the multipolar structure of the primary.  The resulting model therefore spans a 20-dimensional parameter space, which we summarize in Table~\ref{parameterspace} and describe below. The additional parameters are dimensionless quantities that control the deviation of each multipole from the classical Kerr values, namely: $Q = -S^2/M$ and  $Q_{+} = \mathcal{O} = \mathcal{O}_{+} = 0$ (see~\cite{Barack:2003fp, Barack:2006pq, Fransen:2022jtw} for previous parametrizations of some of these quantities).

\begin{table}[t]
\caption{Description of the $d=20$ parameter space of the EMRI model used in this work. We denote by $\mathbf{\lambda}=(\lambda^0\,\ldots,\lambda^{19})$ a vector containing the parameters of an EMRI.}
\centering
\small
\renewcommand{\arraystretch}{1.05}
\begin{tabular}{@{} l c p{6.8cm} @{}}
\toprule
\multicolumn{2}{c}{Parameters} & \multicolumn{1}{c}{Description} \\
\midrule
\midrule
\multicolumn{3}{c}{\textbf{EMRI Standard Parameters}} \\
\midrule
$\lambda^0$  & $t^{}_\mathrm{LSO}$ & Time of LSO crossing (scaled to $1\,$mHz) \\[1mm]

$\lambda^1$  & $\ln \mu$ & Logarithm of the secondary mass in $M^{}_\odot$ \\[1mm]

$\lambda^2$  & $\ln M$ & Logarithm of primary mass in $M^{}_\odot$ \\[1mm]

$\lambda^3$  & $\tilde S$ & Primary dimensionless spin magnitude, $\tilde S \equiv |\mathbf{S}|/M^2$ \\[1mm]

$\lambda^4$  & $e^{}_\mathrm{LSO}$ & Orbital eccentricity at the LSO \\[1mm]

$\lambda^5$  & $\tilde\tilde \gamma_0$ & Angle between $\hat{L}\times\hat{S}$ and the pericenter at the LSO \\[1mm]

$\lambda^6$  & $\Phi^{}_\mathrm{LSO}$ & Mean orbital anomaly at the LSO\\[1mm]

$\lambda^7$  & $\theta^{}_S$ & Polar angle of the EMRI sky location \\[1mm]

$\lambda^8$  & $\phi^{}_S$ & Azimuthal angle of the EMRI sky location \\[1mm]

$\lambda^9$  & $\cos\lambda$ & Cosine of orbital inclination ($\hat{L}\cdot\hat{S}$) \\[1mm]

$\lambda^{10}$ & $\alpha^{}_\mathrm{LSO}$ & Azimuth of the angular momentum $\hat{L}$ at the LSO \\[1mm]

$\lambda^{11}$ & $\theta^{}_K$ & Polar angle of the primary spin orientation \\[1mm]

$\lambda^{12}$ & $\phi^{}_K$ & Azimuthal angle of the primary spin orientation \\[1mm]

$\lambda^{13}$ & $D^{}_L$ & Luminosity distance to the EMRI \\[1mm]
\midrule
\multicolumn{3}{c}{\textbf{Dimensionless Multipole Deviation Parameters}} \\
\midrule
$\lambda^{14}$ & $\tilde Q$ & Spheroidal quadrupole moment, $\tilde Q \equiv Q/M^3$ \\[1mm]

$\lambda^{15}$ & $\tilde Q^{}_{+}$ & Polar nonaxisymmetric quadrupole moment, $\Tilde{Q}^{}_{+} \equiv Q^{}_{+}/M^3$ \\[1mm]

$\lambda^{16}$ & $\psi^{}_{Q}$ & Argument of the nonaxisymmetric quadrupole \\[1mm]

$\lambda^{17}$ & $\tilde{\mathcal O}$ & Spheroidal octupole moment, $\tilde{\mathcal{O}} \equiv \mathcal{O}/M^4$ \\[1mm]

$\lambda^{18}$ & $\tilde{\mathcal O}^{}_{+}$ & Polar nonaxisymmetric octupole moment, $\tilde{\mathcal{O}}^{}_{+} \equiv \mathcal{O}_+/M^4$ \\[1mm]

$\lambda^{19}$ & $\psi^{}_{\mathcal O}$ & Argument of the nonaxisymmetric octupole \\

\bottomrule
\end{tabular}
\label{parameterspace}
\end{table}

A particular EMRI configuration in our 20-dimensional parameter space is specified by a vector $\mathbf{\lambda} = (\lambda^I) = (\lambda^{0},\ldots,\lambda^{19})$, whose components are listed in Table~\ref{parameterspace}. The particular parametrization is chosen to facilitate the parameter-estimation study we present in Sec.~\ref{Sec: results}. The parameter $\lambda^0 = t_\mathrm{LSO}$ is a reference time that corresponds to the instant in which the secondary crosses the last stable orbit (LSO), at which point the orbital evolution transitions from inspiral to plunge. The parameters $(\lambda^1,\lambda^2)$ are associated with the masses $M$ and $\mu$, corresponding to the primary and secondary masses, respectively. The dimensionless spin magnitude of the primary is defined as $\lambda^3 = \Tilde{S} = |\mathbf{S}|/M^2$, where $\mathbf{S}$ is the spin angular momentum. In the nonspinning (Schwarzschild) limit, we have $\Tilde{S} = 0$, while for an extremally rotating Kerr black hole, we have $\Tilde{S} = 1$. 

The orbital parameters $(e^{}_\mathrm{LSO}, \tilde{\gamma}^{}_\mathrm{LSO}, \alpha^{}_\mathrm{LSO}, \lambda, \Phi^{}_\mathrm{LSO})$, associated with parameters $(\lambda^4,\lambda^5,\lambda^{10},\lambda^9,\lambda^8)$, are the values at the LSO (excepting for $\lambda$, which is approximately constant as justified later in this work), of the orbital quantities that describe the orbital trajectory of the secondary around the primary (a graphical representation of the different orbital quantities is provided in Fig. 1 of~\cite{Barack:2003fp}): $e$ is the orbital eccentricity; the angle $\tilde{\gamma}$ measures the direction of the pericenter within the orbital plane, defined as the angle from  $\hat{\mathbf{L}} \times \hat{\mathbf{S}}$ to the pericenter direction. In standard celestial mechanics, this corresponds to the argument of periapsis $\omega$ (see, e.g.~\cite{goldstein:mechanics}); the angle $\alpha$ specifies the orientation of the orbital angular momentum unit vector $\hat{\mathbf{L}}$ around the spin axis $\hat{\mathbf{S}}$. Again, in standard celestial mechanics, this corresponds to the longitude of the ascending node $\Omega$; the inclination angle $\lambda$ is defined as the angle between the orbital angular momentum $\hat{L}$ and the spin $\hat{S}$, and describes the tilt of the orbital plane. Finally, $\Phi$ is the mean anomaly, which measures the orbital phase relative to the pericenter passage. 

The angles ($\theta_S, \phi_S)$ and $(\theta_K, \phi_K)$ [corresponding to parameters $(\lambda^{7},\lambda^{8})$ and $(\lambda^{11},\lambda^{12})$, respectively] specify, in ecliptic-based coordinates, the sky location of the EMRI source and the orientation of the spin of the primary, $\hat{\mathbf{S}}$. Finally, $D_L$ (parameter $\lambda^{13}$) denotes the luminosity distance to the EMRI source in the LISA reference frame.

The remaining parameters describe deviations of the primary from the Kerr geometry.
The parameter $Q$ ($\lambda^{14}$) denotes the axisymmetric (spheroidal) quadrupole moment, corresponding to the harmonic with indices $(\ell,m) = (2,0)$ in the multipolar expansion of the Newtonian gravitational potential [see Eq.~\eqref{potentialnewt}]. Similarly, $\mathcal{O}$ ($\lambda^{17}$) represents the axisymmetric octupole moment with harmonic indices $(\ell,m) = (3,0)$. Non-axisymmetric components are denoted by the subscript “$+$”.  In particular, $Q_+$ (parameter $\lambda^{15}$) corresponds to the polar, nonaxisymmetric quadrupole [harmonic indices $(\ell,m) = (2,2)$], while $\mathcal{O}_+$ (parameter $\lambda^{18}$) denotes the polar, nonaxisymmetric octupole [harmonic indices $(\ell,m) = (3,2)$]. The angles $\psi_Q$ and $\psi_{\mathcal O}$ (parameters $\lambda^{16}$ and $\lambda^{19}$) specify the orientation of these nonaxisymmetric deformations. All these multipolar parameters enter the Lagrangian function [Eq.~\eqref{eq:lagrangian}] that governs the orbital dynamics of the system.

Let us have a look at the physical interpretation of this multipolar parametrization. The dimensionless quantity $\Tilde{Q}$ captures deviations from the Kerr mass quadrupole moment, which in GR is fixed by $Q = -S^2 / M$. In contrast, $\mathcal{O}$ represents the lowest-order odd-parity mass multipole that breaks equatorial symmetry and  vanishes identically for a Kerr black hole~\cite{Fransen:2022jtw}. Since we keep this higher-order mass multipole, our bound on equatorial breaking is weaker (see below), in agreement with the finding of Fransen and Mayerson~\cite{Fransen:2022jtw} that $M_3 \equiv \mathcal{O}$ on its own is poorly constrained. Similarly, $Q_+$ is the lowest-order multipole moment that breaks axial symmetry and is also zero in the Kerr case. The parameter $\mathcal{O}_+$ corresponds to the leading nonaxisymmetric, odd-parity multipole, which simultaneously breaks both equatorial and axial symmetries and is likewise absent in the Kerr geometry.   

In our implementation, deviations from Kerr are therefore controlled by the set of dimensionless multipolar parameters $(\tilde{Q}, \tilde{Q}^{}_+, \tilde{\mathcal{O}}, \tilde{\mathcal{O}}^{}_+)$, together with their associated orientation angles. The Kerr limit is recovered by setting all deviation parameters to zero, i.e., $\tilde{Q}=\tilde{Q}^{}_+=\tilde{\mathcal{O}}=\tilde{\mathcal{O}}^{}_+=0$.

\subsection{Orbital dynamics} \label{Sec:orbital-dynamics}

At the Newtonian level, the equations governing the dynamics of a binary system are: (i) The field equation for the Newtonian gravitational potential, namely the Poisson equation; and (ii) the equations of motion for the components of the binary system. We begin by introducing the equation for the Newtonian gravitational potential $U(t,\mathbf{x})$ sourced by the mass density $\rho(t,\bm{x})$ describing the compact objects. The Poisson equation reads:
\begin{equation}\label{eq:Poisson}
\mathbf{\nabla}^2 U = -4\pi\,\rho\,,\qquad \mathbf{\nabla} = \frac{\partial}{\partial \mathbf{x}} \,,
\end{equation}
where we adopt the sign convention used in ~\cite{Poisson:2014pw}, such that the gravitational acceleration is $\mathbf{a}= \mathbf{\nabla} U$ and the monopole contribution to the gravitational potential takes the form $U=M/r$.

Following the spirit of different approaches to EMRI dynamics (see, e.g.~\cite{LISAConsortiumWaveformWorkingGroup:2023arg,Cardenas-Avendano:2024} and references therein), we describe the EMRI orbital motion within an effective one-body framework. This is the standard procedure in Newtonian gravitational dynamics, where starting from the two-body problem, one separates the motion of the center-of-mass from the relative motion between the two masses, in our case $M$ (primary) and $\mu$ (secondary).  The relative motion, the relevant one for our purposes, can then be described as that of a single body with reduced mass $\mu_{\rm red} = M\mu / (M+\mu)$ moving in the gravitational field generated by the total mass $M+\mu$ (see, e.g., \cite{goldstein:mechanics,Landau:1975pou,Jose:1998bt}). In the extreme-mass-ratio regime relevant for EMRIs, $\mu \ll M$ (i.e., $q\ll 1$), one has $\mu_{\rm red} \simeq \mu$ and $M+\mu \simeq M$. Then, to leading order in the mass ratio, the dynamics is equivalent to that of a particle of mass $\mu$ moving in an external gravitational field generated by the primary.

Then, focusing on the relative motion, the dynamics reduces to an effective one-body problem in which the orbital motion is governed by the exterior gravitational field of the primary. The corresponding gravitational potential, in the vacuum region outside the body, satisfies Laplace's equation and admits a multipolar expansion in terms of spherical harmonics,
\begin{equation}\label{potentialnewt}
U(t,\mathbf{x}) = \sum^{}_{\ell m}\frac{4 \pi}{2\ell + 1}\,
I^{}_{\ell m}(t)\,
\frac{Y^{}_{\ell m}(\theta,\phi)}{r^{\ell + 1}} \,,
\end{equation}
where $r = |\mathbf{x}|$ is the relative distance between the two bodies, and $Y_{\ell m}(\theta,\phi)$ are the scalar spherical harmonics defined on the unit 2-sphere, with $(\theta, \phi)$ being standard spherical coordinates. The coefficients $I_{\ell m}(t)$ are the mass multipole moments of the source, defined as
\begin{equation}\label{eq:Ilmdef}
I^{}_{\ell m}(t) = \int^{}_{\mathcal{V}} d^3\mathbf{x} \;\rho(t,\mathbf{x})\, r^\ell\, Y^\ast_{\ell m}(\theta,\phi)\, \,,
\end{equation}
where $\mathcal{V}$ denotes the volume of the source~\cite{Poisson:2014pw} and the asterisk means complex conjugation. These multipole moments can be defined for any matter distribution with compact support and provide a systematic characterization of deviations from spherical symmetry through higher-order multipoles. The expansion in~\eqref{potentialnewt} therefore expresses the exterior gravitational field as a series in inverse powers of the binary separation $r$.
For a spherically symmetric source configuration, only the monopole term $(\ell,m)=(0,0)$ contributes. Using that $Y_{00}=1/\sqrt{4\pi}$, the corresponding moment reduces to
\begin{equation}
I^{}_{00} = \int^{}_{\mathcal{V}} d^3\mathbf{x} \;\rho\,Y^{}_{00} = \frac{M}{\sqrt{4\pi}}\,,
\end{equation}
where we recall that $M$ is the source (primary) total mass. Substituting this into Eq.~\eqref{potentialnewt}, we recover the familiar Newtonian potential for the exterior of a spherically-symmetric body: $U=M/r\,$.

The equations of motion for the relative dynamics can be derived from the Lagrangian
$\mathcal{L}=T-U$, where we truncate the multipolar expansion of the gravitational potential to include only the contributions discussed above (see Table~\ref{parameterspace}). Using spherical coordinates $(r,\theta,\phi)$, the Lagrangian function takes the form:
\begin{widetext}
\begin{eqnarray}\label{eq:lagrangian}
\mathcal{L}
&=& \frac{1}{2}\mu
\left(\dot r^2 + r^2 \dot\theta^2
+ r^2 \sin^2\theta\,\dot\phi^2 \right)
- \frac{\mu M}{r}
+ \frac{2\mu S \sin^2\theta}{r}\dot\phi
\nonumber \\[2mm]
&&
- \frac{\mu M Q}{2r^3}(1-3\cos^2\theta)
+ \frac{3\mu M Q_+}{2r^3}\sin^2\theta \cos(2\psi_Q)
\nonumber \\[2mm]
&&
+ \frac{\mu M \mathcal O}{2r^4}(5\cos^3\theta-3\cos\theta)
+ \frac{15\mu M \mathcal O_+}{2r^4}
\sin^2\theta\cos\theta \cos(3\psi_{\mathcal O}) \,.
\end{eqnarray}
\end{widetext}
Because the Lagrangian is time independent and invariant under rotations about the $z$-axis, the system admits two conserved quantities~\cite{goldstein:mechanics,Landau:1975pou}. Time-translation invariance implies conservation of the energy, $E$, while axial symmetry implies conservation of the $z$-component of the angular momentum, $L_z$.
The corresponding constants of motion are given by: 
\begin{widetext}
\begin{eqnarray}
E &=&
\frac{1}{2}\mu
\left(\dot r^2 + r^2 \dot\theta^2
+ r^2 \sin^2\theta \dot\phi^2 \right)
- \frac{\mu M}{r}
+ \frac{\mu M Q}{2r^3}(1-3\cos^2\theta)
\nonumber \\[2mm]
&&
- \frac{3\mu M Q_+}{2r^3}
\sin^2\theta \cos(2\psi_Q)
- \frac{\mu M \mathcal O}{2r^4}
(5\cos^3\theta-3\cos\theta)
\nonumber \\[2mm]
&&
- \frac{15\mu M \mathcal O_+}{2r^4}
\sin^2\theta\cos\theta
\cos(3\psi_{\mathcal O}) ,
\label{energy}
\end{eqnarray}
\end{widetext}
\begin{equation}
L_z
=
\mu r^2 \sin^2\theta\,\dot\phi
- \frac{2\mu S \sin^2\theta}{r}.
\label{orbitalmomentumz}
\end{equation}

For convenience, we rescale the constants of motion by the mass of the secondary, defining $\tilde E = E/\mu$ and
$\tilde L_z = L_z/\mu$, and drop the tildes in what follows. The subsequent analysis of the orbital dynamics then proceeds along lines similar to those of~\cite{Hinderer:2010}.

Applying the Euler-Lagrange equations to the Lagrangian in Eq.~\eqref{eq:lagrangian}, one obtains the equations of motion governing the orbital dynamics. In particular, the radial evolution is determined by the equation associated with the coordinate $r$, 
\begin{equation}
\frac{d}{dt}\left(\frac{\partial \mathcal{L}}{\partial \dot{r}}\right)
=
\frac{\partial \mathcal{L}}{\partial r}\,,
\end{equation}
which captures the interplay between the central gravitational attraction, rotational effects, and the multipolar corrections to the potential.  The explicit form of the radial equation of motion is then
\begin{widetext}
\begin{eqnarray}\label{rddot}
\ddot{r}
& = & \frac{L^2}{r^3}
 - \frac{2 L_z S}{r^4}
 - \frac{M}{r^2}
 + \frac{3 M Q}{2 r^4} (1 - 3 \cos^2\theta) \nonumber \\[2mm]
& - & \frac{9 M Q_{+}}{2r^4} \sin^2\theta \cos(2 \psi_Q)
 - \frac{2 M \mathcal{O}}{r^5} (5\cos^3\theta - 3\cos\theta) \nonumber \\[2mm]
& - & \frac{30 M \mathcal{O}_{+}}{r^5} \sin^2\theta \cos\theta \cos(3 \psi_{\mathcal{O}}) \,.
\end{eqnarray}
\end{widetext}
The rich multipolar structure of the primary introduces explicit angular dependence and hence, it introduces couplings between the radial and polar motion, leading to a more complex dynamical structure than in the Keplerian case.

On other hand,  the total orbital angular momentum, which appears in the radial equation of motion [Eq.~\eqref{rddot}] is, to leading order in the spin $S$, given by:
\begin{equation}
L^2 = \mu^2 r^4 \left(\dot{\theta}^2 + \sin^2\theta \dot{\phi}^2 \right).
\end{equation}
In addition, using Eq.~\eqref{orbitalmomentumz} for the conserved quantity $L_z$, we can derive the equation governing the azimuthal motion, 
\begin{equation}\label{phidot}
\dot{\phi} = \frac{L_z}{r^2 \sin^2\theta} + \frac{2 S}{r^3} \,.
\end{equation}
The first term is the standard Newtonian contribution, while the second one arises from the spin-orbit coupling introduced in the Lagrangian [Eq.~\eqref{eq:lagrangian}]. This correction induces an additional precession of the orbital motion around the spin axis of the primary. As a result, even in the absence of higher-order multipolar deformations, the spin of the primary modifies the azimuthal evolution.

Substituting the azimuthal motion [Eq.~\eqref{phidot}] into the definition of $L^2$, we obtain the equation for the polar motion
\begin{equation}\label{thetadot2}
\dot{\theta}^2 = \frac{L^2}{r^4} - \frac{L_z^2}{r^4 \sin^2\theta} - \frac{4 L^{}_z S}{r^5} \,,
\end{equation}
from where we can derive the second derivative of the polar angle: 
\begin{equation}\label{thetaddot}
\ddot{\theta} = \frac{L_z^2 \cos\theta}{r^4 \sin^3\theta} + \frac{2 L_z S}{r^6} - \frac{2 \dot{r} \dot{\theta}}{r} \,.
\end{equation}
The first term is again the standard Newtonian contribution, while the others arise from spin-orbit interactions and the coupling of the radial and polar dynamics. 
Finally, using the expression for the conserved energy $E$ in Eq.~\eqref{energy} and the previous expressions we obtain a first-order equation for the radial motion:
\begin{widetext}
\begin{eqnarray}\label{rdot2}
\dot{r}^2 & = & 2E - \frac{L^2}{r^2}
 + \frac{4 L^{}_z S}{r^3}
 + \frac{2 M}{r} \nonumber \\[2mm]
& - & \frac{M Q}{r^3} (1 - 3 \cos^2\theta)
 + \frac{3 M Q^{}_{+}}{r^3} \sin^2\theta \cos(2 \psi^{}_Q) \nonumber \\[2mm]
& + & \frac{M \mathcal{O}}{r^4} (5\cos^3\theta - 3\cos\theta)
 + \frac{15 M \mathcal{O}^{}_{+}}{r^4} \sin^2\theta \cos\theta \cos(3 \psi^{}_{\mathcal{O}}) \,.
\end{eqnarray}
\end{widetext}

We can now express the constants of motion of the perturbed system in terms of a set of orbital elements. To this end, we parametrize the orbital motion using the true anomaly $\gamma$, the inclination angle $\lambda$, and the argument of periapsis $\tilde \gamma_0$. In particular, the polar angle $\theta$ can be written, at leading (Newtonian) order, as~\cite{goldstein:mechanics}
\begin{equation}\label{relation}
\cos\theta = \sin\lambda \sin\left(\tilde \gamma + \tilde \gamma^{}_0\right)\,.
\end{equation}
For bound motion, the radial coordinate can be described by the equation of a (Keplerian) ellipse,
\begin{equation}
r = \frac{p}{1 + e \cos\tilde \gamma}\,,
\end{equation}
where $p$ is the semilatus rectum and $e$ is the orbital eccentricity. Finally, the inclination angle is related to the conserved angular momenta through
\begin{equation}
\cos\lambda = \frac{L^{}_z}{L}\,,
\end{equation}
where we recall that $L$ is the magnitude of the total orbital angular momentum and $L_z$ its projection along the symmetry axis.

At the radial turning points, corresponding to $\gamma = 0$ (periapsis) and $\gamma = \pi$ (apoapsis), and repeating periodically with period $2\pi$ in the true anomaly, the radial coordinate takes the values:
\begin{equation}
r^{}_{\pm} = \frac{p}{1 \pm e} \,,  
\end{equation}
where $r_-$ and $r_+$ denote the periapsis and apoapsis distances, respectively.

We can now obtain expressions for the constants of motion by 
evaluating the radial equation of motion at the turning points,  $r=r_{\pm}$. Then, imposing $\dot{r}=0$, so that the right-hand side of Eq.\eqref{rdot2} vanishes, yields algebraic relations between the conserved quantities and the orbital elements.
In this way, we derive explicit expressions for the constants of motion in terms of the orbital elements $(p,e,\lambda)$. They are given in Appendix~\ref{derivationapp}, in Eqs.~\eqref{Ldef}, \eqref{Edef}, and \eqref{Lzdef}.

\subsection{Inspiral via adiabatic radiation reaction}\label{sec:inspiralviaRR}

In most \emph{kludge} EMRI waveform schemes, the inspiral is described as a sequence of orbits (bound geodesics when using the Kerr metric to describe the gravitational field of the primary), in such a way that the orbital elements are corrected to account for the GW emission. This evolution is typically implemented using balance laws or, similarly, prescriptions for the loss of energy and angular momentum (and Carter constant for Kerr) carried away by the GWs emitted.  At leading order, the radiation-reaction dynamics is governed by the well-known quadrupole formula of GR~\cite{Misner:1973cw,Maggiore:2007mm}, according to which the dominant contribution comes from the radiative mass quadrupole moment. In this approximation, the GW energy flux is given by ($i,j,k=1,2,3$): 
\begin{equation}\label{fluxformula}
\frac{dE}{dt} =
\frac{1}{5} \left\langle
\dddot{M}^{}_{ij}\dddot{M}^{}_{ij} - \frac{1}{3} (\dddot{M}^{}_{kk})^2 \right\rangle \,,
\end{equation}
where angular brackets denote an average over several orbital cycles.
The corresponding GW angular-momentum flux is
\begin{equation}
\frac{d L^i}{dt} = -\frac{2}{5}\,
\epsilon^{ikl} \left\langle
        \ddot{M}^{}_{k m}\dddot{M}^{}_{l m}
    \right\rangle\,,
\end{equation}
where $\epsilon^{ikl}$ denotes the Levi-Civita symbol. In our coordinate system, the spin of the primary is aligned with the $z$-axis, and we therefore focus on the evolution of $z$-component, $L_z$.

In these expressions, the quantity $M_{ij}$ denotes the mass quadrupole moment of the system. The symmetric trace-free projection required for the description of gravitational radiation in the transverse-traceless gauge of GR, where these expressions have been derived, has been explicitly implemented in the energy-flux formula, Eq.~\eqref{fluxformula}, through the subtraction of the trace term.
For a binary system described in the center-of-mass frame, the quadrupole moment takes the form
\begin{equation}\label{quadsystem}
    M^{}_{ij} = \mu\, x^{}_i x^{}_j \,,
\end{equation}
where we recall that $\mu$ is the mass of the secondary (which coincides with the reduced mass in the extreme-mass-ratio limit adopted here), and $x_i$ are the Cartesian components of the relative separation vector.  In spherical coordinates, the relative separation vector are:
\begin{equation}
x^i =
    \big(
        r \sin\theta \cos\phi,\,
        r \sin\theta \sin\phi,\,
        r \cos\theta
    \big)\,.
\end{equation}
Substituting this expression into Eq.~\eqref{quadsystem} allows us to compute the required third time derivatives of $M_{ij}$ in terms of the time-dependent spherical coordinates $(r,\theta,\phi)$ and their time derivatives.

When computing the time derivatives of the components of the mass quadrupole moment, we systematically eliminate higher-order time derivatives by expressing any occurrences of $\ddot{r}$ and $\ddot{\theta}$ in terms of first-order time derivatives by using Eqs.~\eqref{rddot} and~\eqref{thetaddot}. Likewise, we replace $\dot{r}^2$, $\dot{\theta}^2$, and $\dot{\phi}$ using the substitutions provided by Eqs.~\eqref{rdot2}, \eqref{thetadot2} and \eqref{phidot}. In this way, the GW fluxes can be written completely in terms of $r$, $\theta$, and their first-order derivatives, which can in turn be expressed in terms of the orbital elements.

The calculations of the GW fluxes are performed within a perturbative expansion in the PN parameter $\epsilon$, defined by $\epsilon^2 = M/r$, which characterizes the strength of the gravitational field and/or the typical orbital velocities involved.
We truncate the expansion by retaining terms up to $O(S,\epsilon^8)$, $O(Q,\epsilon^8)$, and $O(Q_{+},\epsilon^8)$, which capture the leading-order contributions associated with the spin and quadrupole moments of the EMRI primary. %
Since we are interested in the mass-multipole sector of our parameter space, we implement mass-quadrupole radiation. However, we note that a fully consistent treatment of the dissipative spin ($S_1$) term would also require the current-quadrupole radiation, which enters at the same PN order~\cite{Fransen:2022jtw,Ryan:1995wh}. We do not include this contribution, in line with the leading mass-quadrupole approximation that we use throughout.
In addition, corrections arising from the mass octupole moments are included at orders $O(\mathcal{O}\,\epsilon^{10})$ and $O(\mathcal{O}_{+}\,\epsilon^{10})$. This truncation provides a consistent hierarchy in which lower multipoles dominate, while higher-order structure enters as progressively smaller corrections.
The resulting expressions for the instantaneous energy and angular-momentum GW fluxes are given in Appendix~\ref{derivationapp}, in Eqs.~\eqref{fluxE} and~\eqref{fluxLz}.

To obtain the secular evolution of the extreme-mass-ratio binary, we employ time-averaged GW fluxes, which are computed by averaging over one radial period using the following prescription:
\begin{equation}
\Big\langle \frac{dE}{dt} \Big\rangle
    =
    \frac{
        \displaystyle \int_{0}^{2\pi}
        \frac{dE}{dt}
        \frac{dt}{d\tilde \gamma}
        \, d\tilde \gamma
    }{
        \displaystyle \int_{0}^{2\pi}
        \frac{dt}{d\tilde \gamma}
        \, d\tilde \gamma
    }\,,
\end{equation}
where $\gamma$ denotes the true anomaly. To evaluate these integrals, we adopt the Newtonian parametrization of the orbital motion, in which
\begin{equation}
    \dot{\tilde \gamma} = \frac{\sqrt{p}}{r^2}\,,
\end{equation}
together with the corresponding Newtonian expression for the radial velocity,
\begin{equation}
\dot{r} = \frac{e \sin\tilde \gamma}{\sqrt{p}}\,,
\end{equation}
and use Eq.~\eqref{relation} to account for the $\dot{r}\dot{\theta}$ terms appearing in Eqs.~\eqref{fluxE} and~\eqref{fluxLz}.

We finally obtain explicit expressions for the time-averaged GW fluxes for $(E,L,L_z)$ by consistently truncating the expressions at linear order in the spin $S$ and the multipolar parameters $Q$, $Q_{+}$, $\mathcal{O}$, and $\mathcal{O}_{+}$. The resulting expressions are shown in Eqs.~\eqref{TAfluxE}, \eqref{TAfluxLz}, and \eqref{TAfluxL} of Appendix~\ref{derivationapp}.

We now derive the secular evolution of the orbital elements from the GW fluxes of the energy and angular momentum. First, to derive the evolution equation for the eccentricity, we use of the Newtonian relation~\cite{Maggiore:2007mm}
\begin{equation}
    e^2 = 1 + \frac{2 E L_z^2}{\mu^3 M^2 \cos^2\lambda}\,.
\end{equation}
Differentiating this expression with respect to time and applying the chain rule, we find
\begin{eqnarray}
\frac{de}{dt}
& = & \frac{\partial e}{\partial E} \frac{dE}{dt}
     + \frac{\partial e}{\partial L_z} \frac{dL_z}{dt} \nonumber \\[2mm]
& = & \frac{L_z^2}{\mu M^2 e \cos^2\lambda}\,\frac{dE}{dt}
     + \frac{2 E L_z}{\mu M^2 e \cos^2\lambda}\,\frac{dL_z}{dt}\,.
\end{eqnarray}
Substituting the expressions for $E$ and $L_z$ in terms of the orbital elements, Eqs.\eqref{Edef} and \eqref{Lzdef}, together with the time-averaged fluxes in Eqs.\eqref{TAfluxE} and \eqref{TAfluxLz}, we obtain the secular (time-averaged) evolution of the eccentricity $e$, as given in Eq.~\eqref{eccentricity} of Appendix~\ref{derivationapp}.

For the derivation of the secular evolution of the orbital frequency, we use the following Newtonian relation~(see, e.g.~\cite{Barack:2003fp})
\begin{equation}
\nu = \frac{1}{2\pi M} \left( -\frac{2E}{\mu} \right)^{3/2}\,.
\end{equation}
Differentiating with respect to time, and noting that $\nu$ depends only on $E$ through this relation, we obtain
\begin{equation}
\frac{d\nu}{dt}
    =
    \frac{d\nu}{dE}\frac{dE}{dt}
    =
    -\frac{3}{\mu}\,
    \nu\,(2\pi M \nu)^{-2/3}
    \frac{dE}{dt}\,.
\end{equation}
Substituting the time-averaged energy flux from Eq.~\eqref{TAfluxE}, we obtain the secular evolution of the orbital frequency, given in Eq.~\eqref{frequency}. In the final step, we use the Newtonian relation (see, e.g.~\cite{Barack:2003fp}):
\begin{equation}
\nu = \frac{1}{2\pi M}\left(\frac{M}{a^{}_{\rm orb}}\right)^{3/2}\,,
\end{equation}
which follows from Kepler’s third law in the Newtonian limit, and where $a_{\rm orb}$ denotes the semimajor axis. 

The apsidal $\tilde \gamma$ and nodal $\alpha$ precession rates characterize the secular precession of orbits induced by deviations from the loss of spherical symmetry in the gravitational field. Specifically, $\tilde \gamma$ describes the precession of the periapsis within the orbital plane, while $\alpha$ measures the precession of the line of nodes about the symmetry axis.
The evolution equations for these angular variables are derived using action-angle variables, following the standard procedure outlined in~\cite{goldstein:mechanics}. The resulting expressions for the apsidal and nodal precession rates are given below in Eqs.~\eqref{apsidal} and~\eqref{nodal}.

\subsection{Secular evolution equations for the orbital dynamics} \label{Sec: equations}
The inspiral trajectory of the secondary in an EMRI is governed by the evolution of the following variables $\Phi$, $\nu$, $\tilde{\gamma}$, $\alpha$, $e$, and $\lambda$. 
In line with previous studies~\cite{Barack:2003fp, Barack:2006pq, Fransen:2022jtw}, we approximate the inclination angle $\lambda$, defined as the relative orientation of the secondary orbital angular momentum and the primary’s spin, to be constant.
This approximation is well justified, as the secular variation of $\lambda$ due to the radiation reaction evolution is negligible for the systems considered here~\cite{Barack:2003fp}. Moreover, including the evolution of $\lambda$ in the model has been shown to have a negligible impact on the main results (see Sec.~3.4 of~\cite{Fransen:2022jtw}). 

Our analysis focuses on the secular evolution of the orbital elements, since their evolution timescale is comparable to the radiation-reaction timescale. 
We define the eccentricity through the mechanical energy $E$ and the axial angular momentum $L_z$ of our dynamics. In this way, its evolution is driven only by the radiative fluxes of $E$ and $L_z$, so that our $de/dt$ is dissipative and includes the leading dissipative contributions of the axisymmetric and nonaxisymmetric multipole moments. The conservative multipolar contribution to $de/dt$ that appears in Fransen and Mayerson~\cite{Fransen:2022jtw} comes from their osculating-element method, in which the geometric eccentricity of the instantaneous ellipse changes slowly under the nonspherical potential even without radiation. This contribution does not appear as a separate term in our derivation. This is the same choice of variables that removes the $1/e$ terms from our $d\nu/dt$. In any case, we have checked that our forecasts agree with those of Fransen and Mayerson (see Sec. \ref{Sec: results}).
In particular, we incorporate the dissipative influence of the quadrupole moment $Q$ on the eccentricity, a contribution that was not considered in previous studies~\cite{Barack:2006pq, Fransen:2022jtw}.
By contrast, following prior works, the corrections to the argument of periapsis $\tilde{\gamma}$ and the nodal precession angle $\alpha$ are treated as conservative, reflecting the multipolar structure of the central object without directly contributing to energy and angular momentum losses.

For the conservative precession, our equations $d\tilde{\gamma}/dt$ and $d\alpha/dt$ coincide with those of Fransen and Mayerson~\cite{Fransen:2022jtw} when we only consider the multipoles in common, and they include the standard general-relativistic periastron advance. They differ from Barack and Cutler~\cite{Barack:2006pq} in the same way that the Fransen and Mayerson equations do. Our differences with respect to Fransen and Mayerson appear only in the definition of the orbital frequency $\nu$ [Eq.~\eqref{frequency}] and in the dissipative sector $de/dt$. As is common in kludge waveform models, our expressions are not obtained at a single fixed PN order, but combine terms of mixed order. We have verified that the resulting forecasts are consistent with both earlier studies.

Finally, we retain only leading-order terms in the mass ratio, $\mu/M \ll 1$, and include contributions linear in the spin and multipolar parameters, $O(S)$, $O(Q)$, $O(\mathcal{O})$, $O(Q_+)$, and $O(\mathcal{O}_+)$, while neglecting higher-order corrections. Under these approximations, the evolution equations for the variables $\Phi$, $\nu$, $\tilde{\gamma}$, $\alpha$, $e$, and $\lambda$ take the following form:

\begin{widetext}
\begin{eqnarray}
\Big\langle  \frac{d \nu}{d t} \Big\rangle & = & \frac{96}{10 \pi} \frac{\mu}{M^3} \frac{(2 \pi M \nu)^{11/3}}{(1 - e^2)^{7/2}} \frac{1}{(1 + e^2)} \bigg[ f^{}_{\nu, 0}(e) 
+ \frac{S}{96}\frac{1}{M^{5/2}} \frac{(2 \pi M \nu)}{(1 - e^2)^{3/2}} f^{}_{\nu, S}(e) \cos\lambda  \nonumber \\[2mm]
& + & \frac{1}{M^{2}} \frac{(2 \pi M \nu)^{4/3}}{(1 - e^2)^{2}} \Big( \frac{Q}{384} f^{}_{\nu, Q, 1}(e) \sin^2\lambda 
 -  \frac{Q}{1536}  f^{}_{\nu, Q, 2}(e) \cos(2 \tilde \gamma) \sin^2\lambda
+ \frac{Q^{}_{+}}{128} f^{}_{\nu, Q^{}_+, 1}(e)  \sin^2\lambda \cos(2 \psi^{}_Q) \nonumber\\[2mm]
&  & + \frac{Q^{}_{+}}{512} e^2 f^{}_{\nu, Q^{}_+, 2}(e) \sin^2\lambda \cos(2 \psi^{}_Q) \cos(2 \tilde \gamma)  \Big) \nonumber \\[2mm]
& + & \frac{e}{M^{3}} \frac{(2 \pi M \nu)^{2}}{(1 - e^2)^{3}} \Big(  \frac{\mathcal{O}}{1024} f^{}_{\nu, \mathcal{O}, 1}(e) \big(\sin\lambda + 5\sin(3\lambda) \big) \sin\tilde \gamma
+ \frac{\mathcal{O}}{768} e^2 f^{}_{\nu, \mathcal{O}, 2}(e) \sin^3\lambda \sin(3 \tilde \gamma) \nonumber \\[2mm]
& - & \frac{\mathcal{O}^{}_{+}}{512} f^{}_{\nu, \mathcal{O}^{}_{+}, 1}(e) \big(5 + 3\cos(2\lambda)\big) \sin\tilde \gamma \cos(3 \psi^{}_{\mathcal{O}})
  - \frac{\mathcal{O}^{}_{+}}{256} e^2 f^{}_{\nu, \mathcal{O}^{}_{+}, 2}(e) \sin^2\lambda \sin(3\tilde \gamma) \cos(3 \psi^{}_{\mathcal{O}})  \Big) \Big] \,, 
\label{frequency}
\end{eqnarray}

\begin{eqnarray}
\Big\langle \frac{d \tilde \tilde \gamma}{d t} \Big\rangle & = & 6 \pi \nu
\frac{(2 \pi M \nu)^{2/3}}{(1 - e^2)} \bigg[ 1 
         + \frac{1}{4} (2 \pi M \nu)^{2/3} (1 - e^2)^{-1} \big(26 - 15 e^2\big) \bigg] \nonumber\\[2mm] 
& + & \pi \nu \bigg[ -12 \big(2 \pi M \nu\big) (1 - e^2)^{-3/2} S \cos\lambda - \frac{3}{4} (2 \pi M \nu)^{4/3} (1 - e^2)^{-2} Q \big(3 + 5 \cos(2\lambda) \big) \nonumber \\[2mm]
& - &  \frac{3}{4} Q^{}_+ \frac{(2 \pi M \nu)^{4/3}}{(1 - e^2)^{2}} \big(11 + 5 \cos(2\lambda) \big) \cos(2 \psi^{}_Q) \nonumber\\[2mm]
& + &  \frac{3}{32\,e} \mathcal{O} \frac{(2 \pi M \nu)^{2}}{(1 - e^2)^{3}}  \Big[\big(5 + 35 e^2\big) \cos(4 \lambda) - 4 \cos(2\lambda) - (1 + 3 e^2)\Big] \csc\lambda \sin\tilde \gamma \nonumber\\[2mm]
& - & \frac{15}{32\,e} \mathcal{O}^{}_+ \frac{(2 \pi M \nu)^{2}}{ (1 - e^2)^{3}}  \Big[\big(3 + 21 e^2\big) \cos(4 \lambda) 
+ \big(4 + 32 e^2\big) \cos(2\lambda) - \big(7 + 21 e^2\big)\Big] \csc\lambda \sin\tilde \gamma \cos(3 \psi^{}_\mathcal{O})  \bigg] \,,
\label{apsidal}          
\end{eqnarray}
\begin{eqnarray}
\Big\langle  \frac{d \alpha}{d t} \Big\rangle & = & \pi \nu \bigg[ 4 (2 \pi M \nu) (1 - e^2)^{-3/2} S  
         + 3 (2 \pi M \nu)^{4/3} (1 - e^2)^{-2} Q \cos\lambda 
        - 3 (2 \pi M \nu)^{4/3} (1 - e^2)^{-2} Q^{}_{+} \cos\lambda \cos(2 \psi^{}_Q) \nonumber \\[2mm]
& - & \frac{3}{8} (2 \pi M \nu)^{2} (1 - e^2)^{-3} \mathcal{O}\, e \cot\lambda \big(-7 + 15 \cos(2\lambda)\big) \sin\tilde \gamma \nonumber \\[2mm]
& + & \frac{15}{16} (2 \pi M \nu)^{2} (1 - e^2)^{-3} \mathcal{O}^{}_+ e \csc\lambda \big(7 \cos\lambda + 9 \cos(3\lambda) \big) \cos(3 \psi^{}_\mathcal{O}) \sin\tilde \gamma  \bigg] \,,
\label{nodal}    
\end{eqnarray}
\begin{eqnarray}
\Big\langle  \frac{d e}{d t} \Big\rangle & = & - \frac{e}{15} \frac{\mu}{M^2}\frac{(2 \pi M \nu)^{8/3}}{(1 - e^2)^{5/2}} \big( 304 + 121e^2 \big) + \frac{\mu}{M^4} \frac{(2 \pi M \nu)^{11/3}}{e (1 - e^2)^{4}} \frac{1}{(1 + e^2)} \bigg[ \frac{S}{15} f^{}_{e, S, 1}(e) \sec\lambda \nonumber \\[2mm]
& - & \frac{S}{15} f^{}_{e, S, 2}(e) \cos\lambda + \frac{S}{15} e^2 f^{}_{e, S, 3}(e) \sin\lambda \tan\lambda \cos(2 \tilde \gamma) + \frac{(2 \pi M \nu)^{-1/3}}{(1 - e^2)^{1/2}} \Big( \frac{Q}{30} f^{}_{e, Q, 1}(e) + \frac{Q}{20}  f^{}_{e, Q, 2}(e) \sin^2\lambda \nonumber \\[2mm]
& - & \frac{Q}{80} f^{}_{e, Q, 3}(e) \sin^2\lambda \cos(2 \tilde \gamma)
         - \frac{Q_+}{20} f^{}_{e, Q_+, 1}(e) \cos(2 \psi^{}_Q) 
         - \frac{Q_+}{20} f^{}_{e, Q_+, 2}(e) \cos(2\lambda) \cos(2 \psi_Q) \nonumber \\[2mm]
& - & \frac{Q_+}{80} e^2 f^{}_{e, Q_+, 3}(e) \sin^2\lambda \cos(2 \tilde \gamma) \cos(2 \psi^{}_Q) \Big) 
+ \frac{e}{M} \frac{(2 \pi M \nu)}{(1 - e^2)^{3/2}} \Big( \frac{\mathcal{O}}{80} f^{}_{e, \mathcal{O}, 1}(e) \sin\lambda \sin\tilde \gamma \nonumber \\[2mm]
& + & \frac{\mathcal{O}}{80} f^{}_{e, \mathcal{O}, 2}(e) \sin\lambda \cos(2\lambda) \sin\tilde \gamma 
+ \frac{\mathcal{O}}{16} f^{}_{e, \mathcal{O}, 3}(e) \sin(3\lambda) \sin\tilde \gamma  + \frac{\mathcal{O}}{24} e^2 f^{}_{e, \mathcal{O}, 4}(e) \sin^3\lambda \sin(3\tilde \gamma) \nonumber \\[2mm]
& + & \frac{\mathcal{O}^{}_{+}}{32} f^{}_{e, \mathcal{O}^{}_+, 1}(e) \sin\lambda \sin\tilde \gamma \cos(3 \psi^{}_{\mathcal{O}}) 
+ \frac{\mathcal{O}^{}_{+}}{16} e^2 f^{}_{e, \mathcal{O}^{}_+, 2}(e) \sin\lambda \sin\tilde \gamma \cos(2 \tilde \gamma) \cos(3 \psi^{}_{\mathcal{O}}) \nonumber \\[2mm]
& + & \frac{\mathcal{O}^{}_+}{32} f^{}_{e, \mathcal{O}^{}_+, 3}(e) \sin\lambda \cos(2 \lambda) \sin\tilde \gamma \cos(3 \psi^{}_{\mathcal{O}}) 
- \frac{\mathcal{O}^{}_+}{16} e^2 f^{}_{e, \mathcal{O}^{}_+, 4}(e) \sin\lambda \cos(2 \lambda) \sin\tilde \gamma \cos(2 \tilde \gamma) \cos(3 \psi^{}_{\mathcal{O}}) \Big) \bigg] \,.
\label{eccentricity}    
\end{eqnarray}
\end{widetext}
The functions $f_{\nu,X}(e)$ in Eq.~\eqref{frequency} and the functions $f_{e,X}(e)$ in Eq.~\eqref{eccentricity} are polynomials in the eccentricity $e$, and their explicit form is given in Appendix~\ref{evolutionapp}.  

The integration of the coupled system of ordinary differential equations governing the orbital elements, namely Eqs.~\eqref{frequency}-\eqref{eccentricity}, together with the equations for the orbital motion, completely determines the EMRI trajectory. 
In practice, we need to take into account some important considerations for a successful and reliable integration. First, we must specify the location of the LSO, which marks the transition from inspiral to plunge and thus sets the endpoint of the evolution described by our system. In this work, as it is done in previous similar studies~\cite{Barack:2006pq, Fransen:2022jtw}, we adopt the Schwarzschild LSO, given by
\begin{equation}
\nu^{}_\mathrm{LSO} = (2 \pi M)^{-1}\; \Big( \frac{1 - e_\mathrm{LSO}^2}{6 + 2 e^{}_\mathrm{LSO}}\Big)^{3/2} \,.
\end{equation}
In this context, it is important to note that adopting a Kerr cutoff (i.e., the Kerr LSO) typically yields tighter parameter-estimation constraints than the Schwarzschild cutoff~\cite{Gair:2017ynp}, leading to improved accuracy in the forecasts for the measurements of the different EMRI parameters.
This effect is particularly pronounced for prograde Kerr orbits, for which the LSO lies closer to the Kerr event horizon, especially for large spin values~\cite{Zi:2021pdp}. 
By contrast, in the Schwarzschild case the constraints remain nearly insensitive to the spin, as illustrated in Fig.~\ref{fig:q_vs_S} and discussed in Sec.~\ref{Sec: results}.

In this sense, and connecting with the astrophysical perspective of EMRIs~\cite{LISA:2022yao}, although not too much is known about the spin distribution of the central massive bodies that inhabit the galactic centres and that can capture stellar-mass compact objects to form EMRIs, there are plausible scenarios in which we can expect high spins for the primary. Therefore, using a more realistic EMRI model that incorporates the Kerr cutoff, together with a consistent treatment of the plunge phase, should lead to significantly improved parameter-estimation performance.

Moreover, rapidly spinning primaries can capture secondaries on orbits with higher initial eccentricities~\cite{Amaro-Seoane:2012jcd}, which likely will translate into larger eccentricities by the time the EMRI system enters the LISA band, and consequently into higher eccentricities at plunge ($e_\mathrm{LSO}$).  This brings another important aspect for the modeling, namely the range of values of $e_\mathrm{LSO}$ that lead to stable and robust evolutions.  
Based on previous reported experience with AK waveform models, as well as our own tests, we restrict ourselves to relatively small eccentricities at the LSO, in the range $0 \leq e_{LSO} \leq 0.3$. Beyond this range, AK models tend to lose reliability when extrapolated to large eccentricities~\cite{Barack:2003fp}.  

\subsection{Numerical validation and verification of the EMRI evolution equations} \label{Subsec: Numerical Validation}

The evolution equations we have derived above contain several terms with factors proportional to $1/e$, which may appear problematic in the limit of low eccentricity, $e \rightarrow 0$. In particular, such terms arise in Eq.~\eqref{apsidal} in association with the octupole moment $\mathcal{O}$, in agreement with what has also been found in the work of Fransen and Mayerson (see~\cite{Fransen:2022jtw}). In our case, similar contributions also appear in connection with the nonaxisymmetric octupole component $\mathcal{O}_+$. By contrast, we do not find $1/e$ terms  in our evolution equation for the orbital frequency, Eq.~\eqref{frequency}, unlike in the model of Fransen and Mayerson~\cite{Fransen:2022jtw}.%
This difference comes from the definition of $\nu$. We define it through the orbital energy, while Fransen and Mayerson~\cite{Fransen:2022jtw} define it through the perihelion-to-perihelion radial period, which diverges as $1/e$. The two definitions are related by a change of variables that does not affect the physical results, as we confirm by the agreement of our constraints with those of~\cite{Fransen:2022jtw}.

To understand better these problematic terms, we have first assessed the impact of the $1/e$ terms associated with the octupole moment $\mathcal{O}$ in the evolution equation for the apsidal precession $\tilde \gamma$ by removing them. We find that the resulting parameter-estimation forecasts remain of the same order, with only minor quantitative differences. We have then performed the same test for the $1/e$ terms associated with the nonaxisymmetric octupole
component $\mathcal{O}_+$, obtaining results that exhibit the same behavior. 

On the other hand, several terms proportional to $1/e$ appear in Eq.~\eqref{eccentricity} in the contributions associated with the spin $S$ and the quadrupole moment $Q$, as well as in its nonaxisymmetric component $Q_+$. Such terms are absent in previous analysis~\cite{Barack:2006pq, Fransen:2022jtw}, as those works included only conservative effects and neglected dissipative contributions from the multipole moments. Once dissipative effects are incorporated, terms of this type naturally arise in the evolution equation for the eccentricity.  

We have performed several consistency checks by removing these $1/e$ terms in Eq.~\eqref{eccentricity} for $S$, $Q$, and $Q_+$,  considering representative values $e_\mathrm{LSO} = \{0.01, 0.1, 0.3\}$. We find that their removal does not significantly affect the resulting parameter estimation constraints. This shows how robust our model predictions are against these contributions. In conclusion, this suggests that the apparent divergences in the low-eccentricity limit are not physical, but rather reflect the choice of orbital elements used to describe nearly circular orbits in a system with a complex dynamics due to the nontrivial multipolar structure~\cite{Fransen:2022jtw}.

We do not include in our EMRI model the PN terms proportional to $(2 \pi M \nu)^{13/3}$ in the equation for the frequency evolution [Eq.~\eqref{frequency}] and to $(2 \pi M \nu)^{10/3}$ in the equation for the eccentricity evolution [Eq.~\eqref{eccentricity}], which were considered both in~\cite{Barack:2006pq} and~\cite{Fransen:2022jtw}. As a check of the robustness of our scheme, we have reintroduced these terms and, by performing a series of numerical experiments with our equations, we have found that their inclusion has an almost negligible impact on the resulting parameter estimation constraints.

In addition, we have performed a PN test by removing the term proportional to $(2 \pi M \nu)^{4/3}$ in the evolution equation for $\tilde \gamma$ [Eq.~\eqref{apsidal}] corresponding to the $Q$ contribution. We find that the parameter-estimation constraint on $\Delta Q$ changes only by a factor of $\sim 1$. This result is consistent with the test performed and reported in~\cite{Fransen:2022jtw}.

\section{Framework for Parameter Estimation Studies} \label{Sec: Framework Parameter Estimation}

In this section, we describe the framework that we use to assess LISA’s capability to measure the EMRI parameters and, in particular, to constrain deviations from the Kerr geometry that we encode in the multipolar structure of the primary. We then present the Fisher-matrix formalism adopted to quantify the information content of the signal and to compute the expected statistical uncertainties on the source parameters. This framework provides the basis for the parameter-estimation results discussed in the following section (Sec.~\ref{Sec: results}).

\subsection{EMRI waveform construction} \label{Sec:waveform-construction}

As in the already classical treatment of the gravitational radiation emitted by a binary system of point masses by Peters and Matthews~\cite{Peters:1963pm}, and also in semianalytic EMRI waveform models, we describe the inspiral as a sequence of Newtonian orbits whose orbital elements/constants of motion evolve adiabatically under the emission of gravitational radiation. In this approximation, the radiation emission is dominated by the time variation of mass quadrupolar moment of the system. At leading order, the gravitational waveform observed in the far zone can be expressed, in the transverse-traceless (TT) gauge, as
\begin{equation}\label{eq:hijTT}
h^{\rm TT}_{ij}(t) =
\frac{2}{D^{}_L}
\left(P^{}_{i}{}^{k}P^{}_{j}{}^{l}-\frac{1}{2}P^{}_{ij}P^{kl}\right)
\;\ddot{\mathcal{I}}^{}_{kl}(t_{\rm ret}) \,,
\end{equation}
where $D_L$ is the luminosity distance to the source, and $t_{\rm ret}=t-D_L$ denotes the retarded time. The tensor $P_{ij}=\delta_{ij}-\hat n_i\hat n_j$ is the projector onto the plane orthogonal to the light of sight, defined by the
unit vector $\hat n$ pointing from the source to the detector. 

The quantity $\mathcal{I}_{ij}$ in Eq.~\eqref{eq:hijTT} denotes the symmetric and trace-free (STF) mass quadrupole moment of the system, and overdots represent derivatives with respect to coordinate time. The TT projection in Eq.~\eqref{eq:hijTT} ensures that only the radiative, transverse degrees of freedom are retained, corresponding to the two physical polarizations of the GW.  In terms of the relative separation vector $\mathbf{x}(t)$, its leading-order expression in the center-of-mass frame is
\begin{equation}\label{eq:STFquadrupole}
\mathcal{I}^{}_{ij}(t) =
\mu\left(x^{}_i x^{}_j-\frac{1}{3}\delta^{}_{ij}r^2\right)\,,
\end{equation}
with $r^2=x^{}_k x^{}_k$. This STF quadrupole moment directly sources the emitted waveform. By contrast, in Sec.~\ref{sec:inspiralviaRR} we introduced the \emph{source} quadrupole $M_{ij}=\mu x_i x_j$ when computing energy and angular momentum fluxes, where the trace-free projection is implemented explicitly, as it can be seen in the GW energy-flux formula in Eq.~\eqref{fluxformula}. 

The detector (LISA) observables are obtained by projecting the TT metric perturbation [Eq.~\eqref{eq:hijTT}] onto the LISA detector using the standard long-wavelength (low frequency) response, which accounts for the time-dependent modulation induced by the orbital motion of the LISA constellation~\cite{Cutler:1998rf,Cornish:2003d,Marsat:2018oam,Marsat:2020rtl} (see also~\cite{Barack:2003fp} for the implementation in the AK model). In this approximation, the GW signal is mapped onto the interferometric data streams through time-dependent antenna pattern functions that encode the detector geometry and its motion around the Sun. For parameter-estimation purposes, we express the signal in terms of the orthogonal TDI variables $(A, E, T)$~\cite{Tinto:2002de,Shaddock:2003dj,Tinto:2020fcc}, which diagonalize the instrumental noise to leading order in the long-wavelength approximation. In this work, we only use the $(A, E)$ channels, as the $T$ channel is significantly less sensitive in the low-frequency regime relevant for EMRIs. These two channels therefore provide the dominant contribution to the signal-to-noise ratio and are the ones we use for our parameter-estimation studies.
The corresponding noise model for the TDI observables is summarized in Appendix~\ref{LISA-noise}.

\subsection{Signal analysis and parameter estimation} \label{Sec:signal-dynamics}

In practice, we follow a procedure similar to the one used in the construction of the LISA response model to EMRI GW signals for the AK models~\cite{Barack:2003fp}. First, we begin by numerically integrating the EMRI evolution equations backwards in time from the LSO, thereby reconstructing the inspiral trajectory. Next, using this trajectory, we compute the waveform amplitude coefficients together with the LISA antenna pattern functions, including the effects of the detector’s orbital motion and the associated Doppler modulation. This yields the LISA response $\mathbf{h} = (h_I(t))$ for $I = A, E$, corresponding to the two orthogonal TDI channels we need for our analysis. With the LISA response in hand, we are now able to carry out the parameter estimation analysis. 

The Fisher information matrix quantifies how sensitively a signal depends on its source parameters and provides an estimate of the expected parameter uncertainties in the high signal-to-noise ratio regime~\cite{Vallisneri:2007ev}.  We assume that the LISA noise is stationary and Gaussian, and can therefore be fully characterized by its one-sided power spectral density (PSD) $S_n(f)$ (see Appendix~\ref{LISA-noise}). Under these assumptions, the components of the Fisher matrix can be written as follows (see, e.g.~\cite{Finn:1992wt,Cutler:1994ys})
\begin{equation}
\tilde \gamma^{}_{\alpha\beta} \;=\;  \Big( \frac{\partial \mathbf{h}}{\partial \lambda^\alpha} \Big| \frac{\partial \mathbf{h}}{\partial \lambda^\beta}\Big) \,,
\label{eq:fisher_def}
\end{equation}
where $\mathbf{h}$ denotes the vector containing the two independent LISA response functions introduced above; $\lambda^\alpha$ are the components of the EMRI parameter vector as described in Table~\ref{tableparameters}; and $\left(\cdot | \cdot\right)$ denotes the noise-weighted inner product defined as
\begin{equation}
\left( \mathbf{a} | \mathbf{b} \right) \equiv 2 \sum^{}_{I=A,E}   
\int_{0}^{\infty} df\;
\frac{ \tilde{a}^{}_I(f)\, \tilde{b}^{\ast}_I(f)+\tilde{a}^{\ast}_I(f)\, \tilde{b}^{}_I(f)}{S^{}_n(f)} \,,
\end{equation}
where the tilde here denotes the Fourier transform of the signal\footnote{We adopt the following convention for the Fourier transform: 
\[ \tilde{s}(f)\equiv \int^\infty_{-\infty} dt\; e^{2\pi i f t} s(t) \,.  \]}, and the sum runs over the two independent TDI channels $(A, E)$.  Moreover, in Eq.~\eqref{eq:fisher_def}, $\partial_{\lambda^\alpha} \mathbf{h}$ denotes the derivative of the LISA response function components with respect to the EMRI parameter $\lambda^\alpha$. 

Since it is not possible to have analytical closed-form expressions for the LISA response functions to EMRI signals, the derivatives with respect to the parameters entering the Fisher matrix must be computed numerically. A particularly accurate numerical approximation is provided by the five-point central difference formula:
\begin{eqnarray}
\frac{dh(x)}{dx} & = & \frac{1}{12 \epsilon} \bigg[ h (x - 2 \epsilon) - h (x + 2 \epsilon) + 8 h (x + \epsilon) 
\nonumber \\[2mm] 
& - &  8 h (x - \epsilon) \bigg]  +  \mathcal{O}(\epsilon^4) \,.
\end{eqnarray}
This formula is fourth-order accurate in the step size $\epsilon$, and significantly reduces truncation errors compared to lower-order finite-difference approximations. In practice, the step size $\epsilon$ is chosen adaptively for each parameter $\lambda^\alpha$ in order to balance truncation and round-off errors: values of $\epsilon$ that are too large can introduce significant truncation errors, while values that are too small amplify round-off numerical noise. To ensure numerical stability of the computation, we have explored a range of values for $\epsilon$ for each parameter and we have selected those that produce the most robust and numerically stable Fisher matrix components. This calibration guarantees that the derivative estimates yield a reliable and accurate numerical computation of the Fisher matrix.

Once the Fisher matrix $\tilde \gamma$ has been computed, the corresponding covariance matrix is obtained as its inverse:
\begin{equation}
\Sigma \equiv \tilde \gamma^{-1} \,.
\end{equation}
The covariance matrix encodes the expected statistical uncertainties and correlations among the parameters listed in Table~\ref{tableparameters}, provided the standard assumptions of the Fisher matrix formalism are met: stationary and Gaussian noise;  sufficiently high signal-to-noise ratio; and that the response function depends approximately linearly on the model parameters over the scale of the uncertainties (see, e.g.,~\cite{Vallisneri:2007ev}).

The $1\sigma$ (standard deviation) statistical uncertainty for each parameter $\lambda^\alpha$ is then given by the square root of the corresponding diagonal element of the covariance matrix:
\begin{equation}
\sigma^{}_\alpha \;=\; \sqrt{\Sigma^{}_{\alpha\alpha}} \,.
\end{equation}
These uncertainties, by virtue of the Cramer-Rao bound~\cite{Kay:1993estimation}, represent a lower bound on the variance of any unbiased estimator. These uncertainties are reported explicitly in Table~\ref{tableparameters} for a representative EMRI configuration, providing a quantitative estimate of the achievable parameter estimation accuracy with the modeled LISA response.

\begin{figure}
\centering
\includegraphics[width=\linewidth]{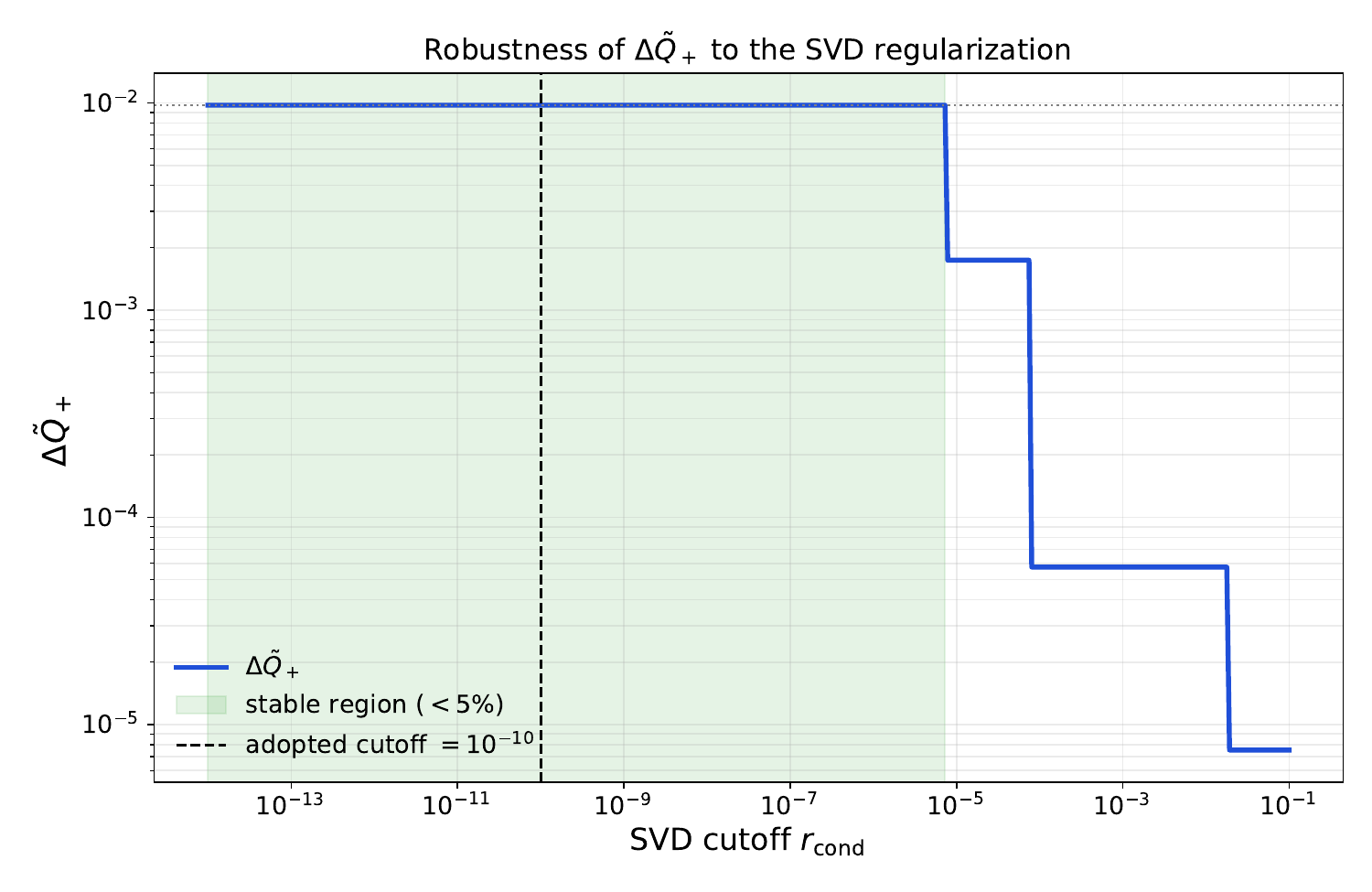}
\caption{Robustness of the inferred $\Delta\tilde{Q}_+$ to the SVD regularization cutoff $r_{\rm cond}$, computed on a well-conditioned Fisher matrix. The cutoff $r_{\rm cond}$ is the relative threshold below which singular values are discarded in the pseudo-inverse. The estimate is flat to within $5\%$ (green band, the stable region) from $r_{\rm cond}=10^{-14}$ to $\sim10^{-5}$, with the adopted cutoff $10^{-10}$ (dashed line) well inside it. Only for $r_{\rm cond}\gtrsim10^{-5}$, where singular values begin to be discarded, does the estimate fall to small values.}
\label{fig:svd_robustness}
\end{figure}

In practice, the inversion of the Fisher matrix requires particular care. Because $\tilde \gamma$ is often ill-conditioned due to strong correlations among parameters, as is the case in our model, the direct inversion can be numerically unstable. To address this, we compute the covariance matrix using a pseudoinverse based on the singular-value decomposition (SVD), introducing a small cutoff to regularize the inversion and suppress poorly constrained directions in parameter space. In particular, singular values below a fixed threshold relative to the largest singular value are discarded,  stabilizing in this way the inversion. 
We have also verified that $\Delta Q_+$ is stable under changes of the SVD cutoff. On a well-conditioned Fisher matrix it varies by less than $5\%$ as $r_{\rm cond}$ goes from $10^{-14}$ to $10^{-5}$ (see Fig.~\ref{fig:svd_robustness}), and our adopted cutoff $10^{-10}$ lies well inside this stable region.

We performed a Fisher analysis in which we first fix the main parameters $(M, \mu, S, e_\mathrm{LSO}, \lambda, \tilde \gamma_\mathrm{LSO})$, following~\cite{Barack:2003fp}, and subsequently introduce the multipole parameters $(Q, Q_+, \psi_Q, \mathcal{O}, \mathcal{O}_+, \psi_{\mathcal{O}})$ one at a time. We find that the constraints on the main parameters are robust as they are not significantly affected by the inclusion or exclusion of the multipole parameters describing non-Kerr deviations, changing only by numerical factors while remaining at the same order of magnitude.

In contrast, as additional multipole moment parameters are incorporated, correlations within the multipole sector increase, leading to growing degeneracies in the Fisher matrix. 
This behavior indicates that the dominant degeneracies are largely confined to the subspace spanned by the multipole moment sector of the parameter space. As expected, these correlations become more pronounced for higher-order multipole moments, with the octupole sector exhibiting stronger degeneracies than the quadrupole sector.

Importantly, the masses and, most notably, the parameter of main interest, namely the amplitude of the nonaxisymmetric mass quadrupole, $Q_+$, belong to a comparatively well-conditioned subset of the parameter space. 
As a result, their inferred uncertainties are stable under the inclusion of additional multipole parameters, and the corresponding Fisher-matrix inversion yields robust and reliable estimates.

\section{LISA Constraints on Nonaxisymmetry} \label{Sec: results}
We now present the results of our parameter-estimation analysis for EMRI systems in which the primary deviates from the standard Kerr geometry. The analysis is based on the framework described in the previous sections and uses one year of synthetic LISA data, corresponding to the final stage of the inspiral prior to the plunge of the secondary compact object into the central supermassive body, the primary.

For consistency and ease of comparison, we normalize all results to a SNR of $30$. This choice corresponds to a conservative (pessimistic) scenario, given that typical EMRI signals are expected to achieve SNRs in the range $\sim 30$-$300$. Consequently, the parameter uncertainties presented below, which already demonstrate a high level of precision, can be regarded as conservative estimates, and can be expected to improve substantially for higher-SNR events in more optimistic scenarios.

As a first step in our analysis, we investigate how the constraint $\Delta Q_+$, which quantifies deviations from axial symmetry, depends on the system parameters $(M, \mu, e_{\mathrm{LSO}}, S)$. To this end, we have performed a series of simulations in which these quantities are systematically varied. Specifically, we consider:
\begin{itemize}
\item $M \in (10^5, 10^6)\, M^{}_{\odot}\,$,
\item $\mu \in (1, 10, 50, 100)\,M^{}_{\odot}\,$,
\item $e^{}_{\mathrm{LSO}} \in (0.01, 0.05, 0.1, 0.15, 0.2, 0.25, 0.3)\,$,
\item $S \in (0, 0.25, 0.5, 0.75)\,$.
\end{itemize}
The angular parameters $(\theta_S, \phi_S, \theta_K, \phi_K)$ are held fixed to the values specified in Table~\ref{tableparameters}. 

\begin{table}[ht]
\centering
\caption{Projected $1\sigma$ measurement uncertainties for the EMRI parameters, normalized to $\mathrm{SNR}=30$, for the reference configuration $M=10^{6} \,M_{\odot}$, $\mu=10 \,M_{\odot}$, $\tilde S=0.25$, $e_{\mathrm{LSO}}=0.1$, inclination $\lambda=\pi/3$, orbital phases $\alpha_{\mathrm{LSO}}=\Phi_{\mathrm{LSO}}=0$ and $\tilde\tilde \gamma_{\mathrm{LSO}}=\pi/2$, and sky/spin angles $(\theta_S,\phi_S,\theta_K,\phi_K)=(2\pi/3,5\pi/3,\pi/2,0)$.}
\begin{tabular}{l|c|c|c|c}
\toprule
Parameter & $\Delta(\ln \mu)$ & $\Delta e_\mathrm{LSO}$ & $\Delta(\cos \lambda)$ & $\Delta \tilde \gamma_{\mathrm{LSO}}$ \\
\midrule
Constraint & $9.1 \times 10^{-3}$ & $5.6 \times 10^{-3}$ & $1.3 \times 10^{-2}$ & $4.3 \times 10^{-2}$ \\
\toprule
Parameter & $\Delta(\ln M)$ & $\Delta S$ & $\Delta Q$ & $\Delta \mathcal{O}$ \\
\hline
Constraint & $5.4 \times 10^{-5}$ & $1.2 \times 10^{-3}$ & $1.1 \times 10^{-2}$ & $5.5 \times 10^{-1}$  \\
\toprule
Parameter & \multicolumn{2}{c|}{\bm{$\Delta Q_+$}} & \multicolumn{2}{c}{$\Delta \mathcal{O}_+$} \\
\hline
Constraint & \multicolumn{2}{c|}{\bm{$1.1 \times 10^{-2}$}} & \multicolumn{2}{c}{$5.0 \times 10^{-2}$} \\
\bottomrule
\end{tabular}
\label{tableparameters}
\end{table}

We have explored the dependence of the parameter constraints on the angular configuration by varying $(\theta_S, \phi_S, \theta_K, \phi_K)$ and the inclination $\lambda$. In particular, we have considered the following cases:
\begin{itemize}
\item $\theta_S \in (\pi/2, 2\pi/3, \pi/6)\,$, 
\item $\phi_S \in (5\pi/3, 2\pi/3)\,$, 
\item $\theta_K \in (\pi/20, 3\pi/4, \pi/2)$, and 
\item $\phi_K \in (0, \pi/2, \pi)$, together with 
\item $\lambda \in (\pi/3, \pi/6, 2\pi/3, 5\pi/6)\,$. 
\end{itemize}
These configurations are consistent with those explored in~\cite{Barack:2006pq, Fransen:2022jtw}. 
For the inclinations considered here, $\lambda\in[\pi/6,5\pi/6]$, the constraints (see Table~\ref{tableparameters}) show only small numerical changes and keep the same order of magnitude. This indicates that, in this range, the measurement accuracy is set mainly by the intrinsic properties of the source and not by its orientation. This near-independence should not, however, be extended to the aligned, nonprecessing limit $\lambda\to0,\pi$. The work of~\cite{Loutrel:2022ant,Loutrel_2024} shows that the detectability of the nonaxisymmetric quadrupole is driven by the conservative modulation of the waveform produced by the precession, and that it degrades quickly towards that limit, where the non-axisymmetry parameters become strongly degenerate. In the EMRI case, pericenter and Lense-Thirring precession are always present, so our configurations lie well inside the precessing regime.

\begin{figure}
    \centering
    \includegraphics[width=1.0\linewidth]{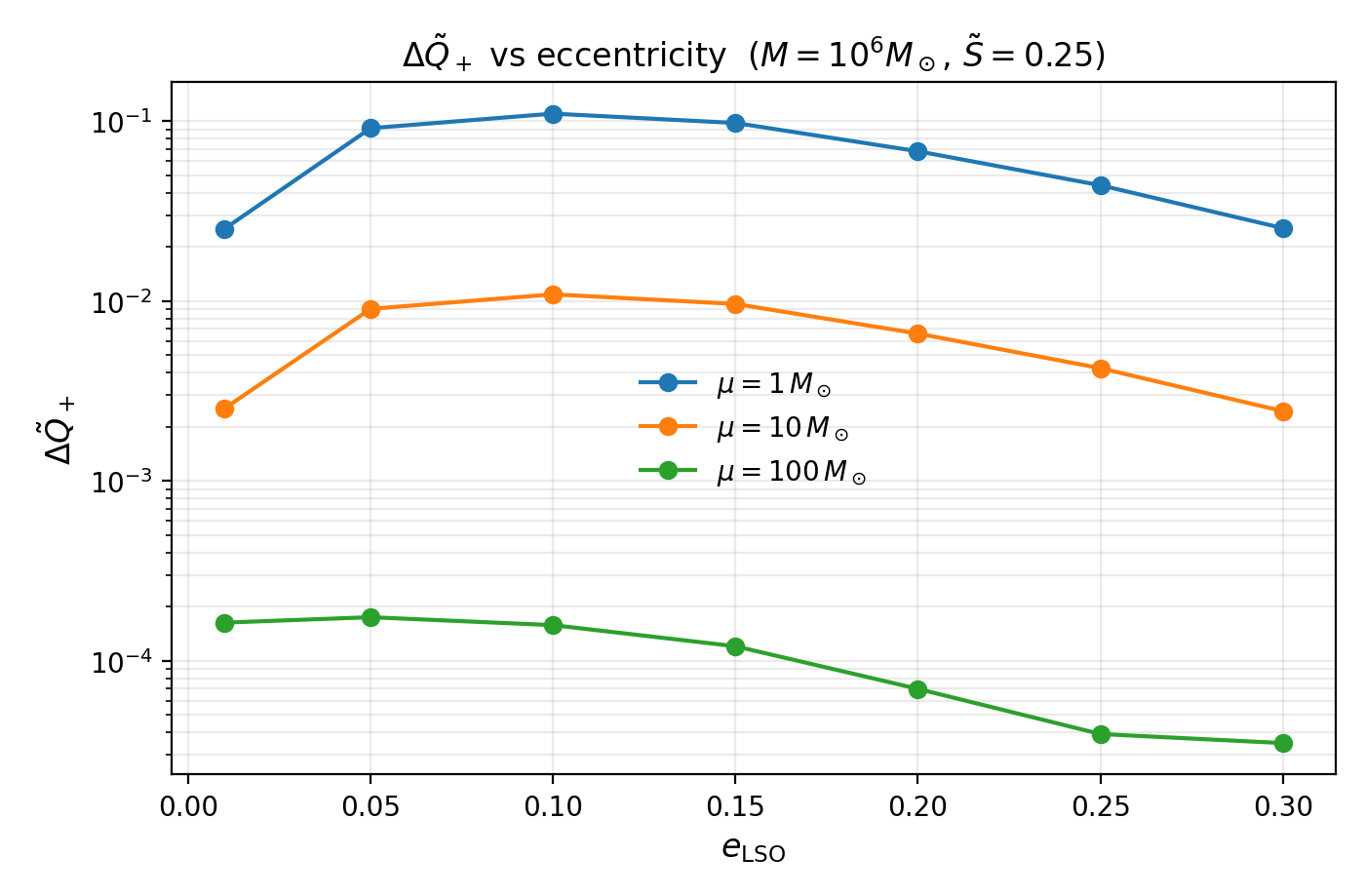}
    \caption{Measurement accuracy $\Delta \tilde{Q}_+$ for detecting axial symmetry breaking, shown as a function of the parameters $\mu$ and $e_{LSO}$. The mass is fixed at $M = 10^6 \,M_{\odot}$ with a spin of $\tilde{S} = 0.25$. All other parameters are held fixed as shown in Table \ref{tableparameters}.}
    \label{fig:q_vs_e}
\end{figure}

In Fig.\ref{fig:q_vs_e}, we show the dependence of $\Delta Q_+$ on the eccentricity $e_{\mathrm{LSO}}$ for different values of the secondary mass, $\mu$. Although the effect is modest, larger eccentricities lead to slightly tighter constraints, without altering the overall scaling. A similar trend has been reported for the axisymmetric quadrupole $\Delta Q$ in~\cite{Barack:2006pq}, and can be understood as a consequence of the enhanced harmonic content of more eccentric orbits, which improves parameter discrimination.

\begin{figure*}
    \centering
    \includegraphics[width=1.0\linewidth]{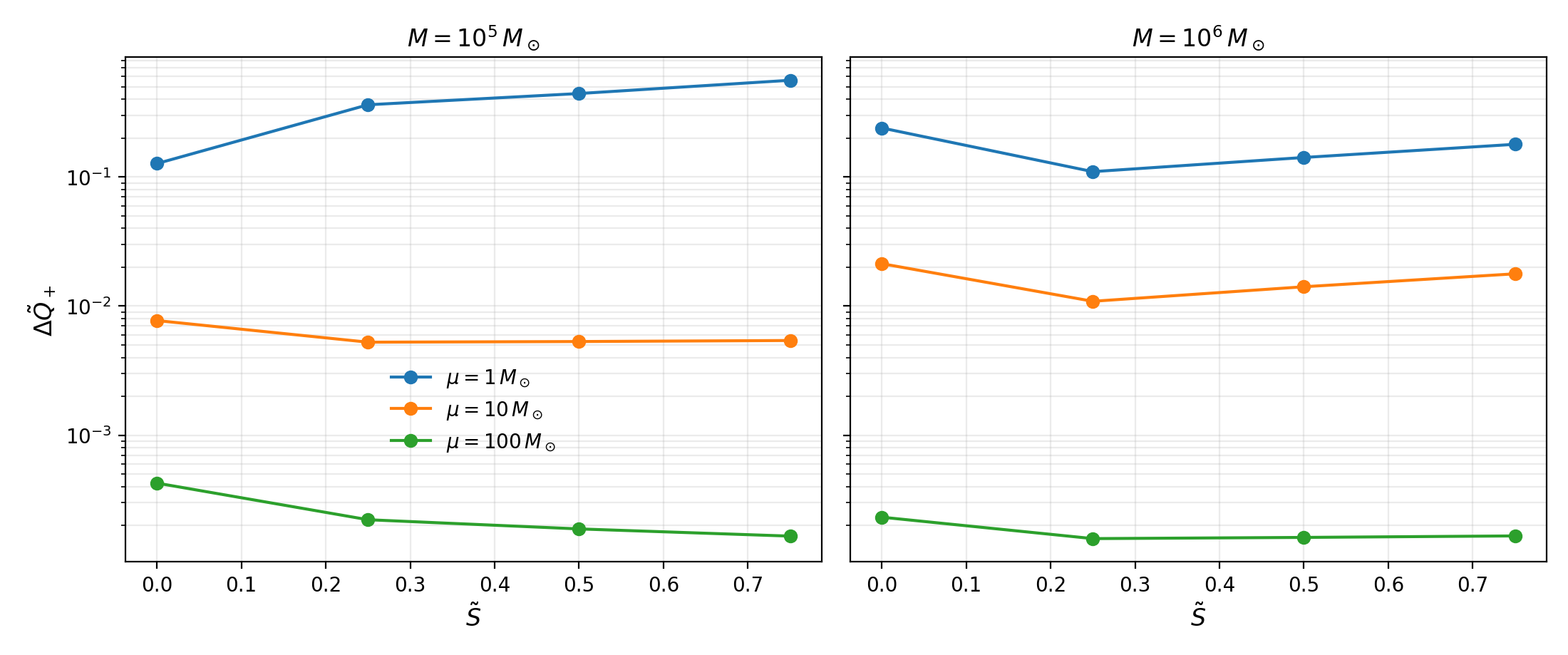}
    \caption{Measurement accuracy of $\Delta \tilde{Q}_+$ shown as a function of the spin $\mu$ and $S$. The left plot corresponds to a mass of $M = 10^5 \,M_{\odot}$, while the right plot shows results for $M = 10^6 \,M_{\odot}$. The eccentricity parameter is fixed at $e_{LSO} = 0.1$. All other parameters are held fixed as shown in Table \ref{tableparameters}.}
    \label{fig:q_vs_S}
\end{figure*}

The impact of the spin $S$ on $\Delta Q_+$ is similarly very weak. As shown in Fig.~\ref{fig:q_vs_S}, varying the spin produces only minor changes in the inferred uncertainty, indicating that the sensitivity to nonaxisymmetric quadrupolar structure is not strongly coupled to the spin sector within the range of values considered here.

Regarding the mass dependence, the constraint tightens strongly as the secondary mass $\mu$ increases and, more weakly, as the primary mass $M$ increases (Figs.~\ref{fig:q_vs_e} and~\ref{fig:q_vs_S}). Both trends are governed by the radiation-reaction timescale and the accumulated number of waveform cycles, which set the information content of the signal (see Eqs.~\eqref{frequency} and~\eqref{eccentricity}).

The dependence on the primary mass $M$ is comparatively mild, as shown in Fig.~\ref{fig:q_vs_S}. On the other hand, variations in the secondary mass $\mu$ have a significantly stronger impact on the constraint on $Q_+$. As shown in Fig.~\ref{fig:q_vs_e}, increasing $\mu$ by one order of magnitude leads to an improvement of approximately two orders of magnitude in the constraint. In particular, the results for $\mu = 10 \,M_{\odot}$ are about two orders of magnitude tighter than those for $\mu = 1 \,M_{\odot}$, and a similar scaling is observed when comparing $\mu = 100 \,M_{\odot}$ to $\mu = 10 \,M_{\odot}$. This strong dependence reflects the increased amplitude of the GW signal and the enhanced information content associated with more massive secondaries.

This implies that stellar-origin BHs (SOBHs), which typically have masses in the range $\sim 5$-$150 \,M_{\odot}$, can probe the multipolar structure of the primary with significantly higher accuracy. Consequently, such systems provide more powerful tests of deviations from GR in the EMRI regime than systems involving neutron stars ($\sim 1.1$-$2.1 \,M_{\odot}$) or white dwarfs ($\sim 0.2$-$1.44 \,M_{\odot}$), whose lower masses lead to weaker constraints.

Our results further show that the constraint on the nonaxisymmetric quadrupole, $Q_+$, is roughly one-to-two orders of magnitude smaller than that on the axisymmetric octupole, $\mathcal{O}$, as shown in Table~\ref{tableparameters}. This indicates that LISA will be considerably more sensitive to deviations from axial symmetry than to higher-order deformations associated with equatorial symmetry breaking. The main reason is that, within the mass-multipole sector considered in this work, axial symmetry is already broken at quadrupole order, by $Q_+$, whereas equatorial symmetry is broken only at octupole order, by the odd-parity mass octupole $\mathcal{O}$. Axial-symmetry breaking is therefore probed at a lower multipole order, and is naturally better constrained.

In addition, the axial deviation enters through $Q_+\cos(2\psi_Q)$, which is maximal in the symmetric configuration and therefore leaves a clear imprint on the waveform. The equatorial odd-parity octupole, instead, enters through $\mathcal{O}\,\sin\gamma$, which vanishes in the nonprecessing limit and only becomes measurable once the orbit precesses.

It is also worth noting that the nonaxisymmetric octupole $\mathcal{O}_+$ is better constrained ($\Delta\mathcal{O}_+\simeq5\times10^{-2}$) than the axisymmetric octupole $\mathcal{O}$ ($\Delta\mathcal{O}\simeq5\times10^{-1}$), even though it breaks an additional symmetry. The reason behind this is related to the strength of the coupling. The component $\mathcal{O}_+$ enters the GW fluxes with a much larger coefficient. In the leading term of the energy flux [Eq.~\eqref{TAfluxE}] the effective prefactors are $16880/512\approx33$ for $\mathcal{O}_+$ and $2896/1024\approx2.8$ for $\mathcal{O}$. Their ratio, about $12$, is close to the ratio $\Delta\mathcal{O}/\Delta\mathcal{O}_+\approx11$ that we observe. This happens because the $m=2$ azimuthal modulation of $\mathcal{O}_+$ couples directly to the orbital phase, while the $m=0$ odd-parity structure of $\mathcal{O}$ leaves a weaker imprint.

Moreover, across a broader region of the parameter space, we find that $\Delta Q_+$ yields constraints comparable to those obtained for the axisymmetric quadrupole $\Delta Q$. This indicates that tests of axial symmetry are well within the reach of EMRI observations and do not pose a significant challenge. By contrast, the measurement of higher-order multipoles such as $\mathcal{O}$ lies closer to the sensitivity threshold, where parameter correlations and reduced signal amplitude begin to degrade accuracy.
Furthermore, the octupole enters the secular precession only through terms proportional to $\sin\tilde\gamma$, so in a kludge treatment the result depends on how $\tilde\gamma$ is treated. We obtain $\Delta\mathcal{O}\simeq0.5$ when $\sin\tilde\gamma$ is evaluated at the fiducial value $\tilde\gamma_0=\pi/2$, but the effect largely averages out when $\tilde\gamma$ precesses freely. For this reason, we quote the octupole numbers as indicative, and we emphasize that the robust, model-independent result of this work is the constraint on axial-symmetry breaking, $\Delta Q_+\sim10^{-2}$.

On the other hand, the orientation angles $\psi_Q$ and $\psi_{\mathcal O}$ of the nonaxisymmetric deformations cannot be measured independently. Since $Q_+$ enters the secular dynamics only through the combination $Q_+\cos(2\psi_Q)$, and $\mathcal{O}_+$ only through $\mathcal{O}_+\cos(3\psi_{\mathcal O})$, each phase is completely degenerate with its amplitude, and the Fisher analysis gives an essentially unbounded $\Delta\psi_Q$. What LISA constrains is the effective amplitude $Q_+\cos(2\psi_Q)$ of the nonaxisymmetric deformation, and not its orientation. For this reason we do not include $\psi_Q$ and $\psi_{\mathcal O}$ in Table~\ref{tableparameters}.

As discussed in Sec.~\ref{Sec: parameterspace}, our baseline analysis assumes vanishing multipole deformation parameters, that is, we recover the Kerr limit. To assess the impact of non-Kerr signatures, we also consider configurations in which these dimensionless parameters are set to unity, effectively modeling a deformed (non-Kerr) central object (the primary). In this case, setting $\tilde{Q}_+ = 1$ leads, for representative configurations such as $M = 10^6 \,M_{\odot}$ and $\mu = 10 \,M_{\odot}$, to a modest, order-unity improvement in the constraint on $\tilde{Q}_+$. Similar trends are observed for other multipole moments across the parameter space.

Finally, we have explored the possibility of treating $Q_+$ and $\mathcal{O}+$ as a single effective parameter encoding axial symmetry breaking. This approach is analogous to that adopted in~\cite{Fransen:2022jtw} for equatorial symmetry breaking, where the current quadrupole $S_2$ and the mass octupole $\mathcal{O}$ were linked through a common parametrization. Specifically, the authors assumed $\Tilde{M}_1 = \Tilde{S}_2 = \Tilde{M}_3$, thereby parametrizing equatorial symmetry breaking with a single parameter.
Following the same strategy, we impose $\tilde{Q}_+ = \tilde{\mathcal{O}}_+$. We find that our results remain unchanged to within $\sim10\%$. However, when $Q_+$ and $\mathcal{O}_+$ are measured independently, the constraint on $Q_+$ is consistently tighter over a wide region of the parameter space. This indicates that $Q_+$ alone provides a sufficiently sensitive probe of axial symmetry breaking, and can therefore serve as the primary parameter in future analyses.

\section{Conclusion and Discussion} \label{Sec: conclusion}

In this work, we have shown that LISA observations of EMRIs will be capable of constraining axial-symmetry breaking with high accuracy, at the level of $10^{-4}$-$10^{-2}$.
EMRIs therefore provide a uniquely powerful tool for testing the nature of BHs by probing their strong-field gravity. 

These estimates are obtained using a kludge-based waveform model combined with a Fisher-matrix analysis, and are largely independent of the central object (EMRI primary) mass and spin for EMRIs with SNR$=30$. Since EMRI signals are expected to reach SNRs in the range $\sim 30$-$300$, stronger (higher-SNR) sources should allow for considerably improved constraints. In particular, for a favorable “golden” EMRI, we can anticipate especially stringent tests that may challenge our current understanding of BHs (the Kerr paradigm) and other possible (exotic) compact objects. 

From a theoretical perspective, the parameters $Q_+$ and $\mathcal{O}_+$ provide direct measures of symmetry breaking in the multipolar structure of the central object. This makes them especially relevant for testing exotic compact-object models. For instance, \emph{fuzzball} geometries, a prime example of exotic compact object motivated by string theory, are expected to violate both axial and equatorial symmetries, and would therefore manifest through nonvanishing values of these parameters (see~\cite{Muguruza:2026zcs} for a detailed analysis and discussion).
In this sense, EMRIs offer a promising observational window into possible deviations from the Kerr solution and, more generally, from GR 

There are different avenues to continue this work that can also complement other works in the literature going in this line (see~\cite{Cardenas-Avendano:2024} and references therein). 
These extensions include both methodological developments towards realizing tests of fundamental physics with the future LISA EMRI observations~\cite{Sopuerta:2010fte} and also developments that incorporate additional physical effects that we have not yet been included or have not been studied enough. They include from further modifications to the geometry of the primary, modifications to the geometry of the secondary, environmental effects, and beyond GR modifications.

A natural extension of the analysis presented here is to investigate a broader class of nonaxisymmetric deformations, such as multipolar configurations with different azimuthal indices (e.g., $m=1$ modes).  
Since axial symmetry is broken whenever $m \neq 0$, one may expect constraints of comparable order to those obtained here for $m=2$, although a more detailed analysis is required to quantify potential differences in the EMRI waveform.

On the other hand, our analysis, as previous studies in this line~\cite{Barack:2006pq, Fransen:2022jtw}, adopts a Schwarzschild cutoff (LSO) as the evolution termination criterion. 
As discussed earlier in Sec.~\ref{Sec: equations}, employing a Kerr LSO is expected to yield tighter constraints, particularly for rapidly spinning primaries. Quantifying the impact of this choice within the present framework constitutes an important direction for future work.

Further progress will require advances in EMRI waveform modeling~\cite{Khalvati:2024tzz,Speri:2026ade} and data-analysis techniques (see~\cite{Khalvati:2025znb,Strub:2025dfs,Cole:2025sqo}). 
On the one hand, kludge-based approaches, such as those employed here and other works in the literature, provide a computationally efficient framework that captures the essential physics and is well-suited for exploratory studies of parameter sensitivity to assess the scientific capabilities of space-based detectors like LISA to address revolutionary fundamental physics questions like the nature of BHs and the validity of GR. 
These models can be systematically improved by incorporating additional physical effects (e.g., higher multipoles, environmental interactions, beyond-GR corrections, etc.) and by refining their treatment of the EMRI relativistic dynamics.
Examples include the numerical kludge (NK)~\cite{Babak:2006uv} (see also~\cite{Sopuerta:2011te,Sopuerta:2012de} for a similar formalism that uses an approximate post-Minkowkian self-force), which employs Kerr geodesics, and augmented AK (AAK) models, which incorporate relativistic frequency information from Kerr geodesics to improve phase accuracy.
One can go beyond this and use relativistic waveform models based on BH perturbation theory, which offer higher fidelity descriptions of EMRI signals. In particular, recent developments such as the fast EMRI waveforms (FEW) framework~\cite{Chua:2020stf, Katz:2021yft, Speri:2023jte, Chapman-Bird:2025xtd} enable accurate waveform generation at significantly reduced computational cost. Incorporating such models, together with Bayesian inference techniques, will be very relevant to validate and refine the conclusions presented here.

Our work in this paper considers a wide range of possible deviations from the Kerr paradigm in EMRIs by modifying the multipolar structure of the primary, but we have not explicitly introduced modifications to GR nor environmental effects. Regarding non-GR modifications, there is a extensive literature where EMRIs are used for tests of modifications to GR~\cite{Gair:2011ym,Kumar:2024utz}, sometimes including parameter-estimation studies in connection to LISA and other space-based GW projected observatories~\cite{Zi:2021pdp}, such as the presence of fundamental scalar~\cite{Maselli:2020zgv,Maselli:2021men,DellaRocca:2024pnm,Speri:2024qak,Zi:2025lio} and vector fields~\cite{Zi:2025qos}, scalar Gauss-Bonnet gravity~\cite{Zi:2026zpw}, Chern-Simons modified gravity~\cite{Sopuerta:2009iy,Pani:2011xj,Canizares:2012he}, scalar-tensor theories~\cite{Yunes:2011aa}, combinations of higher-curvature terms~\cite{Daniel:2024lev}, extra spatial dimensions~\cite{Rahman:2022fay,Zi:2024dpi,Kumar:2025jsi}, etc. 

On the other hand, environmental effects are becoming more and more important in GW astronomy as the precision of future detectors will increase and we need to account for effects that can degrade such precision~\cite{Barausse:2014tra,Zwick:2022dih,Cole:2022yzw,Khalvati:2024tzz,Shen:2025svs}. There many studies exploring the impact of accretion disks and matter distributions around the primary~\cite{Lyu:2024gnk,Li:2025zgo,Speri:2022upm,Duque:2024mfw,HegadeKR:2025dur,HegadeKR:2025rpr,Vicente:2025gsg,Duque:2025yfm,Strateny:2025zkd,Zeng:2026ydj}, including several forms of dark matter~\cite{Brito:2023pyl, Zhang:2024ugv,Gliorio:2025cbh,Mitra:2025tag,Li:2025ffh,Dyson:2025dlj,Kakehi:2025peb,Das:2025eiv,Feng:2025fkc,Zhao:2026yis,Li:2026uva}. 

An important open question that we need to address is the degeneracy between deviations from the geometry of the primary/secondary and/or GR and environmental effects~\cite{Kejriwal:2023djc,Zi:2026zpw}, that is, how to disentangle beyond-GR signatures from astrophysical elements (e.g., accretion disks, dark-matter spikes, etc.) that may be present around an EMRI and affect its evolution and GW emission. 
While further study is required to address this problem in depth, one can argue that such environmental effects are expected to vary from source to source, whereas genuine deviations from the Kerr geometry would be in principle universal. 
Discriminating these possibilities will likely require a systematic population-based analysis combining multiple EMRI observations. 
However, since the LISA EMRI event rates remains uncertain~\cite{Babak:2017tow} (see also~\cite{Mancieri:2025cmx,Zhang:2025jmm}), it is crucial to address this open question in future investigations.

Ultimately, high-precision EMRI waveforms are needed for the scientific exploitation of the future LISA data. They can be obtained within the self-force program based on BH perturbation theory~\cite{Barack:2018yvs}, but this is still under development for the standard EMRI scenario. Therefore, there are important challenges and a lot of work to be done to incorporate the physical effects that we have studied in this paper into the waveform of the self-force program. Nevertheless, it is very important to make progress in this direction to ensure that during the LISA era, revolutionary research and discoveries in fundamental physics will be made.

In summary, our results highlight the extraordinary potential of EMRIs as precision probes of the multipolar structure of compact objects. With the advent of the LISA mission, these systems may provide strong direct evidence of departures from the Kerr hypothesis, offering a unique observational window onto the fundamental nature of gravity in the strong-field regime. 
Nevertheless, in order to realize the enormous potential of EMRI observations with LISA more developments on the modeling front are needed.

\begin{acknowledgments}
We thank Iv\'an Mart\'{\i}n V\'{\i}lchez and Alejandro C\'ardenas-Avenda\~no  for fruitful discussions and valuable insights. P.F.M. and C.F.S. are supported by Contracts No. PID2022-137674NB-I00/AEI/10.13039/501100011033 (Spanish Ministry of Science and Innovation) and No. 2017-SGR-1469 (AGAUR, Generalitat de Catalunya). This work was also partly supported by the Spanish program Unidad de Excelencia María de Maeztu CEX2020-001058-M, financed by MCIN/AEI/10.13039/501100011033, and by the MaX-CSIC Excellence Award MaX4-SOMMA-ICE. P.F.M. also acknowledges financial support from Spanish Grant No. PRE2022-101913 funded by Spanish Ministry of Science and Innovation.
\end{acknowledgments}

\begin{center}
\textbf{DATA AVAILABILITY}
\end{center}

There are no publicly available research data or software supporting this manuscript. Requests for further information or data should be sent to the authors.

\appendix

\section{CONSTANTS OF MOTION AND GW FLUXES} \label{derivationapp}
The constants of motion of the EMRI dynamical system we have developed above can be written in terms of the orbital parameters as follows: 

\begin{widetext}
\begin{eqnarray}
L & = & \sqrt{M p} \Bigg[ \Bigg. 1 + \frac{2 S \cos\lambda}{\sqrt{M} p^{3/2}} \frac{(1 + 3 e^2)}{(1 + e^2)} \nonumber \\[2mm]
& - & \frac{Q}{2 p^2}(1 - 3 \cos^2\theta)\frac{(1 + 3 e^2)}{(1 + e^2)} 
+ \frac{3 Q^{}_{+}}{2 p^2} \sin^2\theta \cos\left(2 \psi^{}_Q\right)\frac{(1 + 3 e^2)}{(1 + e^2)} \nonumber \\[2mm]
& + &\frac{\mathcal{O}}{2 p^3}(5 \cos^3\theta - 3 \cos\theta)\frac{(1 + 6 e^2 + e^4)}{(1 + e^2)} + \frac{15 \mathcal{O}^{}_{+}}{2 p^3} \sin^2\theta \cos\theta \cos\left(3 \psi^{}_{\mathcal{O}}\right) \frac{(1 + 6 e^2 + e^4)}{(1 + e^2)} \Bigg. \Bigg] \,, \label{Ldef}
\end{eqnarray}

\begin{eqnarray}
E & = & -\frac{M(1 - e^2)}{2 p} \Bigg[ \Bigg. 1 + \frac{4 S \cos\lambda}{\sqrt{M} p^{3/2}} \frac{(1 + 3 e^2)}{(1 - e^2)} \nonumber \\[2mm]
& - & \frac{Q}{p^2}(1 - 3 \cos^2\theta)\frac{(1 + 3 e^2)}{(1 - e^2)} + \frac{3 Q^{}_{+}}{p^2} \sin^2\theta \cos\left(2 \psi^{}_Q\right)\frac{(1 + 3 e^2)}{(1 - e^2)} \nonumber \\[2mm]
& + & \frac{\mathcal{O}}{p^3}(5 \cos^3\theta - 3 \cos\theta)\frac{(1 + 6 e^2 + e^4)}{(1 - e^2)} + \frac{15 \mathcal{O}^{}_{+}}{p^3} \sin^2\theta \cos\theta \cos\left(3 \psi^{}_{\mathcal{O}}\right)\frac{(1 + 6 e^2 + e^4)}{(1 - e^2)} \Bigg. \Bigg] \,,
\label{Edef}
\end{eqnarray}

\begin{eqnarray}
L^{}_z & = & \sqrt{M p} \cos\lambda \Bigg[ \Bigg. 1 + \frac{2 S \cos\lambda}{\sqrt{M} p^{3/2}} \frac{(1 + 3 e^2)}{(1 + e^2)} \nonumber \\[2mm]
& - & \frac{Q}{2 p^2}(1 - 3 \cos^2\theta)\frac{(1 + 3 e^2)}{(1 + e^2)} 
+ \frac{3 Q^{}_{+}}{2 p^2} \sin^2\theta \cos\left(2 \psi_Q\right)\frac{(1 + 3 e^2)}{(1 + e^2)} \nonumber \\[2mm]
& + & \frac{\mathcal{O}}{2 p^3}(5 \cos^3\theta - 3 \cos\theta)\frac{(1 + 6 e^2 + e^4)}{(1 + e^2)} + \frac{15 \mathcal{O}^{}_{+}}{2 p^3} \sin^2\theta \cos\theta \cos\left(3 \psi_{\mathcal{O}}\right) \frac{(1 + 6 e^2 + e^4)}{(1 + e^2)} \Bigg. \Bigg] \,.
\label{Lzdef}
\end{eqnarray}
\end{widetext}

We have also computed the instantaneous fluxes associated with the constants of motion, keeping only first-order contributions in $O(S)$, $O(Q)$, $O(Q_+)$, $O(\mathcal{O})$, and $O(\mathcal{O}_{+})$,

\begin{widetext}
\begin{eqnarray}
\frac{d E}{d t} & = & - \frac{8}{15 r^6} \mu^2 (11 L^2 M^2 + 2 M^3 r + 2 E M^2 r^2) \nonumber \\[2mm] 
& + & \frac{16 S L^{}_z}{15 r^8} \mu^2 \left(2 M^2 r - 89 L^2 M + 22 E M r^2 \right) \nonumber \\[2mm]
& + & \frac{4 Q}{15 r^8} \mu^2 \left[-39 L^2 M^2 + 5 M^3 r + 6  E M^2 r^2 + (15 M^3 r - 117 L^2 M^2 + 18 E M^2 r^2) \cos(2\theta)  + 6 M^2 r^3 \sin(2\theta) \dot{r}\dot{\theta} \right] \nonumber \\[2mm]
& + & \frac{4 Q^{}_{+}}{5 r^8} \mu^2 \left[- 39 L^2 M^2 + 5 M^3 r + 6 E M^2 r^2 + (- 5 M^3 r + 39 L^2 M^2 + 6 E M^2 r^2) \cos(2\theta) - 2 M^2 r^3 \sin(2\theta) \dot{r}\dot{\theta} \right] \cos\left(2 \psi_Q\right) \nonumber \\[2mm]
& + & \frac{2 \mathcal{O}}{15 r^9} \mu^2 \left[ \left(- 168 L^2 M^2 + 45 M^3 r + 48 E M^2 r^2\right) \cos(\theta) + \left(- 280 L^2 M^2 + 75 M^3 r + 80 E M^2 r^2\right) \cos(3 \theta) \right] \nonumber \\[2mm]
& + & \frac{4 \mathcal{O}}{5 r^9} \mu^2 \left(M^2 r^3 \sin(\theta) + 5 M^2 r^3 \sin(3 \theta) \right) \dot{r}\dot{\theta} \nonumber \\[2mm]
& + & \frac{4 \mathcal{O}^{}_{+}}{15 r^9} \mu^2 \left[ \left(- 1680 L^2 M^2 + 450 M^3 r + 480 E M^2 r^2\right) \cos(\theta) \sin^2(\theta) + 15 r^3 \left(\sin\theta - 3 \sin[3 \theta]\right) \dot{r}\dot{\theta} \right] \cos\left(3 \psi_{\mathcal{O}}\right) \,,
\label{fluxE}
\end{eqnarray}
\begin{eqnarray}
\frac{d L^{}_z}{d t} & = & - \frac{8 L^{}_z}{5 r^5} \mu^2 \left(3 L^2 M - 2 E M r^2\right) \nonumber \\[2mm] 
& + & \frac{4 S}{5 r^7} \mu^2 \left(-15 L^4 - 68 L^2 L_z^2 M + 5 L^2 M r + 34 L_z^2 M r + 18 E L^2 r^2 + 4 E M r^3 \right) \nonumber \\[2mm]
& + & \frac{4 S}{5 r^7} \mu^2 \left[ - \left(3 L^4 + L^2 M r + 6 E L^2 r^2 + M^2 r^2 + 4 E M r^3\right) \cos(2\theta) + \left(3 L^2 r^3 + 2 M r^4 + 6 E r^5\right) \sin(2\theta) \dot{r}\dot{\theta}  \right] \nonumber \\[2mm]
& + & \frac{2 Q L^{}_z}{5 r^7} \mu^2 \left[ -15 L^2 M + 8 M^2 r + 18 E M r^2 + \left(24 M^2 r - 45 L^2 M + 54 E M r^2\right) \cos(2\theta)  + 6 M r^3 \sin(2\theta) \dot{r}\dot{\theta} \right] \nonumber \\[2mm]
& + & \frac{6 Q^{}_{+} L^{}_z}{5 r^7} \mu^2 \left[- 15 L^2 M + 8 M^2 r + 18 E M r^2 + \left(15 L^2 M - 8 M^2 r - 18 E M r^2\right) \cos(2\theta) - 2 M r^3 \sin(2\theta) \dot{r}\dot{\theta} \right] \cos\left(2 \psi_Q\right) \nonumber \\[2mm]
& + & \frac{2 \mathcal{O} L^{}_z}{5 r^8} \mu^2 \left[ \left(27 M^2 r - 36 L^2 M + 48 E M r^2\right) \cos(\theta) + \left(- 60 L^2 M + 45 M^2 r + 80 E M r^2\right) \cos(3 \theta) \right] \nonumber \\[2mm]
& + & \frac{2 \mathcal{O} L^{}_z}{5 r^8} \mu^2 \left(3 M r^3 \sin(\theta) + 15 M r^3 \sin(3 \theta) \right) \dot{r}\dot{\theta} \nonumber \\[2mm]
& + & \frac{2 \mathcal{O}^{}_{+} L^{}_z}{5 r^8} \mu^2 \left[ \left(- 720 L^2 M + 540 M^2 r + 960 E M r^2\right) \cos(\theta) \sin^2(\theta) + 15 M r^3 \left(\sin\theta - 3 \sin[3 \theta]\right) \dot{r}\dot{\theta} \right] \cos\left(3 \psi^{}_{\mathcal{O}}\right) \,.
\label{fluxLz}
\end{eqnarray}
\end{widetext}

We have also computed the time-averaged fluxes to first order in $S$, $Q$, $Q_+$, $\mathcal{O}$, and $\mathcal{O}_{+}$. The expressions we have obtained are

\begin{widetext}
\begin{eqnarray}
\Big\langle  \frac{d E}{d t} \Big\rangle & = & - \frac{32}{5} \frac{1}{p^5}\frac{(1 - e^2)^{3/2}}{ (1 + e^2)} M^{3} \mu^2 \Bigg[ \Bigg. 1 + \frac{97}{24}e^2 + \frac{329}{96}e^4 + \frac{37}{96}e^6 \nonumber \\[2mm] 
& + & \frac{S}{96 p^{3/2}} M^{-1} \left(1888 + 14408 e^2 + 21500 e^4 + 8215 e^6 + 379 e^8  \right) \cos\lambda \nonumber \\[2mm]
& + & \frac{Q}{384 p^{2}} \left(752 + 5952 e^2 + 9180 e^4 + 3585 e^6 + 177 e^8 \right) \sin^2\lambda \nonumber \\[2mm]
& - & \frac{Q}{1536 p^{2}}  \left(640 + 12784 e^2 + 23848 e^4 + 9953 e^6 + 537 e^8\right) \cos\left(2 \tilde \gamma\right) \sin^2\lambda \nonumber \\[2mm]
& + & \frac{Q^{}_{+}}{128 p^{2}} \left(664 + 5424 e^2 + 8355 e^4 + 3486 e^6 + 177 e^8 \right)  \sin^2\lambda \cos\left(2 \psi^{}_Q\right) \nonumber \\[2mm]
& + & \frac{Q^{}_{+}}{512 p^{2}} e^2 \left(9360 + 19176 e^2 + 9473 e^4 + 537 e^6\right)  \sin^2\lambda \cos\left(2 \psi^{}_Q\right) \cos\left(2 \tilde \gamma\right)  \nonumber \\[2mm]
& - & \frac{\mathcal{O}}{1024 p^{3}} e\, \left(2896 + 13932 e^2 + 15966 e^4 + 5311 e^6 + 205 e^8\right) (\sin\lambda + 5\sin[3\lambda]) \sin\tilde \gamma \nonumber \\[2mm]
& - & \frac{\mathcal{O}}{768 p^{3}} e^3 \left(17620 + 31595  e^2 + 13730  e^4 + 635  e^6\right) \sin^3\lambda \sin\left(3 \tilde \gamma\right) \nonumber \\[2mm]
& + & \frac{\mathcal{O}^{}_{+}}{512 p^{3}} e\, \left(16880 + 85800 e^2 + 92670 e^4 + 28295 e^6 + 1025 e^8\right) \left(5 + 3\cos\left(2\lambda\right)\right) \sin\tilde \gamma \cos\left(3 \psi^{}_{\mathcal{O}}\right) \nonumber \\[2mm]
& + & \frac{\mathcal{O}_{+}}{256 p^{3}} e^3 \left(18280 + 35555 e^2 + 14390 e^4 + 635 e^6 \right) \sin^2\lambda \sin\left(3\tilde \gamma\right) \cos\left(3 \psi^{}_{\mathcal{O}}\right) \Bigg. \Bigg] \,,
\label{TAfluxE}
\end{eqnarray}
\begin{eqnarray}
\Big\langle \frac{d L^{}_z}{d t} \Big\rangle & = & - \frac{32}{5} \frac{1}{p^{7/2}}\frac{(1 - e^2)^{3/2}}{ (1 + e^2)} M^{5/2} \mu^2 \cos\lambda \Bigg[ \Bigg. 1 + \frac{15}{8}e^2 + \frac{7}{8}e^4  \nonumber \\[2mm] 
& - & \frac{S}{32 p^{3/2}} M^{-1} \left[80 + 328 e^2 + 295 e^4 + 47 e^6 + \left(496 + 2148 e^2 + 1939 e^4 + 207  e^6\right) \cos^2\lambda \right] \sec\lambda \nonumber \\[2mm]
& + & \frac{S}{32 p^{3/2}} M^{-1} e^2 \left(28 + 55 e^2 + 27 e^4\right) \sin\lambda \tan\lambda \cos\left(2 \tilde \gamma\right) \nonumber \\[2mm]
& + & \frac{Q}{128 p^{2}} \left[ 448 + 1952 e^2 + 1770 e^4 + 186 e^6 - \left(672 + 1641 e^2 + 2655 e^4 + 279 e^6\right)\sin^2\lambda \right] \nonumber \\[2mm]
& - & \frac{Q}{64 p^{2}} \left(144 + 993 e^2 + 1029 e^4 + 120 e^6 \right) \cos\left(2 \tilde \gamma\right) \sin^2\lambda \nonumber \\[2mm]
& - & \frac{3 Q^{}_{+}}{256 p^{2}} \left(168 + 740 e^2 + 681 e^4 + 93 e^6\right)  \left(3 + \cos\left(2\lambda\right)\right) \cos\left(2 \psi^{}_Q \right) \nonumber \\[2mm]
& + & \frac{ \mathcal{O}}{1024 p^{3}} e\, \left(1668 + 5502 e^2 + 3525 e^4 + 435 e^6 \right) \left(6\sin\left(2\lambda\right) + 5\sin\left(4\lambda\right)\right) \sin\tilde \gamma \nonumber \\[2mm]
& + & \frac{ \mathcal{O}}{512 p^{3}} e^3 \left(3700 + 4935 e^2 + 995 e^4\right) \cos\lambda \sin^3\lambda \sin\left(3 \tilde \gamma\right) \nonumber \\[2mm]
& + & \frac{ \mathcal{O}^{}_{+}}{512 p^{3}} e\, \left(18000 - 41520 e^2 + 39465 e^4 + 6825 e^6 \right) \sin\tilde \gamma \sin^3\lambda \cos\left(3 \psi^{}_{\mathcal{O}}\right) \nonumber \\[2mm]
& - & \frac{ \mathcal{O}^{}_{+}}{64 p^{3}} e\, \left(3000 - 5220 e^2 + 8145 e^4 + 1485 e^6 \right)\sin\tilde \gamma \sin\lambda \cos\left(3 \psi^{}_{\mathcal{O}}\right) \nonumber \\[2mm]
& - & \frac{ \mathcal{O}^{}_{+}}{256 p^{3}} e^3 \left(10200 + 9405 e^2 + 2085 e^6\right) \cos\left(2 \tilde \gamma\right) \sin\tilde \gamma \sin^3\lambda \cos\left(3 \psi^{}_{\mathcal{O}}\right) \Bigg. \Bigg] \,,
\label{TAfluxLz}
\end{eqnarray}
\begin{eqnarray}
\Big\langle  \frac{d L}{d t} \Big\rangle & = & - \frac{32}{5} \frac{1}{p^{7/2}}\frac{(1 - e^2)^{3/2}}{ (1 + e^2)} M^{5/2} \mu^2 \Bigg[ \Bigg. 1 + \frac{15}{8}e^2 + \frac{7}{8}e^4 \nonumber \\[2mm] 
& - & \frac{S}{32 p^{3/2}} M^{-1} \left[80 + 328 e^2 + 295 e^4 + 47 e^6 + \left(496 + 2148 e^2 + 1939 e^4 + 207  e^6\right) \cos^2\lambda \right] \sec\lambda \nonumber \\[2mm]
& + & \frac{S}{32 p^{3/2}} M^{-1} e^2 \left(28 + 55 e^2 + 27 e^4\right) \sin\lambda \tan\lambda \cos\left(2 \tilde \gamma\right) \nonumber \\[2mm]
& + & \frac{Q}{128 p^{2}} \left[448 + 1952 e^2 + 1770 e^4 + 186 e^6 - \left(672 + 1641 e^2 + 2655 e^4 + 279 e^6\right)\sin^2\lambda \right] \nonumber \\[2mm]
& - & \frac{Q}{64 p^{2}} \left(144 + 993 e^2 + 1029 e^4 + 120 e^6 \right) \cos\left(2 \tilde \gamma\right) \sin^2\lambda \nonumber \\[2mm]
& - & \frac{3 Q^{}_{+}}{256 p^{2}} \left(168 + 740 e^2 + 681 e^4 + 93 e^6\right) \left(3 + \cos\left(2\lambda\right)\right) \cos\left(2 \psi^{}_Q\right) \nonumber \\[2mm]
& + & \frac{ \mathcal{O}}{1024 p^{3}} e\, \left(1668 + 5502 e^2 + 3525 e^4 + 435 e^6 \right) \left(6\sin\left(2\lambda\right) + 5\sin\left(4\lambda\right)\right) \sin\tilde \gamma \nonumber \\[2mm]
& + & \frac{ \mathcal{O}}{512 p^{3}} e^3 \left(3700 + 4935 e^2 + 995 e^4\right) \cos\lambda \sin^3\lambda \sin\left(3 \tilde \gamma\right) \nonumber \\[2mm]
& + & \frac{ \mathcal{O}^{}_{+}}{512 p^{3}} e\, \left(18000 - 41520 e^2 + 39465 e^4 + 6825 e^6 \right) \sin\tilde \gamma \sin^3\lambda \cos\left(3 \psi^{}_{\mathcal{O}}\right) \nonumber \\[2mm]
& - & \frac{ \mathcal{O}^{}_{+}}{64 p^{3}} e\, \left(3000 - 5220 e^2 + 8145 e^4 + 1485 e^6 \right)\sin\tilde \gamma \sin\lambda \cos\left(3 \psi^{}_{\mathcal{O}}\right) \nonumber \\[2mm]
& - & \frac{ \mathcal{O}^{}_{+}}{256 p^{3}} e^3 \left(10200 + 9405 e^2 + 2085 e^6\right) \cos\left(2 \tilde \gamma\right) \sin\tilde \gamma \sin^3\lambda \cos\left(3 \psi^{}_{\mathcal{O}}\right) \Bigg. \Bigg] \,.
\label{TAfluxL}
\end{eqnarray}
\end{widetext}

\section{LISA NOISE MODEL} \label{LISA-noise}

The noise PSDs of the LISA $A$ and $E$ TDI channels, $S_n^{A,E}(f)$, are obtained using second-generation TDI, and are given by~\cite{Bayle2022LISANode}
\begin{eqnarray}
S_n^{A,E}(f) & = & 32 \sin^2 x \, \sin^2 (2x) 
\Big[ (2 + \cos x)\, S^{}_I(f) \nonumber \\[2mm]
& + & 2 \big(3 + 2 \cos x + \cos 2x \big)\, S^{}_{II}(f) \Big] \,,
\end{eqnarray}
where $L = 2.5 \times 10^6\,\mathrm{km}$ is the LISA constellation arm length, $x = 2\pi f L/c$, and $A$, $E$ are two orthogonal TDI channels. This noise model is the one adopted in~\cite{Babak:2021mhe}.
The functions $S_I(f)$ and $S_{II}(f)$ are defined as follows:
\begin{equation}
S^{}_I(f) = S^{}_{\mathrm{oms}} 
\left( \frac{2\pi f}{c} \right)^2 
\left[ 1 + \left( \frac{2\,\mathrm{mHz}}{f} \right)^4 \right],
\end{equation}
\begin{eqnarray}
S^{}_{II}(f) & = & \frac{S^{}_{\mathrm{acc}}}{(2\pi c f)^2}
\left[ 1 + \left( \frac{0.4\,\mathrm{mHz}}{f} \right)^2 \right] \nonumber \\[2mm]
& \times & 
\left[ 1 + \left( \frac{f}{8\,\mathrm{mHz}} \right)^4 \right]
\end{eqnarray}
where $S_{\mathrm{oms}}$ is the amplitude of the Optical Metrology System noise and $S_{\mathrm{acc}}$ is the amplitude of the differential acceleration noise. 

In this work, we adopt the following values~\cite{LISADataWG}:
\begin{eqnarray}
\sqrt{S^{}_{\mathrm{oms}}} & = & 7.9\;\frac{\mathrm{pm}}{\sqrt{\mathrm{Hz}}}\,, \\[2mm] 
\sqrt{S_{\mathrm{acc}}} & = & 2.4\;\frac{\mathrm{fm\,s^{-2}}}{\sqrt{\mathrm{Hz}}}\,.
\end{eqnarray}

\section{POLYNOMIALS OF THE ECCENTRICITY IN THE EVOLUTION EQUATIONS} \label{evolutionapp}

The functions $f_{\nu,X}(e)$ that appear in Eq.~\ref{frequency} for the evolution of the orbital frequency $\nu$ are given by
\begin{equation}
f^{}_{\nu, 0}(e) = 1 + \frac{97}{24}e^2 + \frac{329}{96}e^4 + \frac{37}{96}e^6\,,
\end{equation}
\begin{equation}
f^{}_{\nu, S}(e) = 1888 + 14408 e^2 + 21500 e^4 + 8215 e^6 + 379 e^8 \,,
\end{equation}
\begin{equation}
f^{}_{\nu, Q, 1}(e) = 752 + 5952 e^2 + 9180 e^4 + 3585 e^6 + 177 e^8\,,
\end{equation}
\begin{equation}
f^{}_{\nu, Q, 2}(e) = 640 + 12784 e^2 + 23848 e^4 + 9953 e^6 + 537 e^8\,,
\end{equation}
\begin{equation}
f^{}_{\nu, Q^{}_+, 1}(e) = 664 + 5424 e^2 + 8355 e^4 + 3486 e^6 + 177 e^8\,,
\end{equation}
\begin{equation}
f^{}_{\nu, Q^{}_+, 2}(e) = 9360 + 19176 e^2 + 9473 e^4 + 537 e^6\,,
\end{equation}
\begin{equation}
f^{}_{\nu, \mathcal{O}, 1}(e) = 2896 + 13932 e^2 + 15966 e^4 + 5311 e^6 + 205 e^8\,,
\end{equation}
\begin{equation}
f^{}_{\nu, \mathcal{O}, 2}(e) = 17620 + 31595  e^2 + 13730  e^4 + 635  e^6\,,
\end{equation}
\begin{equation}
f^{}_{\nu, \mathcal{O}^{}_+, 1}(e) = 16880 + 85800 e^2 + 92670 e^4 + 28295 e^6 + 1025 e^8\,,
\end{equation}
\begin{equation}
f^{}_{\nu, \mathcal{O}^{}_+, 2}(e) = 18280 + 35555 e^2 + 14390 e^4 + 635 e^6\,.
\end{equation}
On the other hand, the functions $f_{e,X}(e)$ in Eq.~\eqref{eccentricity} for the evolution of the eccentricity $e$ are given by
\begin{equation}
f^{}_{e, S, 1}(e) = 240 + 744 e^2 - 99 e^4 - 744 e^6 - 141 e^8\,,
\end{equation}
\begin{equation}
f^{}_{e, S, 2}(e) = 400 + 9680 e^2 + 22265 e^4 + 13183 e^6 + 862 e^8\,,
\end{equation}
\begin{equation}
f^{}_{e, S, 3}(e) = 84 + 81 e^2 - 84 e^4 - 81 e^6\,,
\end{equation}
\begin{equation}
f^{}_{e, Q, 1}(e) = 80 + 3696 e^2 + 9453 e^4 + 5961 e^6 + 456 e^8\,,
\end{equation}
\begin{equation}
f^{}_{e, Q, 2}(e) = 80 + 4983 e^2 + 8166 e^4 + 5961 e^6 + 456 e^8\,,
\end{equation}
\begin{eqnarray}
f^{}_{e, Q, 3}(e) & = & -512 + 5992 e^2 + 23560 e^4 + 17225 e^6 \nonumber\\[2mm]
& + & 1497 e^8\,,
\end{eqnarray}
\begin{equation}
f^{}_{e, Q^{}_+, 1}(e) = 240 + 5562 e^2 + 12798 e^4 + 7875 e^6 + 684 e^8\,,
\end{equation}
\begin{equation}
f^{}_{e, Q^{}_+, 2}(e) = 80 + 1854 e^2 + 4266 e^4 + 2625 e^6 + 228 e^8\,,
\end{equation}
\begin{equation}
f^{}_{e, Q^{}_+, 3}(e) = 9360 + 19176 e^2 + 9473 e^4 + 537 e^6\,,
\end{equation}
\begin{eqnarray}
f^{}_{e, \mathcal{O}, 1}(e) & = & -7112 - 9072 e^2 + 27828 e^4 + 23851 e^6 \nonumber \\[2mm] 
& + &  2815 e^8\,,
\end{eqnarray}
\begin{eqnarray}
f^{}_{e, \mathcal{O}, 2}(e) & = & -1668 - 3834 e^2 + 1977 e^4 + 3090 e^6 \nonumber \\[2mm] 
& + & 435 e^8\,,
\end{eqnarray}
\begin{eqnarray}
f^{}_{e, \mathcal{O}, 3}(e) & = & 2896 + 13932 e^2 + 15966 e^4 + 5311 e^6 \nonumber \\[2mm]
& + & 205 e^8 \,,
\end{eqnarray}
\begin{equation}
f^{}_{e, \mathcal{O}, 4}(e) = 4828 + 11897  e^2 + 7856  e^4 + 851  e^6\,,
\end{equation}
\begin{eqnarray}
f^{}_{e, \mathcal{O}^{}_+, 1}(e) & = & 14680 + 118144 e^2 + 149237 e^4 + 115818 e^6 \nonumber \\[2mm] 
& + & 10653  e^8\,,
\end{eqnarray}
\begin{equation}
f^{}_{e, \mathcal{O}^{}_+, 2}(e) = 4864 + 12077 e^2 + 7676 e^4 + 815 e^6\,,
\end{equation}
\begin{eqnarray}
f^{}_{e, \mathcal{O}^{}_+, 3}(e) & = & 8808 + 63104 e^2 + 121051 e^4 + 12358 e^6 \nonumber \\[2mm] 
& - & 893  e^8\,,
\end{eqnarray}
\begin{equation}
f^{}_{e, \mathcal{O}^{}_+, 4}(e) = 4864 + 12077 e^2 + 7676 e^4 + 815 e^6 \,.
\end{equation}
%


\begin{thebibliography}{189}%
\makeatletter
\providecommand \@ifxundefined [1]{%
 \@ifx{#1\undefined}
}%
\providecommand \@ifnum [1]{%
 \ifnum #1\expandafter \@firstoftwo
 \else \expandafter \@secondoftwo
 \fi
}%
\providecommand \@ifx [1]{%
 \ifx #1\expandafter \@firstoftwo
 \else \expandafter \@secondoftwo
 \fi
}%
\providecommand \natexlab [1]{#1}%
\providecommand \enquote  [1]{``#1''}%
\providecommand \bibnamefont  [1]{#1}%
\providecommand \bibfnamefont [1]{#1}%
\providecommand \citenamefont [1]{#1}%
\providecommand \href@noop [0]{\@secondoftwo}%
\providecommand \href [0]{\begingroup \@sanitize@url \@href}%
\providecommand \@href[1]{\@@startlink{#1}\@@href}%
\providecommand \@@href[1]{\endgroup#1\@@endlink}%
\providecommand \@sanitize@url [0]{\catcode `\\12\catcode `\$12\catcode
  `\&12\catcode `\#12\catcode `\^12\catcode `\_12\catcode `\%12\relax}%
\providecommand \@@startlink[1]{}%
\providecommand \@@endlink[0]{}%
\providecommand \url  [0]{\begingroup\@sanitize@url \@url }%
\providecommand \@url [1]{\endgroup\@href {#1}{\urlprefix }}%
\providecommand \urlprefix  [0]{URL }%
\providecommand \Eprint [0]{\href }%
\providecommand \doibase [0]{https://doi.org/}%
\providecommand \selectlanguage [0]{\@gobble}%
\providecommand \bibinfo  [0]{\@secondoftwo}%
\providecommand \bibfield  [0]{\@secondoftwo}%
\providecommand \translation [1]{[#1]}%
\providecommand \BibitemOpen [0]{}%
\providecommand \bibitemStop [0]{}%
\providecommand \bibitemNoStop [0]{.\EOS\space}%
\providecommand \EOS [0]{\spacefactor3000\relax}%
\providecommand \BibitemShut  [1]{\csname bibitem#1\endcsname}%
\let\auto@bib@innerbib\@empty
\bibitem [{\citenamefont {Abbott}\ \emph {et~al.}(2016)\citenamefont {Abbott}
  \emph {et~al.}}]{LIGOScientific:2016aoc}%
  \BibitemOpen
  \bibfield  {author} {\bibinfo {author} {\bibfnamefont {B.~P.}\ \bibnamefont
  {Abbott}} \emph {et~al.} (\bibinfo {Collaboration} {LIGO Scientific and
  Virgo Collaborations}),\ }\bibfield  {title} {\bibinfo {title} {{Observation of
  gravitational waves from a binary black hole nerger}},\ }\href
  {https://doi.org/10.1103/PhysRevLett.116.061102} {\bibfield  {journal}
  {\bibinfo  {journal} {{\PRL}}\ }\textbf {\bibinfo {volume} {116}},\ \bibinfo
  {pages} {061102} (\bibinfo {year} {2016})},\ \Eprint
  {https://arxiv.org/abs/1602.03837} {arXiv:1602.03837 [gr-qc]} \BibitemShut
  {NoStop}%
\bibitem [{\citenamefont {Abbott}\ \emph {et~al.}(2019)\citenamefont {Abbott}
  \emph {et~al.}}]{LIGOScientific:2018mvr}%
  \BibitemOpen
  \bibfield  {author} {\bibinfo {author} {\bibfnamefont {B.~P.}\ \bibnamefont
  {Abbott}} \emph {et~al.} (\bibinfo {Collaboration} {LIGO Scientific and
  Virgo Collaborations}),\ }\bibfield  {title} {\bibinfo {title} {{GWTC-1: A
  gravitational-wave transient catalog of compact binary mergers observed by
  LIGO and Virgo during the first and second observing runs}},\ }\href
  {https://doi.org/10.1103/PhysRevX.9.031040} {\bibfield  {journal} {\bibinfo
  {journal} {{\PRX}}\ }\textbf {\bibinfo {volume} {9}},\ \bibinfo {pages}
  {031040} (\bibinfo {year} {2019})},\ \Eprint
  {https://arxiv.org/abs/1811.12907} {arXiv:1811.12907 [astro-ph.HE]}
  \BibitemShut {NoStop}%
\bibitem [{\citenamefont {Abbott}\ \emph {et~al.}(2024)\citenamefont {Abbott}
  \emph {et~al.}}]{LIGOScientific:2021usb}%
  \BibitemOpen
  \bibfield  {author} {\bibinfo {author} {\bibfnamefont {R.}~\bibnamefont
  {Abbott}} \emph {et~al.} (\bibinfo {Collaboration} {LIGO Scientific and
  VIRGO Collaborations}),\ }\bibfield  {title} {\bibinfo {title} {{GWTC-2.1: Deep extended
  catalog of compact binary coalescences observed by LIGO and Virgo during the
  first half of the third observing run}},\ }\href
  {https://doi.org/10.1103/PhysRevD.109.022001} {\bibfield  {journal} {\bibinfo
   {journal} {{\PRD}}\ }\textbf {\bibinfo {volume} {109}},\ \bibinfo {pages}
  {022001} (\bibinfo {year} {2024})},\ \Eprint
  {https://arxiv.org/abs/2108.01045} {arXiv:2108.01045 [gr-qc]} \BibitemShut
  {NoStop}%
\bibitem [{\citenamefont {Abbott}\ \emph {et~al.}(2023)\citenamefont {Abbott}
  \emph {et~al.}}]{LIGOScientific:2021djp}%
  \BibitemOpen
  \bibfield  {author} {\bibinfo {author} {\bibfnamefont {R.}~\bibnamefont
  {Abbott}} \emph {et~al.} (\bibinfo {Collaboration} {KAGRA, VIRGO,  and LIGO
  Scientific Collaborations}),\ }\bibfield  {title} {\bibinfo {title} {{GWTC-3: Compact binary
  coalescences observed by LIGO and Virgo during the second part of the third
  observing run}},\ }\href {https://doi.org/10.1103/PhysRevX.13.041039}
  {\bibfield  {journal} {\bibinfo  {journal} {{\PRX}}\ }\textbf {\bibinfo
  {volume} {13}},\ \bibinfo {pages} {041039} (\bibinfo {year} {2023})},\
  \Eprint {https://arxiv.org/abs/2111.03606} {arXiv:2111.03606 [gr-qc]}
  \BibitemShut {NoStop}%
\bibitem [{\citenamefont {Abac}\ \emph
  {et~al.}(2026{\natexlab{a}})\citenamefont {Abac} \emph
  {et~al.}}]{LIGOScientific:2025slb}%
  \BibitemOpen
  \bibfield  {author} {\bibinfo {author} {\bibfnamefont {A.~G.}\ \bibnamefont
  {Abac}} \emph {et~al.} (\bibinfo {Collaboration} {LIGO Scientific, VIRGO,
  and KAGRA Collaborations}),\ }\bibfield  {title} {\bibinfo {title} {{GWTC-4.0: Updating the
  gravitational-wave transient catalog with observations from the first part of
  the fourth LIGO-Virgo-KAGRA observing run}},\ }\href
  {https://doi.org/10.3847/2041-8213/ae2c74} {\bibfield  {journal} {\bibinfo
  {journal} {Astrophys. J. Lett.}\ }\textbf {\bibinfo {volume} {1004}},\
  \bibinfo {pages} {L22} (\bibinfo {year} {2026}{\natexlab{a}})},\ \Eprint
  {https://arxiv.org/abs/2508.18082} {arXiv:2508.18082 [gr-qc]} \BibitemShut
  {NoStop}%
\bibitem [{\citenamefont {Punturo}\ \emph {et~al.}(2010)\citenamefont {Punturo}
  \emph {et~al.}}]{Punturo:2010zz}%
  \BibitemOpen
  \bibfield  {author} {\bibinfo {author} {\bibfnamefont {M.}~\bibnamefont
  {Punturo}} \emph {et~al.},\ }\bibfield  {title} {\bibinfo {title} {{The
  Einstein Telescope: A third-generation gravitational wave observatory}},\
  }\href {https://doi.org/10.1088/0264-9381/27/19/194002} {\bibfield  {journal}
  {\bibinfo  {journal} {Classical Quantum Gravity}\ }\textbf {\bibinfo {volume}
  {27}},\ \bibinfo {pages} {194002} (\bibinfo {year} {2010})}\BibitemShut
  {NoStop}%
\bibitem [{\citenamefont {Reitze}\ \emph {et~al.}(2019)\citenamefont {Reitze}
  \emph {et~al.}}]{Reitze:2019iox}%
  \BibitemOpen
  \bibfield  {author} {\bibinfo {author} {\bibfnamefont {D.}~\bibnamefont
  {Reitze}} \emph {et~al.},\ }\bibfield  {title} {\bibinfo {title} {{Cosmic
  Explorer: The U.S. contribution to gravitational-wave astronomy beyond
  LIGO}},\ }\href@noop {} {\bibfield  {journal} {\bibinfo  {journal} {Bull. Am.
  Astron. Soc.}\ }\textbf {\bibinfo {volume} {51}},\ \bibinfo {pages} {035}
  (\bibinfo {year} {2019})},\ \Eprint {https://arxiv.org/abs/1907.04833}
  {arXiv:1907.04833 [astro-ph.IM]} \BibitemShut {NoStop}%
\bibitem [{\citenamefont {Badurina}\ \emph {et~al.}(2020)\citenamefont
  {Badurina} \emph {et~al.}}]{Badurina:2019hst}%
  \BibitemOpen
  \bibfield  {author} {\bibinfo {author} {\bibfnamefont {L.}~\bibnamefont
  {Badurina}} \emph {et~al.},\ }\bibfield  {title} {\bibinfo {title} {{AION: An
  atom interferometer observatory and network}},\ }\href
  {https://doi.org/10.1088/1475-7516/2020/05/011} {\bibfield  {journal}
  {\bibinfo  {journal} {{JCAP}}\ }\textbf {\bibinfo {volume} {05}},\ \bibinfo
  {pages} {011} (\bibinfo {year} {2020})},\ \Eprint
  {https://arxiv.org/abs/1911.11755} {arXiv:1911.11755 [astro-ph.CO]}
  \BibitemShut {NoStop}%
\bibitem [{\citenamefont {Canuel}\ \emph {et~al.}(2020)\citenamefont {Canuel}
  \emph {et~al.}}]{Canuel:2019abg}%
  \BibitemOpen
  \bibfield  {author} {\bibinfo {author} {\bibfnamefont {B.}~\bibnamefont
  {Canuel}} \emph {et~al.},\ }\bibfield  {title} {\bibinfo {title}
  {{ELGAR\textemdash{}a European Laboratory for gravitation and
  atom-interferometric research}},\ }\href
  {https://doi.org/10.1088/1361-6382/aba80e} {\bibfield  {journal} {\bibinfo
  {journal} {{\CQG}}\ }\textbf {\bibinfo {volume} {37}},\ \bibinfo {pages}
  {225017} (\bibinfo {year} {2020})},\ \Eprint
  {https://arxiv.org/abs/1911.03701} {arXiv:1911.03701 [physics.atom-ph]}
  \BibitemShut {NoStop}%
\bibitem [{\citenamefont {Amaro-Seoane}\ \emph {et~al.}(2017)\citenamefont
  {Amaro-Seoane} \emph {et~al.}}]{LISA:2017pwj}%
  \BibitemOpen
  \bibfield  {author} {\bibinfo {author} {\bibfnamefont {P.}~\bibnamefont
  {Amaro-Seoane}} \emph {et~al.} (\bibinfo {Collaboration} {LISA Collaboration}),\ }\bibfield
   {title} {\bibinfo {title} {{Laser Interferometer Space Antenna}},\
  }\href@noop {} {\  (\bibinfo {year} {2017})},\ \Eprint
  {https://arxiv.org/abs/1702.00786} {arXiv:1702.00786 [astro-ph.IM]}
  \BibitemShut {NoStop}%
\bibitem [{\citenamefont {Armano}\ \emph {et~al.}(2016)\citenamefont {Armano}
  \emph {et~al.}}]{Armano:2016bkm}%
  \BibitemOpen
  \bibfield  {author} {\bibinfo {author} {\bibfnamefont {M.}~\bibnamefont
  {Armano}} \emph {et~al.},\ }\bibfield  {title} {\bibinfo {title} {{Sub-Femto-
  g free fall for space-based gravitational wave observatories: LISA Pathfinder
  results}},\ }\href {https://doi.org/10.1103/PhysRevLett.116.231101}
  {\bibfield  {journal} {\bibinfo  {journal} {{\PRL}}\ }\textbf {\bibinfo
  {volume} {116}},\ \bibinfo {pages} {231101} (\bibinfo {year}
  {2016})}\BibitemShut {NoStop}%
\bibitem [{\citenamefont {Armano}\ \emph {et~al.}(2018)\citenamefont {Armano}
  \emph {et~al.}}]{Armano:2018kix}%
  \BibitemOpen
  \bibfield  {author} {\bibinfo {author} {\bibfnamefont {M.}~\bibnamefont
  {Armano}} \emph {et~al.},\ }\bibfield  {title} {\bibinfo {title} {{Beyond the
  required LISA free-fall performance: New LISA Pathfinder results down to 20
  $\mu$Hz}},\ }\href {https://doi.org/10.1103/PhysRevLett.120.061101}
  {\bibfield  {journal} {\bibinfo  {journal} {{\PRL}}\ }\textbf {\bibinfo
  {volume} {120}},\ \bibinfo {pages} {061101} (\bibinfo {year}
  {2018})}\BibitemShut {NoStop}%
\bibitem [{\citenamefont {Armano}\ \emph {et~al.}(2024)\citenamefont {Armano}
  \emph {et~al.}}]{LISAPathfinder:2024ucp}%
  \BibitemOpen
  \bibfield  {author} {\bibinfo {author} {\bibfnamefont {M.}~\bibnamefont
  {Armano}} \emph {et~al.} (\bibinfo {Collaboration} {LISA Pathfinder Collaboration}),\
  }\bibfield  {title} {\bibinfo {title} {{In-depth analysis of LISA Pathfinder
  performance results: Time evolution, noise projection, physical models, and
  implications for LISA}},\ }\href
  {https://doi.org/10.1103/PhysRevD.110.042004} {\bibfield  {journal} {\bibinfo
   {journal} {Phys. Rev. D}\ }\textbf {\bibinfo {volume} {110}},\ \bibinfo
  {pages} {042004} (\bibinfo {year} {2024})},\ \Eprint
  {https://arxiv.org/abs/2405.05207} {arXiv:2405.05207 [astro-ph.IM]}
  \BibitemShut {NoStop}%
\bibitem [{\citenamefont {Amaro-Seoane}\ \emph
  {et~al.}(2013{\natexlab{a}})\citenamefont {Amaro-Seoane} \emph
  {et~al.}}]{eLISA:2013xep}%
  \BibitemOpen
  \bibfield  {author} {\bibinfo {author} {\bibfnamefont {P.}~\bibnamefont
  {Amaro-Seoane}} \emph {et~al.} (\bibinfo {Collaboration} {eLISA Collaboration}),\
  }\bibfield  {title} {\bibinfo {title} {{The gravitational Universe}},\
  }\href@noop {} {\  (\bibinfo {year} {2013}{\natexlab{a}})},\ \Eprint
  {https://arxiv.org/abs/1305.5720} {arXiv:1305.5720 [astro-ph.CO]}
  \BibitemShut {NoStop}%
\bibitem [{\citenamefont {Colpi}\ \emph {et~al.}(2024)\citenamefont {Colpi}
  \emph {et~al.}}]{Colpi:2024xhw}%
  \BibitemOpen
  \bibfield  {author} {\bibinfo {author} {\bibfnamefont {M.}~\bibnamefont
  {Colpi}} \emph {et~al.},\ }\bibfield  {title} {\bibinfo {title} {{LISA
  definition study report}},\ }\href {https://arxiv.org/abs/2402.07571} {\
  (\bibinfo {year} {2024})},\ \Eprint {https://arxiv.org/abs/2402.07571}
  {arXiv:2402.07571 [astro-ph.CO]} \BibitemShut {NoStop}%
\bibitem [{\citenamefont {Abac}\ \emph {et~al.}(2025)\citenamefont {Abac} \emph
  {et~al.}}]{LIGOScientific:2025rid}%
  \BibitemOpen
  \bibfield  {author} {\bibinfo {author} {\bibfnamefont {A.~G.}\ \bibnamefont
  {Abac}} \emph {et~al.} (\bibinfo {Collaboration} {LIGO Scientific, Virgo, and
  KAGRA Collaborations}),\ }\bibfield  {title} {\bibinfo {title} {{GW250114: Testing
  Hawking{\textquoteright}s area law and the Kerr nature of black holes}},\
  }\href {https://doi.org/10.1103/kw5g-d732} {\bibfield  {journal} {\bibinfo
  {journal} {{\PRL}}\ }\textbf {\bibinfo {volume} {135}},\ \bibinfo {pages}
  {111403} (\bibinfo {year} {2025})},\ \Eprint
  {https://arxiv.org/abs/2509.08054} {arXiv:2509.08054 [gr-qc]} \BibitemShut
  {NoStop}%
\bibitem [{\citenamefont {Abac}\ \emph
  {et~al.}(2026{\natexlab{b}})\citenamefont {Abac} \emph
  {et~al.}}]{LIGOScientific:2025wao}%
  \BibitemOpen
  \bibfield  {author} {\bibinfo {author} {\bibfnamefont {A.~G.}\ \bibnamefont
  {Abac}} \emph {et~al.} (\bibinfo {Collaboration} {LIGO Scientific, Virgo,
  and KAGRA Collaborations}),\ }\bibfield  {title} {\bibinfo {title} {{Black hole spectroscopy and
  tests of general relativity with GW250114}},\ }\href
  {https://doi.org/10.1103/6c61-fm1n} {\bibfield  {journal} {\bibinfo
  {journal} {{\PRL}}\ }\textbf {\bibinfo {volume} {136}},\ \bibinfo {pages}
  {041403} (\bibinfo {year} {2026}{\natexlab{b}})},\ \Eprint
  {https://arxiv.org/abs/2509.08099} {arXiv:2509.08099 [gr-qc]} \BibitemShut
  {NoStop}%
\bibitem [{\citenamefont {Seoane}\ \emph {et~al.}(2023)\citenamefont {Seoane}
  \emph {et~al.}}]{LISA:2022yao}%
  \BibitemOpen
  \bibfield  {author} {\bibinfo {author} {\bibfnamefont {P.~A.}\ \bibnamefont
  {Seoane}} \emph {et~al.} (\bibinfo {Collaboration} {LISA Collaboration}),\ }\bibfield
  {title} {\bibinfo {title} {{Astrophysics with the Laser Interferometer Space
  Antenna}},\ }\href {https://doi.org/10.1007/s41114-022-00041-y} {\bibfield
  {journal} {\bibinfo  {journal} {{\LRR}}\ }\textbf {\bibinfo {volume} {26}},\
  \bibinfo {pages} {2} (\bibinfo {year} {2023})},\ \Eprint
  {https://arxiv.org/abs/2203.06016} {arXiv:2203.06016 [gr-qc]} \BibitemShut
  {NoStop}%
\bibitem [{\citenamefont {Laghi}\ \emph {et~al.}(2021)\citenamefont {Laghi},
  \citenamefont {Tamanini}, \citenamefont {Del~Pozzo}, \citenamefont {Sesana},
  \citenamefont {Gair}, \citenamefont {Babak},\ and\ \citenamefont
  {Izquierdo-Villalba}}]{Laghi:2021pqk}%
  \BibitemOpen
  \bibfield  {author} {\bibinfo {author} {\bibfnamefont {D.}~\bibnamefont
  {Laghi}}, \bibinfo {author} {\bibfnamefont {N.}~\bibnamefont {Tamanini}},
  \bibinfo {author} {\bibfnamefont {W.}~\bibnamefont {Del~Pozzo}}, \bibinfo
  {author} {\bibfnamefont {A.}~\bibnamefont {Sesana}}, \bibinfo {author}
  {\bibfnamefont {J.}~\bibnamefont {Gair}}, \bibinfo {author} {\bibfnamefont
  {S.}~\bibnamefont {Babak}},\ and\ \bibinfo {author} {\bibfnamefont
  {D.}~\bibnamefont {Izquierdo-Villalba}},\ }\bibfield  {title} {\bibinfo
  {title} {{Gravitational-wave cosmology with extreme mass-ratio inspirals}},\
  }\href {https://doi.org/10.1093/mnras/stab2741} {\bibfield  {journal}
  {\bibinfo  {journal} {{\MNRAS}}\ }\textbf {\bibinfo {volume} {508}},\
  \bibinfo {pages} {4512} (\bibinfo {year} {2021})},\ \Eprint
  {https://arxiv.org/abs/2102.01708} {arXiv:2102.01708 [astro-ph.CO]}
  \BibitemShut {NoStop}%
\bibitem [{\citenamefont {Auclair}\ \emph {et~al.}(2023)\citenamefont {Auclair}
  \emph {et~al.}}]{LISACosmologyWorkingGroup:2022jok}%
  \BibitemOpen
  \bibfield  {author} {\bibinfo {author} {\bibfnamefont {P.}~\bibnamefont
  {Auclair}} \emph {et~al.} (\bibinfo {Collaboration} {LISA Cosmology Working
  Group Collaboration}),\ }\bibfield  {title} {\bibinfo {title} {{Cosmology with the Laser
  Interferometer Space Antenna}},\ }\href
  {https://doi.org/10.1007/s41114-023-00045-2} {\bibfield  {journal} {\bibinfo
  {journal} {{\LRR}}\ }\textbf {\bibinfo {volume} {26}},\ \bibinfo {pages} {5}
  (\bibinfo {year} {2023})},\ \Eprint {https://arxiv.org/abs/2204.05434}
  {arXiv:2204.05434 [astro-ph.CO]} \BibitemShut {NoStop}%
\bibitem [{\citenamefont {Gair}\ \emph {et~al.}(2013)\citenamefont {Gair},
  \citenamefont {Vallisneri}, \citenamefont {Larson},\ and\ \citenamefont
  {Baker}}]{Gair:2012nm}%
  \BibitemOpen
  \bibfield  {author} {\bibinfo {author} {\bibfnamefont {J.~R.}\ \bibnamefont
  {Gair}}, \bibinfo {author} {\bibfnamefont {M.}~\bibnamefont {Vallisneri}},
  \bibinfo {author} {\bibfnamefont {S.~L.}\ \bibnamefont {Larson}},\ and\
  \bibinfo {author} {\bibfnamefont {J.~G.}\ \bibnamefont {Baker}},\ }\bibfield
  {title} {\bibinfo {title} {{Testing general relativity with low-frequency,
  space-based gravitational-wave detectors}},\ }\href
  {https://doi.org/10.12942/lrr-2013-7} {\bibfield  {journal} {\bibinfo
  {journal} {{\LRR}}\ }\textbf {\bibinfo {volume} {16}},\ \bibinfo {pages} {7}
  (\bibinfo {year} {2013})},\ \Eprint {https://arxiv.org/abs/1212.5575}
  {arXiv:1212.5575 [gr-qc]} \BibitemShut {NoStop}%
\bibitem [{\citenamefont {Berti}\ \emph {et~al.}(2019)\citenamefont {Berti}
  \emph {et~al.}}]{Berti:2019xgr}%
  \BibitemOpen
  \bibfield  {author} {\bibinfo {author} {\bibfnamefont {E.}~\bibnamefont
  {Berti}} \emph {et~al.},\ }\bibfield  {title} {\bibinfo {title} {{Tests of
  general relativity and fundamental physics with space-based gravitational
  wave detectors}},\ }\href@noop {} {\bibfield  {journal} {\bibinfo  {journal}
  {{\BAAS}}\ }\textbf {\bibinfo {volume} {51}},\ \bibinfo {pages} {32}
  (\bibinfo {year} {2019})},\ \Eprint {https://arxiv.org/abs/1903.02781}
  {arXiv:1903.02781 [astro-ph.HE]} \BibitemShut {NoStop}%
\bibitem [{\citenamefont {Barausse}\ \emph {et~al.}(2020)\citenamefont
  {Barausse} \emph {et~al.}}]{Barausse:2020rsu}%
  \BibitemOpen
  \bibfield  {author} {\bibinfo {author} {\bibfnamefont {E.}~\bibnamefont
  {Barausse}} \emph {et~al.},\ }\bibfield  {title} {\bibinfo {title}
  {{Prospects for fundamental physics with LISA}},\ }\href
  {https://doi.org/10.1007/s10714-020-02691-1} {\bibfield  {journal} {\bibinfo
  {journal} {{\GRG}}\ }\textbf {\bibinfo {volume} {52}},\ \bibinfo {pages} {81}
  (\bibinfo {year} {2020})},\ \Eprint {https://arxiv.org/abs/2001.09793}
  {arXiv:2001.09793 [gr-qc]} \BibitemShut {NoStop}%
\bibitem [{\citenamefont {Arun}\ \emph {et~al.}(2022)\citenamefont {Arun} \emph
  {et~al.}}]{LISA:2022kgy}%
  \BibitemOpen
  \bibfield  {author} {\bibinfo {author} {\bibfnamefont {K.~G.}\ \bibnamefont
  {Arun}} \emph {et~al.} (\bibinfo {Collaboration} {LISA Collaboration}),\ }\bibfield
  {title} {\bibinfo {title} {{New horizons for fundamental physics with
  LISA}},\ }\href {https://doi.org/10.1007/s41114-022-00036-9} {\bibfield
  {journal} {\bibinfo  {journal} {{\LRR}}\ }\textbf {\bibinfo {volume} {25}},\
  \bibinfo {pages} {4} (\bibinfo {year} {2022})},\ \Eprint
  {https://arxiv.org/abs/2205.01597} {arXiv:2205.01597 [gr-qc]} \BibitemShut
  {NoStop}%
\bibitem [{\citenamefont {Kerr}(1963)}]{Kerr:1963ud}%
  \BibitemOpen
  \bibfield  {author} {\bibinfo {author} {\bibfnamefont {R.~P.}\ \bibnamefont
  {Kerr}},\ }\bibfield  {title} {\bibinfo {title} {{Gravitational field of a
  spinning mass as an example of algebraically special metrics}},\ }\href
  {https://doi.org/10.1103/PhysRevLett.11.237} {\bibfield  {journal} {\bibinfo
  {journal} {{\PRL}}\ }\textbf {\bibinfo {volume} {11}},\ \bibinfo {pages}
  {237} (\bibinfo {year} {1963})}\BibitemShut {NoStop}%
\bibitem [{\citenamefont {Cardoso}\ \emph {et~al.}(2016)\citenamefont
  {Cardoso}, \citenamefont {Hopper}, \citenamefont {Macedo}, \citenamefont
  {Palenzuela},\ and\ \citenamefont {Pani}}]{Cardoso:2016oxy}%
  \BibitemOpen
  \bibfield  {author} {\bibinfo {author} {\bibfnamefont {V.}~\bibnamefont
  {Cardoso}}, \bibinfo {author} {\bibfnamefont {S.}~\bibnamefont {Hopper}},
  \bibinfo {author} {\bibfnamefont {C.~F.~B.}\ \bibnamefont {Macedo}}, \bibinfo
  {author} {\bibfnamefont {C.}~\bibnamefont {Palenzuela}},\ and\ \bibinfo
  {author} {\bibfnamefont {P.}~\bibnamefont {Pani}},\ }\bibfield  {title}
  {\bibinfo {title} {{Gravitational-wave signatures of exotic compact objects
  and of quantum corrections at the horizon scale}},\ }\href
  {https://doi.org/10.1103/PhysRevD.94.084031} {\bibfield  {journal} {\bibinfo
  {journal} {{\PRD}}\ }\textbf {\bibinfo {volume} {94}},\ \bibinfo {pages}
  {084031} (\bibinfo {year} {2016})},\ \Eprint
  {https://arxiv.org/abs/1608.08637} {arXiv:1608.08637 [gr-qc]} \BibitemShut
  {NoStop}%
\bibitem [{\citenamefont {Cardoso}\ and\ \citenamefont
  {Pani}(2019)}]{Cardoso:2019rvt}%
  \BibitemOpen
  \bibfield  {author} {\bibinfo {author} {\bibfnamefont {V.}~\bibnamefont
  {Cardoso}}\ and\ \bibinfo {author} {\bibfnamefont {P.}~\bibnamefont {Pani}},\
  }\bibfield  {title} {\bibinfo {title} {{Testing the nature of dark compact
  objects: a status report}},\ }\href
  {https://doi.org/10.1007/s41114-019-0020-4} {\bibfield  {journal} {\bibinfo
  {journal} {{\LRR}}\ }\textbf {\bibinfo {volume} {22}},\ \bibinfo {pages} {4}
  (\bibinfo {year} {2019})},\ \Eprint {https://arxiv.org/abs/1904.05363}
  {arXiv:1904.05363 [gr-qc]} \BibitemShut {NoStop}%
\bibitem [{\citenamefont {Amaro-Seoane}\ \emph {et~al.}(2007)\citenamefont
  {Amaro-Seoane}, \citenamefont {Gair}, \citenamefont {Freitag}, \citenamefont
  {Coleman~Miller}, \citenamefont {Mandel}, \citenamefont {Cutler},\ and\
  \citenamefont {Babak}}]{Amaro-Seoane:2007osp}%
  \BibitemOpen
  \bibfield  {author} {\bibinfo {author} {\bibfnamefont {P.}~\bibnamefont
  {Amaro-Seoane}}, \bibinfo {author} {\bibfnamefont {J.~R.}\ \bibnamefont
  {Gair}}, \bibinfo {author} {\bibfnamefont {M.}~\bibnamefont {Freitag}},
  \bibinfo {author} {\bibfnamefont {M.}~\bibnamefont {Coleman~Miller}},
  \bibinfo {author} {\bibfnamefont {I.}~\bibnamefont {Mandel}}, \bibinfo
  {author} {\bibfnamefont {C.~J.}\ \bibnamefont {Cutler}},\ and\ \bibinfo
  {author} {\bibfnamefont {S.}~\bibnamefont {Babak}},\ }\bibfield  {title}
  {\bibinfo {title} {{Astrophysics, detection and science applications of
  intermediate- and extreme mass-ratio inspirals}},\ }\href
  {https://doi.org/10.1088/0264-9381/24/17/R01} {\bibfield  {journal} {\bibinfo
   {journal} {{\CQG}}\ }\textbf {\bibinfo {volume} {24}},\ \bibinfo {pages}
  {R113} (\bibinfo {year} {2007})},\ \Eprint
  {https://arxiv.org/abs/astro-ph/0703495} {arXiv:astro-ph/0703495}
  \BibitemShut {NoStop}%
\bibitem [{\citenamefont {Amaro-Seoane}(2018)}]{Amaro-Seoane:2012lgq}%
  \BibitemOpen
  \bibfield  {author} {\bibinfo {author} {\bibfnamefont {P.}~\bibnamefont
  {Amaro-Seoane}},\ }\bibfield  {title} {\bibinfo {title} {{Relativistic
  dynamics and extreme mass ratio inspirals}},\ }\href
  {https://doi.org/10.1007/s41114-018-0013-8} {\bibfield  {journal} {\bibinfo
  {journal} {{\LRR}}\ }\textbf {\bibinfo {volume} {21}},\ \bibinfo {pages} {4}
  (\bibinfo {year} {2018})},\ \Eprint {https://arxiv.org/abs/1205.5240}
  {arXiv:1205.5240 [astro-ph.CO]} \BibitemShut {NoStop}%
\bibitem [{\citenamefont {Amaro-Seoane}\ \emph {et~al.}(2015)\citenamefont
  {Amaro-Seoane}, \citenamefont {Gair}, \citenamefont {Pound}, \citenamefont
  {Hughes},\ and\ \citenamefont {Sopuerta}}]{Amaro-Seoane:2014ela}%
  \BibitemOpen
  \bibfield  {author} {\bibinfo {author} {\bibfnamefont {P.}~\bibnamefont
  {Amaro-Seoane}}, \bibinfo {author} {\bibfnamefont {J.~R.}\ \bibnamefont
  {Gair}}, \bibinfo {author} {\bibfnamefont {A.}~\bibnamefont {Pound}},
  \bibinfo {author} {\bibfnamefont {S.~A.}\ \bibnamefont {Hughes}},\ and\
  \bibinfo {author} {\bibfnamefont {C.~F.}\ \bibnamefont {Sopuerta}},\
  }\bibfield  {title} {\bibinfo {title} {{Research update on extreme-mass-ratio
  inspirals}},\ }\href {https://doi.org/10.1088/1742-6596/610/1/012002}
  {\bibfield  {journal} {\bibinfo  {journal} {{\JPCS}}\ }\textbf {\bibinfo
  {volume} {610}},\ \bibinfo {pages} {012002} (\bibinfo {year} {2015})},\
  \Eprint {https://arxiv.org/abs/1410.0958} {arXiv:1410.0958 [astro-ph.CO]}
  \BibitemShut {NoStop}%
\bibitem [{\citenamefont {{Amaro-Seoane}}(2022)}]{2020arXiv201103059A}%
  \BibitemOpen
  \bibfield  {author} {\bibinfo {author} {\bibfnamefont {P.}~\bibnamefont
  {{Amaro-Seoane}}},\ }\bibfield  {title} {\bibinfo {title} {{The gravitational
  capture of compact objects by massive black holes}},\ }in\ \href
  {https://doi.org/10.1007/978-981-15-4702-7\_17-1} {\emph {\bibinfo
  {booktitle} {Handbook of Gravitational Wave astronomy}}},\ \bibinfo {editor}
  {edited by\ \bibinfo {editor} {\bibfnamefont {C.}~\bibnamefont {{Bambi}}},
  \bibinfo {editor} {\bibfnamefont {S.}~\bibnamefont {{Katsanevas}}},\ and\
  \bibinfo {editor} {\bibfnamefont {K.~D.}\ \bibnamefont {{Kokkotas}}}}\
  (\bibinfo  {publisher} {{Springer}},\ \bibinfo {year} {2022})\ p.~\bibinfo
  {pages} {17},\ \Eprint {https://arxiv.org/abs/2011.03059} {arXiv:2011.03059
  [gr-qc]} \BibitemShut {NoStop}%
\bibitem [{\citenamefont {Babak}\ \emph {et~al.}(2017)\citenamefont {Babak},
  \citenamefont {Gair}, \citenamefont {Sesana}, \citenamefont {Barausse},
  \citenamefont {Sopuerta}, \citenamefont {Berry}, \citenamefont {Berti},
  \citenamefont {Amaro-Seoane}, \citenamefont {Petiteau},\ and\ \citenamefont
  {Klein}}]{Babak:2017tow}%
  \BibitemOpen
  \bibfield  {author} {\bibinfo {author} {\bibfnamefont {S.}~\bibnamefont
  {Babak}}, \bibinfo {author} {\bibfnamefont {J.}~\bibnamefont {Gair}},
  \bibinfo {author} {\bibfnamefont {A.}~\bibnamefont {Sesana}}, \bibinfo
  {author} {\bibfnamefont {E.}~\bibnamefont {Barausse}}, \bibinfo {author}
  {\bibfnamefont {C.~F.}\ \bibnamefont {Sopuerta}}, \bibinfo {author}
  {\bibfnamefont {C.~P.~L.}\ \bibnamefont {Berry}}, \bibinfo {author}
  {\bibfnamefont {E.}~\bibnamefont {Berti}}, \bibinfo {author} {\bibfnamefont
  {P.}~\bibnamefont {Amaro-Seoane}}, \bibinfo {author} {\bibfnamefont
  {A.}~\bibnamefont {Petiteau}},\ and\ \bibinfo {author} {\bibfnamefont
  {A.}~\bibnamefont {Klein}},\ }\bibfield  {title} {\bibinfo {title} {{Science
  with the space-based interferometer LISA. V: Extreme mass-ratio inspirals}},\
  }\href {https://doi.org/10.1103/PhysRevD.95.103012} {\bibfield  {journal}
  {\bibinfo  {journal} {{\PRD}}\ }\textbf {\bibinfo {volume} {95}},\ \bibinfo
  {pages} {103012} (\bibinfo {year} {2017})},\ \Eprint
  {https://arxiv.org/abs/1703.09722} {arXiv:1703.09722 [gr-qc]} \BibitemShut
  {NoStop}%
\bibitem [{\citenamefont {Gair}\ \emph {et~al.}(2017)\citenamefont {Gair},
  \citenamefont {Babak}, \citenamefont {Sesana}, \citenamefont {Amaro-Seoane},
  \citenamefont {Barausse}, \citenamefont {Berry}, \citenamefont {Berti},\ and\
  \citenamefont {Sopuerta}}]{Gair:2017ynp}%
  \BibitemOpen
  \bibfield  {author} {\bibinfo {author} {\bibfnamefont {J.~R.}\ \bibnamefont
  {Gair}}, \bibinfo {author} {\bibfnamefont {S.}~\bibnamefont {Babak}},
  \bibinfo {author} {\bibfnamefont {A.}~\bibnamefont {Sesana}}, \bibinfo
  {author} {\bibfnamefont {P.}~\bibnamefont {Amaro-Seoane}}, \bibinfo {author}
  {\bibfnamefont {E.}~\bibnamefont {Barausse}}, \bibinfo {author}
  {\bibfnamefont {C.~P.~L.}\ \bibnamefont {Berry}}, \bibinfo {author}
  {\bibfnamefont {E.}~\bibnamefont {Berti}},\ and\ \bibinfo {author}
  {\bibfnamefont {C.}~\bibnamefont {Sopuerta}},\ }\bibfield  {title} {\bibinfo
  {title} {{Prospects for observing extreme-mass-ratio inspirals with LISA}},\
  }\href {https://doi.org/10.1088/1742-6596/840/1/012021} {\bibfield  {journal}
  {\bibinfo  {journal} {{\JPCS}}\ }\textbf {\bibinfo {volume} {840}},\ \bibinfo
  {pages} {012021} (\bibinfo {year} {2017})},\ \Eprint
  {https://arxiv.org/abs/1704.00009} {arXiv:1704.00009 [astro-ph.GA]}
  \BibitemShut {NoStop}%
\bibitem [{\citenamefont {Miller}\ \emph {et~al.}(2005)\citenamefont {Miller},
  \citenamefont {Freitag}, \citenamefont {Hamilton},\ and\ \citenamefont
  {Lauburg}}]{ColemanMiller:2005rm}%
  \BibitemOpen
  \bibfield  {author} {\bibinfo {author} {\bibfnamefont {M.~C.}\ \bibnamefont
  {Miller}}, \bibinfo {author} {\bibfnamefont {M.}~\bibnamefont {Freitag}},
  \bibinfo {author} {\bibfnamefont {D.~P.}\ \bibnamefont {Hamilton}},\ and\
  \bibinfo {author} {\bibfnamefont {V.~M.}\ \bibnamefont {Lauburg}},\
  }\bibfield  {title} {\bibinfo {title} {{Binary encounters with supermassive
  black holes: Zero-eccentricity LISA events}},\ }\href
  {https://doi.org/10.1086/497335} {\bibfield  {journal} {\bibinfo  {journal}
  {{\APJL}}\ }\textbf {\bibinfo {volume} {631}},\ \bibinfo {pages} {L117}
  (\bibinfo {year} {2005})},\ \Eprint {https://arxiv.org/abs/astro-ph/0507133}
  {arXiv:astro-ph/0507133} \BibitemShut {NoStop}%
\bibitem [{\citenamefont {Di~Stefano}\ \emph {et~al.}(2001)\citenamefont
  {Di~Stefano}, \citenamefont {Greiner}, \citenamefont {Murray},\ and\
  \citenamefont {Garcia}}]{DiStefano:2001ci}%
  \BibitemOpen
  \bibfield  {author} {\bibinfo {author} {\bibfnamefont {R.}~\bibnamefont
  {Di~Stefano}}, \bibinfo {author} {\bibfnamefont {J.}~\bibnamefont {Greiner}},
  \bibinfo {author} {\bibfnamefont {S.}~\bibnamefont {Murray}},\ and\ \bibinfo
  {author} {\bibfnamefont {M.}~\bibnamefont {Garcia}},\ }\bibfield  {title}
  {\bibinfo {title} {{A new way to detect massive black holes in galaxies: the
  stellar remnants of tidal disruption}},\ }\href
  {https://doi.org/10.1086/319835} {\bibfield  {journal} {\bibinfo  {journal}
  {{\APJL}}\ }\textbf {\bibinfo {volume} {551}},\ \bibinfo {pages} {L37}
  (\bibinfo {year} {2001})},\ \Eprint {https://arxiv.org/abs/astro-ph/0112434}
  {arXiv:astro-ph/0112434} \BibitemShut {NoStop}%
\bibitem [{\citenamefont {Davies}\ and\ \citenamefont
  {King}(2005)}]{Davies:2005tc}%
  \BibitemOpen
  \bibfield  {author} {\bibinfo {author} {\bibfnamefont {M.~B.}\ \bibnamefont
  {Davies}}\ and\ \bibinfo {author} {\bibfnamefont {A.~R.}\ \bibnamefont
  {King}},\ }\bibfield  {title} {\bibinfo {title} {{The Stars of the Galactic
  Center}},\ }\href {https://doi.org/10.1086/430308} {\bibfield  {journal}
  {\bibinfo  {journal} {{\APJL}}\ }\textbf {\bibinfo {volume} {624}},\ \bibinfo
  {pages} {L25} (\bibinfo {year} {2005})},\ \Eprint
  {https://arxiv.org/abs/astro-ph/0503441} {arXiv:astro-ph/0503441}
  \BibitemShut {NoStop}%
\bibitem [{\citenamefont {Levin}(2003)}]{Levin:2003ej}%
  \BibitemOpen
  \bibfield  {author} {\bibinfo {author} {\bibfnamefont {Y.}~\bibnamefont
  {Levin}},\ }\bibfield  {title} {\bibinfo {title} {{Formation of massive stars
  and black holes in selfgravitating AGN discs, and gravitational waves in LISA
  band}},\ }\href@noop {} {\  (\bibinfo {year} {2003})},\ \Eprint
  {https://arxiv.org/abs/astro-ph/0307084} {arXiv:astro-ph/0307084}
  \BibitemShut {NoStop}%
\bibitem [{\citenamefont {Levin}(2007)}]{Levin:2006uc}%
  \BibitemOpen
  \bibfield  {author} {\bibinfo {author} {\bibfnamefont {Y.}~\bibnamefont
  {Levin}},\ }\bibfield  {title} {\bibinfo {title} {{Starbursts near
  supermassive black holes: young stars in the Galactic Center, and
  gravitational waves in LISA band}},\ }\href
  {https://doi.org/10.1111/j.1365-2966.2006.11155.x} {\bibfield  {journal}
  {\bibinfo  {journal} {{\MNRAS}}\ }\textbf {\bibinfo {volume} {374}},\
  \bibinfo {pages} {515} (\bibinfo {year} {2007})},\ \Eprint
  {https://arxiv.org/abs/astro-ph/0603583} {arXiv:astro-ph/0603583}
  \BibitemShut {NoStop}%
\bibitem [{\citenamefont {Secunda}\ \emph {et~al.}(2021)\citenamefont
  {Secunda}, \citenamefont {Hernandez}, \citenamefont {Goodman}, \citenamefont
  {Leigh}, \citenamefont {McKernan}, \citenamefont {Ford},\ and\ \citenamefont
  {Adorno}}]{Secunda:2020cdw}%
  \BibitemOpen
  \bibfield  {author} {\bibinfo {author} {\bibfnamefont {A.}~\bibnamefont
  {Secunda}}, \bibinfo {author} {\bibfnamefont {B.}~\bibnamefont {Hernandez}},
  \bibinfo {author} {\bibfnamefont {J.}~\bibnamefont {Goodman}}, \bibinfo
  {author} {\bibfnamefont {N.~W.~C.}\ \bibnamefont {Leigh}}, \bibinfo {author}
  {\bibfnamefont {B.}~\bibnamefont {McKernan}}, \bibinfo {author}
  {\bibfnamefont {K.~E.~S.}\ \bibnamefont {Ford}},\ and\ \bibinfo {author}
  {\bibfnamefont {J.~I.}\ \bibnamefont {Adorno}},\ }\bibfield  {title}
  {\bibinfo {title} {{Evolution of retrograde orbiters in an active galactic
  nucleus disk}},\ }\href {https://doi.org/10.3847/2041-8213/abe11d} {\bibfield
   {journal} {\bibinfo  {journal} {{\APJL}}\ }\textbf {\bibinfo {volume}
  {908}},\ \bibinfo {pages} {L27} (\bibinfo {year} {2021})},\ \Eprint
  {https://arxiv.org/abs/2009.03910} {arXiv:2009.03910 [astro-ph.HE]}
  \BibitemShut {NoStop}%
\bibitem [{\citenamefont {Pan}\ and\ \citenamefont {Yang}(2021)}]{Pan:2021ksp}%
  \BibitemOpen
  \bibfield  {author} {\bibinfo {author} {\bibfnamefont {Z.}~\bibnamefont
  {Pan}}\ and\ \bibinfo {author} {\bibfnamefont {H.}~\bibnamefont {Yang}},\
  }\bibfield  {title} {\bibinfo {title} {{Formation rate of extreme mass ratio
  inspirals in active galactic nuclei}},\ }\href
  {https://doi.org/10.1103/PhysRevD.103.103018} {\bibfield  {journal} {\bibinfo
   {journal} {{\PRD}}\ }\textbf {\bibinfo {volume} {103}},\ \bibinfo {pages}
  {103018} (\bibinfo {year} {2021})},\ \Eprint
  {https://arxiv.org/abs/2101.09146} {arXiv:2101.09146 [astro-ph.HE]}
  \BibitemShut {NoStop}%
\bibitem [{\citenamefont {Pan}\ \emph {et~al.}(2021)\citenamefont {Pan},
  \citenamefont {Lyu},\ and\ \citenamefont {Yang}}]{Pan:2021oob}%
  \BibitemOpen
  \bibfield  {author} {\bibinfo {author} {\bibfnamefont {Z.}~\bibnamefont
  {Pan}}, \bibinfo {author} {\bibfnamefont {Z.}~\bibnamefont {Lyu}},\ and\
  \bibinfo {author} {\bibfnamefont {H.}~\bibnamefont {Yang}},\ }\bibfield
  {title} {\bibinfo {title} {{Wet extreme mass ratio inspirals may be more
  common for spaceborne gravitational wave detection}},\ }\href
  {https://doi.org/10.1103/PhysRevD.104.063007} {\bibfield  {journal} {\bibinfo
   {journal} {{\PRD}}\ }\textbf {\bibinfo {volume} {104}},\ \bibinfo {pages}
  {063007} (\bibinfo {year} {2021})},\ \Eprint
  {https://arxiv.org/abs/2104.01208} {arXiv:2104.01208 [astro-ph.HE]}
  \BibitemShut {NoStop}%
\bibitem [{\citenamefont {Bortolas}\ and\ \citenamefont
  {Mapelli}(2019)}]{Bortolas:2019sif}%
  \BibitemOpen
  \bibfield  {author} {\bibinfo {author} {\bibfnamefont {E.}~\bibnamefont
  {Bortolas}}\ and\ \bibinfo {author} {\bibfnamefont {M.}~\bibnamefont
  {Mapelli}},\ }\bibfield  {title} {\bibinfo {title} {{Can supernova kicks
  trigger EMRIs in the galactic Centre?}},\ }\href
  {https://doi.org/10.1093/mnras/stz440} {\bibfield  {journal} {\bibinfo
  {journal} {{\MNRAS}}\ }\textbf {\bibinfo {volume} {485}},\ \bibinfo {pages}
  {2125} (\bibinfo {year} {2019})},\ \Eprint {https://arxiv.org/abs/1902.04581}
  {arXiv:1902.04581 [astro-ph.GA]} \BibitemShut {NoStop}%
\bibitem [{\citenamefont {Berry}\ \emph {et~al.}(2019)\citenamefont {Berry},
  \citenamefont {Hughes}, \citenamefont {Sopuerta}, \citenamefont {Chua},
  \citenamefont {Heffernan}, \citenamefont {Holley-Bockelmann}, \citenamefont
  {Mihaylov}, \citenamefont {Miller},\ and\ \citenamefont
  {Sesana}}]{Berry:2019wgg}%
  \BibitemOpen
  \bibfield  {author} {\bibinfo {author} {\bibfnamefont {C.~P.~L.}\
  \bibnamefont {Berry}}, \bibinfo {author} {\bibfnamefont {S.~A.}\ \bibnamefont
  {Hughes}}, \bibinfo {author} {\bibfnamefont {C.~F.}\ \bibnamefont
  {Sopuerta}}, \bibinfo {author} {\bibfnamefont {A.~J.~K.}\ \bibnamefont
  {Chua}}, \bibinfo {author} {\bibfnamefont {A.}~\bibnamefont {Heffernan}},
  \bibinfo {author} {\bibfnamefont {K.}~\bibnamefont {Holley-Bockelmann}},
  \bibinfo {author} {\bibfnamefont {D.~P.}\ \bibnamefont {Mihaylov}}, \bibinfo
  {author} {\bibfnamefont {M.~C.}\ \bibnamefont {Miller}},\ and\ \bibinfo
  {author} {\bibfnamefont {A.}~\bibnamefont {Sesana}},\ }\bibfield  {title}
  {\bibinfo {title} {{The unique potential of extreme mass-ratio inspirals for
  gravitational-wave astronomy}},\ }\href@noop {} {\bibfield  {journal}
  {\bibinfo  {journal} {{\BAAS}}\ }\textbf {\bibinfo {volume} {51}},\ \bibinfo
  {pages} {42} (\bibinfo {year} {2019})},\ \Eprint
  {https://arxiv.org/abs/1903.03686} {arXiv:1903.03686 [astro-ph.HE]}
  \BibitemShut {NoStop}%
\bibitem [{\citenamefont {Datta}\ and\ \citenamefont
  {Bose}(2019)}]{Datta:2019euh}%
  \BibitemOpen
  \bibfield  {author} {\bibinfo {author} {\bibfnamefont {S.}~\bibnamefont
  {Datta}}\ and\ \bibinfo {author} {\bibfnamefont {S.}~\bibnamefont {Bose}},\
  }\bibfield  {title} {\bibinfo {title} {{Probing the nature of central objects
  in extreme-mass-ratio inspirals with gravitational waves}},\ }\href
  {https://doi.org/10.1103/PhysRevD.99.084001} {\bibfield  {journal} {\bibinfo
  {journal} {{\PRD}}\ }\textbf {\bibinfo {volume} {99}},\ \bibinfo {pages}
  {084001} (\bibinfo {year} {2019})},\ \Eprint
  {https://arxiv.org/abs/1902.01723} {arXiv:1902.01723 [gr-qc]} \BibitemShut
  {NoStop}%
\bibitem [{\citenamefont {Cárdenas-Avendaño}\ and\ \citenamefont
  {Sopuerta}(2024)}]{Cardenas-Avendano:2024}%
  \BibitemOpen
  \bibfield  {author} {\bibinfo {author} {\bibfnamefont {A.}~\bibnamefont
  {Cárdenas-Avendaño}}\ and\ \bibinfo {author} {\bibfnamefont {C.~F.}\
  \bibnamefont {Sopuerta}},\ }\bibinfo {title} {Testing gravity with
  extreme-mass-ratio inspirals},\ in\ \href
  {https://doi.org/10.1007/978-981-97-2871-8_8} {\emph {\bibinfo {booktitle}
  {Recent Progress on Gravity Tests}}}\ (\bibinfo  {publisher} {Springer Nature
  Singapore},\ \bibinfo {year} {2024})\ p.\ \bibinfo {pages}
  {275-359}\BibitemShut {NoStop}%
\bibitem [{\citenamefont {Sopuerta}(2010)}]{Sopuerta:2010fte}%
  \BibitemOpen
  \bibfield  {author} {\bibinfo {author} {\bibfnamefont {C.~F.}\ \bibnamefont
  {Sopuerta}},\ }\bibfield  {title} {\bibinfo {title} {{Gravitational Waves
  Notes, Issue \#4 : 'A Roadmap to Fundamental Physics from LISA EMRI
  Observations'}}\ }(\bibinfo  {publisher} {{Max Planck Institute for
  Gravitational Physics. Albert Einstein Institute}},\ \bibinfo {year} {2010})\
  \Eprint {https://arxiv.org/abs/1009.1402} {arXiv:1009.1402 [astro-ph.CO]}
  \BibitemShut {NoStop}%
\bibitem [{\citenamefont {Sopuerta}(2013)}]{Sopuerta:2012hg}%
  \BibitemOpen
  \bibfield  {author} {\bibinfo {author} {\bibfnamefont {C.~F.}\ \bibnamefont
  {Sopuerta}},\ }\bibfield  {title} {\bibinfo {title} {{Probing the strong
  gravity regime with eLISA: Progress on EMRIs}},\ }\href@noop {} {\bibfield
  {journal} {\bibinfo  {journal} {{\ACS}}\ }\textbf {\bibinfo {volume} {467}},\
  \bibinfo {pages} {69} (\bibinfo {year} {2013})},\ \Eprint
  {https://arxiv.org/abs/1210.0156} {arXiv:1210.0156 [gr-qc]} \BibitemShut
  {NoStop}%
\bibitem [{\citenamefont {Hawking}\ and\ \citenamefont
  {Ellis}(2011)}]{Hawking:1973uf}%
  \BibitemOpen
  \bibfield  {author} {\bibinfo {author} {\bibfnamefont {S.~W.}\ \bibnamefont
  {Hawking}}\ and\ \bibinfo {author} {\bibfnamefont {G.~F.~R.}\ \bibnamefont
  {Ellis}},\ }\href {https://doi.org/10.1017/CBO9780511524646} {\emph {\bibinfo
  {title} {{The Large Scale Structure of Space-Time}}}},\ Cambridge Monographs
  on Mathematical Physics\ (\bibinfo  {publisher} {Cambridge University
  Press},\ \bibinfo {year} {2011})\BibitemShut {NoStop}%
\bibitem [{\citenamefont {Ryan}(1995{\natexlab{a}})}]{Ryan:1995wh}%
  \BibitemOpen
  \bibfield  {author} {\bibinfo {author} {\bibfnamefont {F.~D.}\ \bibnamefont
  {Ryan}},\ }\bibfield  {title} {\bibinfo {title} {{Gravitational waves from
  the inspiral of a compact object into a massive, axisymmetric body with
  arbitrary multipole moments}},\ }\href
  {https://doi.org/10.1103/PhysRevD.52.5707} {\bibfield  {journal} {\bibinfo
  {journal} {{\PRD}}\ }\textbf {\bibinfo {volume} {52}},\ \bibinfo {pages}
  {5707} (\bibinfo {year} {1995}{\natexlab{a}})}\BibitemShut {NoStop}%
\bibitem [{\citenamefont {Ryan}(1995{\natexlab{b}})}]{Ryan:1995zm}%
  \BibitemOpen
  \bibfield  {author} {\bibinfo {author} {\bibfnamefont {F.~D.}\ \bibnamefont
  {Ryan}},\ }\bibfield  {title} {\bibinfo {title} {{Effect of gravitational
  radiation reaction on circular orbits around a spinning black hole}},\ }\href
  {https://doi.org/10.1103/PhysRevD.52.R3159} {\bibfield  {journal} {\bibinfo
  {journal} {{\PRD}}\ }\textbf {\bibinfo {volume} {52}},\ \bibinfo {pages}
  {3159} (\bibinfo {year} {1995}{\natexlab{b}})},\ \Eprint
  {https://arxiv.org/abs/gr-qc/9506023} {arXiv:gr-qc/9506023} \BibitemShut
  {NoStop}%
\bibitem [{\citenamefont {Ryan}(1996)}]{Ryan:1995xi}%
  \BibitemOpen
  \bibfield  {author} {\bibinfo {author} {\bibfnamefont {F.~D.}\ \bibnamefont
  {Ryan}},\ }\bibfield  {title} {\bibinfo {title} {{Effect of gravitational
  radiation reaction on nonequatorial orbits around a Kerr black hole}},\
  }\href {https://doi.org/10.1103/PhysRevD.53.3064} {\bibfield  {journal}
  {\bibinfo  {journal} {{\PRD}}\ }\textbf {\bibinfo {volume} {53}},\ \bibinfo
  {pages} {3064} (\bibinfo {year} {1996})},\ \Eprint
  {https://arxiv.org/abs/gr-qc/9511062} {arXiv:gr-qc/9511062} \BibitemShut
  {NoStop}%
\bibitem [{\citenamefont {Ryan}(1997{\natexlab{a}})}]{Ryan:1997hg}%
  \BibitemOpen
  \bibfield  {author} {\bibinfo {author} {\bibfnamefont {F.~D.}\ \bibnamefont
  {Ryan}},\ }\bibfield  {title} {\bibinfo {title} {{Accuracy of estimating the
  multipole moments of a massive body from the gravitational waves of a binary
  inspiral}},\ }\href {https://doi.org/10.1103/PhysRevD.56.1845} {\bibfield
  {journal} {\bibinfo  {journal} {{\PRD}}\ }\textbf {\bibinfo {volume} {56}},\
  \bibinfo {pages} {1845} (\bibinfo {year} {1997}{\natexlab{a}})}\BibitemShut
  {NoStop}%
\bibitem [{\citenamefont {Ryan}(1997{\natexlab{b}})}]{Ryan:1997kh}%
  \BibitemOpen
  \bibfield  {author} {\bibinfo {author} {\bibfnamefont {F.~D.}\ \bibnamefont
  {Ryan}},\ }\bibfield  {title} {\bibinfo {title} {{Scalar waves produced by a
  scalar charge orbiting a massive body with arbitrary multipole moments}},\
  }\href {https://doi.org/10.1103/PhysRevD.56.7732} {\bibfield  {journal}
  {\bibinfo  {journal} {{\PRD}}\ }\textbf {\bibinfo {volume} {56}},\ \bibinfo
  {pages} {7732} (\bibinfo {year} {1997}{\natexlab{b}})}\BibitemShut {NoStop}%
\bibitem [{\citenamefont {Geroch}(1970)}]{Geroch:1970rg}%
  \BibitemOpen
  \bibfield  {author} {\bibinfo {author} {\bibfnamefont {R.}~\bibnamefont
  {Geroch}},\ }\bibfield  {title} {\bibinfo {title} {Multipole moments. ii.
  Curved space},\ }\href@noop {} {\bibfield  {journal} {\bibinfo  {journal}
  {{\JMP}}\ }\textbf {\bibinfo {volume} {11}},\ \bibinfo {pages} {2580}
  (\bibinfo {year} {1970})}\BibitemShut {NoStop}%
\bibitem [{\citenamefont {Hansen}(1974)}]{Hansen:1974ro}%
  \BibitemOpen
  \bibfield  {author} {\bibinfo {author} {\bibfnamefont {R.~O.}\ \bibnamefont
  {Hansen}},\ }\bibfield  {title} {\bibinfo {title} {Multipole moments of
  stationary space-times},\ }\href@noop {} {\bibfield  {journal} {\bibinfo
  {journal} {{\JMP}}\ }\textbf {\bibinfo {volume} {15}},\ \bibinfo {pages}
  {4652} (\bibinfo {year} {1974})}\BibitemShut {NoStop}%
\bibitem [{\citenamefont {Thorne}(1980)}]{Thorne:1980rm}%
  \BibitemOpen
  \bibfield  {author} {\bibinfo {author} {\bibfnamefont {K.~S.}\ \bibnamefont
  {Thorne}},\ }\bibfield  {title} {\bibinfo {title} {Multipole expansions of
  gravitational radiation},\ }\href@noop {} {\bibfield  {journal} {\bibinfo
  {journal} {{\RMP}}\ }\textbf {\bibinfo {volume} {52}},\ \bibinfo {pages}
  {299} (\bibinfo {year} {1980})}\BibitemShut {NoStop}%
\bibitem [{\citenamefont {Fodor}\ \emph {et~al.}(1989)\citenamefont {Fodor},
  \citenamefont {Hoenselaers},\ and\ \citenamefont {Perj\'es}}]{Fodor:1989gf}%
  \BibitemOpen
  \bibfield  {author} {\bibinfo {author} {\bibfnamefont {G.}~\bibnamefont
  {Fodor}}, \bibinfo {author} {\bibfnamefont {C.}~\bibnamefont {Hoenselaers}},\
  and\ \bibinfo {author} {\bibfnamefont {Z.}~\bibnamefont {Perj\'es}},\
  }\bibfield  {title} {\bibinfo {title} {Multipole moments of axisymmetric
  systems in relativity},\ }\href@noop {} {\bibfield  {journal} {\bibinfo
  {journal} {{\JMP}}\ }\textbf {\bibinfo {volume} {30}},\ \bibinfo {pages}
  {2252} (\bibinfo {year} {1989})}\BibitemShut {NoStop}%
\bibitem [{\citenamefont {Ashtekar}\ \emph {et~al.}(2004)\citenamefont
  {Ashtekar}, \citenamefont {Engle}, \citenamefont {Pawlowski},\ and\
  \citenamefont {Van Den~Broeck}}]{Ashtekar:2004gp}%
  \BibitemOpen
  \bibfield  {author} {\bibinfo {author} {\bibfnamefont {A.}~\bibnamefont
  {Ashtekar}}, \bibinfo {author} {\bibfnamefont {J.}~\bibnamefont {Engle}},
  \bibinfo {author} {\bibfnamefont {T.}~\bibnamefont {Pawlowski}},\ and\
  \bibinfo {author} {\bibfnamefont {C.}~\bibnamefont {Van Den~Broeck}},\
  }\bibfield  {title} {\bibinfo {title} {{Multipole moments of isolated
  horizons}},\ }\href {https://doi.org/10.1088/0264-9381/21/11/003} {\bibfield
  {journal} {\bibinfo  {journal} {{\CQG}}\ }\textbf {\bibinfo {volume} {21}},\
  \bibinfo {pages} {2549} (\bibinfo {year} {2004})},\ \Eprint
  {https://arxiv.org/abs/gr-qc/0401114} {arXiv:gr-qc/0401114} \BibitemShut
  {NoStop}%
\bibitem [{\citenamefont {Ashtekar}\ \emph {et~al.}(2022)\citenamefont
  {Ashtekar}, \citenamefont {Khera}, \citenamefont {Kolanowski},\ and\
  \citenamefont {Lewandowski}}]{Ashtekar:2021wld}%
  \BibitemOpen
  \bibfield  {author} {\bibinfo {author} {\bibfnamefont {A.}~\bibnamefont
  {Ashtekar}}, \bibinfo {author} {\bibfnamefont {N.}~\bibnamefont {Khera}},
  \bibinfo {author} {\bibfnamefont {M.}~\bibnamefont {Kolanowski}},\ and\
  \bibinfo {author} {\bibfnamefont {J.}~\bibnamefont {Lewandowski}},\
  }\bibfield  {title} {\bibinfo {title} {{Non-expanding horizons: Multipoles
  and the symmetry group}},\ }\href {https://doi.org/10.1007/JHEP01(2022)028}
  {\bibfield  {journal} {\bibinfo  {journal} {{JHEP}}\ }\textbf {\bibinfo
  {volume} {01}},\ \bibinfo {pages} {028} (\bibinfo {year} {2022})},\ \Eprint
  {https://arxiv.org/abs/2111.07873} {arXiv:2111.07873 [gr-qc]} \BibitemShut
  {NoStop}%
\bibitem [{\citenamefont {Gourgoulhon}\ \emph {et~al.}(2026)\citenamefont
  {Gourgoulhon}, \citenamefont {Tiec},\ and\ \citenamefont
  {Casals}}]{Gourgoulhon:2026nes}%
  \BibitemOpen
  \bibfield  {author} {\bibinfo {author} {\bibfnamefont {E.}~\bibnamefont
  {Gourgoulhon}}, \bibinfo {author} {\bibfnamefont {A.~L.}\ \bibnamefont
  {Tiec}},\ and\ \bibinfo {author} {\bibfnamefont {M.}~\bibnamefont {Casals}},\
  }\bibfield  {title} {\bibinfo {title} {{Horizon multipole moments of a Kerr
  black hole}},\ }\href {https://doi.org/10.1103/rgbl-yk9n} {\bibfield
  {journal} {\bibinfo  {journal} {Phys. Rev. D}\ }\textbf {\bibinfo {volume}
  {114}},\ \bibinfo {pages} {024014} (\bibinfo {year} {2026})},\ \Eprint
  {https://arxiv.org/abs/2602.05823} {arXiv:2602.05823 [gr-qc]} \BibitemShut
  {NoStop}%
\bibitem [{\citenamefont {Israel}(1966)}]{Israel:1966rt}%
  \BibitemOpen
  \bibfield  {author} {\bibinfo {author} {\bibfnamefont {W.}~\bibnamefont
  {Israel}},\ }\bibfield  {title} {\bibinfo {title} {{Singular hypersurfaces
  and thin shells in general relativity}},\ }\href
  {https://doi.org/10.1007/BF02710419} {\bibfield  {journal} {\bibinfo
  {journal} {{\NCB}}\ }\textbf {\bibinfo {volume} {44S10}},\ \bibinfo {pages}
  {1} (\bibinfo {year} {1966})},\ \bibinfo {note} {[Erratum: Nuovo Cim.B 48,
  463 (1967)]}\BibitemShut {NoStop}%
\bibitem [{\citenamefont {Carter}(1971)}]{Carter:1971zc}%
  \BibitemOpen
  \bibfield  {author} {\bibinfo {author} {\bibfnamefont {B.}~\bibnamefont
  {Carter}},\ }\bibfield  {title} {\bibinfo {title} {{Axisymmetric black hole
  has only two degrees of freedom}},\ }\href
  {https://doi.org/10.1103/PhysRevLett.26.331} {\bibfield  {journal} {\bibinfo
  {journal} {{\PRL}}\ }\textbf {\bibinfo {volume} {26}},\ \bibinfo {pages}
  {331} (\bibinfo {year} {1971})}\BibitemShut {NoStop}%
\bibitem [{\citenamefont {Hawking}(1972)}]{Hawking:1972}%
  \BibitemOpen
  \bibfield  {author} {\bibinfo {author} {\bibfnamefont {S.~W.}\ \bibnamefont
  {Hawking}},\ }\bibfield  {title} {\bibinfo {title} {Black holes in general
  relativity},\ }\href {https://doi.org/10.1007/BF01877517} {\bibfield
  {journal} {\bibinfo  {journal} {{\CMP}}\ }\textbf {\bibinfo {volume} {25}},\
  \bibinfo {pages} {152} (\bibinfo {year} {1972})}\BibitemShut {NoStop}%
\bibitem [{\citenamefont {Chruściel}\ \emph {et~al.}(2012)\citenamefont
  {Chruściel}, \citenamefont {Costa},\ and\ \citenamefont
  {Heusler}}]{Chruciel:2012}%
  \BibitemOpen
  \bibfield  {author} {\bibinfo {author} {\bibfnamefont {P.~T.}\ \bibnamefont
  {Chruściel}}, \bibinfo {author} {\bibfnamefont {J.~L.}\ \bibnamefont
  {Costa}},\ and\ \bibinfo {author} {\bibfnamefont {M.}~\bibnamefont
  {Heusler}},\ }\bibfield  {title} {\bibinfo {title} {Stationary black holes:
  Uniqueness and beyond},\ }\bibfield  {journal} {\bibinfo  {journal} {Living
  Reviews in Relativity}\ }\textbf {\bibinfo {volume} {15}},\ \href
  {https://doi.org/10.12942/lrr-2012-7} {10.12942/lrr-2012-7} (\bibinfo {year}
  {2012})\BibitemShut {NoStop}%
\bibitem [{\citenamefont {Danzmann}(1996)}]{Danzmann:1996da}%
  \BibitemOpen
  \bibfield  {author} {\bibinfo {author} {\bibfnamefont {K.}~\bibnamefont
  {Danzmann}},\ }\bibfield  {title} {\bibinfo {title} {{LISA: Laser
  interferometer space antenna for gravitational wave measurements}},\ }\href
  {https://doi.org/10.1088/0264-9381/13/11A/033} {\bibfield  {journal}
  {\bibinfo  {journal} {Classical Quantum Gravity}\ }\textbf {\bibinfo {volume}
  {13}},\ \bibinfo {pages} {A247} (\bibinfo {year} {1996})}\BibitemShut
  {NoStop}%
\bibitem [{\citenamefont {Cutler}(1998)}]{Cutler:1998rf}%
  \BibitemOpen
  \bibfield  {author} {\bibinfo {author} {\bibfnamefont {C.}~\bibnamefont
  {Cutler}},\ }\bibfield  {title} {\bibinfo {title} {{Angular resolution of the
  LISA gravitational wave detector}},\ }\href
  {https://doi.org/10.1103/PhysRevD.57.7089} {\bibfield  {journal} {\bibinfo
  {journal} {{\PRD}}\ }\textbf {\bibinfo {volume} {57}},\ \bibinfo {pages}
  {7089} (\bibinfo {year} {1998})},\ \Eprint
  {https://arxiv.org/abs/gr-qc/9703068} {arXiv:gr-qc/9703068} \BibitemShut
  {NoStop}%
\bibitem [{\citenamefont {Prince}\ \emph {et~al.}(2002)\citenamefont {Prince},
  \citenamefont {Tinto}, \citenamefont {Larson},\ and\ \citenamefont
  {Armstrong}}]{Prince:2002hp}%
  \BibitemOpen
  \bibfield  {author} {\bibinfo {author} {\bibfnamefont {T.~A.}\ \bibnamefont
  {Prince}}, \bibinfo {author} {\bibfnamefont {M.}~\bibnamefont {Tinto}},
  \bibinfo {author} {\bibfnamefont {S.~L.}\ \bibnamefont {Larson}},\ and\
  \bibinfo {author} {\bibfnamefont {J.~W.}\ \bibnamefont {Armstrong}},\
  }\bibfield  {title} {\bibinfo {title} {{The LISA optimal sensitivity}},\
  }\href {https://doi.org/10.1103/PhysRevD.66.122002} {\bibfield  {journal}
  {\bibinfo  {journal} {Phys. Rev. D}\ }\textbf {\bibinfo {volume} {66}},\
  \bibinfo {pages} {122002} (\bibinfo {year} {2002})},\ \Eprint
  {https://arxiv.org/abs/gr-qc/0209039} {arXiv:gr-qc/0209039} \BibitemShut
  {NoStop}%
\bibitem [{\citenamefont {Sotiriou}\ and\ \citenamefont
  {Apostolatos}(2005)}]{Sotiriou:2004bm}%
  \BibitemOpen
  \bibfield  {author} {\bibinfo {author} {\bibfnamefont {T.~P.}\ \bibnamefont
  {Sotiriou}}\ and\ \bibinfo {author} {\bibfnamefont {T.~A.}\ \bibnamefont
  {Apostolatos}},\ }\bibfield  {title} {\bibinfo {title} {{Tracing the geometry
  around a massive, axisymmetric body to measure, through gravitational waves,
  its mass moments and electromagnetic moments}},\ }\href
  {https://doi.org/10.1103/PhysRevD.71.044005} {\bibfield  {journal} {\bibinfo
  {journal} {{\PRD}}\ }\textbf {\bibinfo {volume} {71}},\ \bibinfo {pages}
  {044005} (\bibinfo {year} {2005})},\ \Eprint
  {https://arxiv.org/abs/gr-qc/0410102} {arXiv:gr-qc/0410102} \BibitemShut
  {NoStop}%
\bibitem [{\citenamefont {Brink}(2008)}]{Brink:2008xx}%
  \BibitemOpen
  \bibfield  {author} {\bibinfo {author} {\bibfnamefont {J.}~\bibnamefont
  {Brink}},\ }\bibfield  {title} {\bibinfo {title} {{Spacetime encodings I- A
  spacetime reconstruction problem}},\ }\href
  {https://doi.org/10.1103/PhysRevD.78.102001} {\bibfield  {journal} {\bibinfo
  {journal} {{\PRD}}\ }\textbf {\bibinfo {volume} {78}},\ \bibinfo {pages}
  {102001} (\bibinfo {year} {2008})},\ \Eprint
  {https://arxiv.org/abs/0807.1178} {arXiv:0807.1178 [gr-qc]} \BibitemShut
  {NoStop}%
\bibitem [{\citenamefont {Kastha}\ \emph {et~al.}(2018)\citenamefont {Kastha},
  \citenamefont {Gupta}, \citenamefont {Arun}, \citenamefont {Sathyaprakash},\
  and\ \citenamefont {Van Den~Broeck}}]{Kastha:2018bcr}%
  \BibitemOpen
  \bibfield  {author} {\bibinfo {author} {\bibfnamefont {S.}~\bibnamefont
  {Kastha}}, \bibinfo {author} {\bibfnamefont {A.}~\bibnamefont {Gupta}},
  \bibinfo {author} {\bibfnamefont {K.~G.}\ \bibnamefont {Arun}}, \bibinfo
  {author} {\bibfnamefont {B.~S.}\ \bibnamefont {Sathyaprakash}},\ and\
  \bibinfo {author} {\bibfnamefont {C.}~\bibnamefont {Van Den~Broeck}},\
  }\bibfield  {title} {\bibinfo {title} {{Testing the multipole structure of
  compact binaries using gravitational wave observations}},\ }\href
  {https://doi.org/10.1103/PhysRevD.98.124033} {\bibfield  {journal} {\bibinfo
  {journal} {{\PRD}}\ }\textbf {\bibinfo {volume} {98}},\ \bibinfo {pages}
  {124033} (\bibinfo {year} {2018})},\ \Eprint
  {https://arxiv.org/abs/1809.10465} {arXiv:1809.10465 [gr-qc]} \BibitemShut
  {NoStop}%
\bibitem [{\citenamefont {Datta}\ \emph {et~al.}(2020)\citenamefont {Datta},
  \citenamefont {Brito}, \citenamefont {Bose}, \citenamefont {Pani},\ and\
  \citenamefont {Hughes}}]{Datta:2019epe}%
  \BibitemOpen
  \bibfield  {author} {\bibinfo {author} {\bibfnamefont {S.}~\bibnamefont
  {Datta}}, \bibinfo {author} {\bibfnamefont {R.}~\bibnamefont {Brito}},
  \bibinfo {author} {\bibfnamefont {S.}~\bibnamefont {Bose}}, \bibinfo {author}
  {\bibfnamefont {P.}~\bibnamefont {Pani}},\ and\ \bibinfo {author}
  {\bibfnamefont {S.~A.}\ \bibnamefont {Hughes}},\ }\bibfield  {title}
  {\bibinfo {title} {{Tidal heating as a discriminator for horizons in extreme
  mass ratio inspirals}},\ }\href {https://doi.org/10.1103/PhysRevD.101.044004}
  {\bibfield  {journal} {\bibinfo  {journal} {{\PRD}}\ }\textbf {\bibinfo
  {volume} {101}},\ \bibinfo {pages} {044004} (\bibinfo {year} {2020})},\
  \Eprint {https://arxiv.org/abs/1910.07841} {arXiv:1910.07841 [gr-qc]}
  \BibitemShut {NoStop}%
\bibitem [{\citenamefont {{Manko}}\ and\ \citenamefont
  {{Novikov}}(1992)}]{Manko:1992mn}%
  \BibitemOpen
  \bibfield  {author} {\bibinfo {author} {\bibfnamefont {V.~S.}\ \bibnamefont
  {{Manko}}}\ and\ \bibinfo {author} {\bibfnamefont {I.~D.}\ \bibnamefont
  {{Novikov}}},\ }\bibfield  {title} {\bibinfo {title} {{Generalizations of the
  Kerr and Kerr-Newman metrics possessing an arbitrary set of mass-multipole
  moments}},\ }\href {https://doi.org/10.1088/0264-9381/9/11/013} {\bibfield
  {journal} {\bibinfo  {journal} {{\CQG}}\ }\textbf {\bibinfo {volume} {9}},\
  \bibinfo {pages} {2477} (\bibinfo {year} {1992})}\BibitemShut {NoStop}%
\bibitem [{\citenamefont {Chowdhuri}\ \emph {et~al.}(2024)\citenamefont
  {Chowdhuri}, \citenamefont {Bhattacharyya},\ and\ \citenamefont
  {Kumar}}]{AbhishekChowdhuri:2023gvu}%
  \BibitemOpen
  \bibfield  {author} {\bibinfo {author} {\bibfnamefont {A.}~\bibnamefont
  {Chowdhuri}}, \bibinfo {author} {\bibfnamefont {A.}~\bibnamefont
  {Bhattacharyya}},\ and\ \bibinfo {author} {\bibfnamefont {S.}~\bibnamefont
  {Kumar}},\ }\bibfield  {title} {\bibinfo {title} {{Prospects of detecting
  deviations to Kerr geometry with radiation reaction effects in EMRIs}},\
  }\href {https://doi.org/10.1088/1475-7516/2024/04/001} {\bibfield  {journal}
  {\bibinfo  {journal} {{JCAP}}\ }\textbf {\bibinfo {volume} {04}},\ \bibinfo
  {pages} {001} (\bibinfo {year} {2024})},\ \Eprint
  {https://arxiv.org/abs/2311.05983} {arXiv:2311.05983 [gr-qc]} \BibitemShut
  {NoStop}%
\bibitem [{\citenamefont {Collins}\ and\ \citenamefont
  {Hughes}(2004)}]{Collins:2004na}%
  \BibitemOpen
  \bibfield  {author} {\bibinfo {author} {\bibfnamefont {N.~A.}\ \bibnamefont
  {Collins}}\ and\ \bibinfo {author} {\bibfnamefont {S.~A.}\ \bibnamefont
  {Hughes}},\ }\bibfield  {title} {\bibinfo {title} {{Towards a formalism for
  mapping the spacetimes of massive compact objects: Bumpy black holes and
  their orbits}},\ }\href {https://doi.org/10.1103/PhysRevD.69.124022}
  {\bibfield  {journal} {\bibinfo  {journal} {{\PRD}}\ }\textbf {\bibinfo
  {volume} {69}},\ \bibinfo {pages} {124022} (\bibinfo {year} {2004})},\
  \Eprint {https://arxiv.org/abs/gr-qc/0402063} {arXiv:gr-qc/0402063}
  \BibitemShut {NoStop}%
\bibitem [{\citenamefont {Vigeland}\ and\ \citenamefont
  {Hughes}(2010)}]{Vigeland:2009pr}%
  \BibitemOpen
  \bibfield  {author} {\bibinfo {author} {\bibfnamefont {S.~J.}\ \bibnamefont
  {Vigeland}}\ and\ \bibinfo {author} {\bibfnamefont {S.~A.}\ \bibnamefont
  {Hughes}},\ }\bibfield  {title} {\bibinfo {title} {{Spacetime and orbits of
  bumpy black holes}},\ }\href {https://doi.org/10.1103/PhysRevD.81.024030}
  {\bibfield  {journal} {\bibinfo  {journal} {{\PRD}}\ }\textbf {\bibinfo
  {volume} {81}},\ \bibinfo {pages} {024030} (\bibinfo {year} {2010})},\
  \Eprint {https://arxiv.org/abs/0911.1756} {arXiv:0911.1756 [gr-qc]}
  \BibitemShut {NoStop}%
\bibitem [{\citenamefont {Vigeland}(2010)}]{Vigeland:2010xe}%
  \BibitemOpen
  \bibfield  {author} {\bibinfo {author} {\bibfnamefont {S.~J.}\ \bibnamefont
  {Vigeland}},\ }\bibfield  {title} {\bibinfo {title} {{Multipole moments of
  bumpy black holes}},\ }\href {https://doi.org/10.1103/PhysRevD.82.104041}
  {\bibfield  {journal} {\bibinfo  {journal} {{\PRD}}\ }\textbf {\bibinfo
  {volume} {82}},\ \bibinfo {pages} {104041} (\bibinfo {year} {2010})},\
  \Eprint {https://arxiv.org/abs/1008.1278} {arXiv:1008.1278 [gr-qc]}
  \BibitemShut {NoStop}%
\bibitem [{\citenamefont {Vigeland}\ \emph {et~al.}(2011)\citenamefont
  {Vigeland}, \citenamefont {Yunes},\ and\ \citenamefont
  {Stein}}]{Vigeland:2011ji}%
  \BibitemOpen
  \bibfield  {author} {\bibinfo {author} {\bibfnamefont {S.}~\bibnamefont
  {Vigeland}}, \bibinfo {author} {\bibfnamefont {N.}~\bibnamefont {Yunes}},\
  and\ \bibinfo {author} {\bibfnamefont {L.}~\bibnamefont {Stein}},\ }\bibfield
   {title} {\bibinfo {title} {{Bumpy black holes in alternate theories of
  gravity}},\ }\href {https://doi.org/10.1103/PhysRevD.83.104027} {\bibfield
  {journal} {\bibinfo  {journal} {{\PRD}}\ }\textbf {\bibinfo {volume} {83}},\
  \bibinfo {pages} {104027} (\bibinfo {year} {2011})},\ \Eprint
  {https://arxiv.org/abs/1102.3706} {arXiv:1102.3706 [gr-qc]} \BibitemShut
  {NoStop}%
\bibitem [{\citenamefont {Moore}\ \emph {et~al.}(2017)\citenamefont {Moore},
  \citenamefont {Chua},\ and\ \citenamefont {Gair}}]{Moore:2017lxy}%
  \BibitemOpen
  \bibfield  {author} {\bibinfo {author} {\bibfnamefont {C.~J.}\ \bibnamefont
  {Moore}}, \bibinfo {author} {\bibfnamefont {A.~J.~K.}\ \bibnamefont {Chua}},\
  and\ \bibinfo {author} {\bibfnamefont {J.~R.}\ \bibnamefont {Gair}},\
  }\bibfield  {title} {\bibinfo {title} {{Gravitational waves from extreme mass
  ratio inspirals around bumpy black holes}},\ }\href
  {https://doi.org/10.1088/1361-6382/aa85fa} {\bibfield  {journal} {\bibinfo
  {journal} {{\CQG}}\ }\textbf {\bibinfo {volume} {34}},\ \bibinfo {pages}
  {195009} (\bibinfo {year} {2017})},\ \Eprint
  {https://arxiv.org/abs/1707.00712} {arXiv:1707.00712 [gr-qc]} \BibitemShut
  {NoStop}%
\bibitem [{\citenamefont {Xin}\ \emph {et~al.}(2019)\citenamefont {Xin},
  \citenamefont {Han},\ and\ \citenamefont {Yang}}]{Xin:2018urr}%
  \BibitemOpen
  \bibfield  {author} {\bibinfo {author} {\bibfnamefont {S.}~\bibnamefont
  {Xin}}, \bibinfo {author} {\bibfnamefont {W.-B.}\ \bibnamefont {Han}},\ and\
  \bibinfo {author} {\bibfnamefont {S.-C.}\ \bibnamefont {Yang}},\ }\bibfield
  {title} {\bibinfo {title} {{Gravitational waves from extreme-mass-ratio
  inspirals using general parametrized metrics}},\ }\href
  {https://doi.org/10.1103/PhysRevD.100.084055} {\bibfield  {journal} {\bibinfo
   {journal} {{\PRD}}\ }\textbf {\bibinfo {volume} {100}},\ \bibinfo {pages}
  {084055} (\bibinfo {year} {2019})},\ \Eprint
  {https://arxiv.org/abs/1812.04185} {arXiv:1812.04185 [gr-qc]} \BibitemShut
  {NoStop}%
\bibitem [{\citenamefont {Datta}\ and\ \citenamefont
  {Mukherjee}(2021)}]{Datta:2020axm}%
  \BibitemOpen
  \bibfield  {author} {\bibinfo {author} {\bibfnamefont {S.}~\bibnamefont
  {Datta}}\ and\ \bibinfo {author} {\bibfnamefont {S.}~\bibnamefont
  {Mukherjee}},\ }\bibfield  {title} {\bibinfo {title} {{Possible connection
  between the reflection symmetry and existence of equatorial circular
  orbit}},\ }\href {https://doi.org/10.1103/PhysRevD.103.104032} {\bibfield
  {journal} {\bibinfo  {journal} {Phys. Rev. D}\ }\textbf {\bibinfo {volume}
  {103}},\ \bibinfo {pages} {104032} (\bibinfo {year} {2021})},\ \Eprint
  {https://arxiv.org/abs/2010.12387} {arXiv:2010.12387 [gr-qc]} \BibitemShut
  {NoStop}%
\bibitem [{\citenamefont {LaHaye}\ \emph {et~al.}(2026)\citenamefont {LaHaye},
  \citenamefont {Weller}, \citenamefont {Li}, \citenamefont {Bourg},
  \citenamefont {Chen},\ and\ \citenamefont {Yang}}]{LaHaye:2025ley}%
  \BibitemOpen
  \bibfield  {author} {\bibinfo {author} {\bibfnamefont {M.}~\bibnamefont
  {LaHaye}}, \bibinfo {author} {\bibfnamefont {C.}~\bibnamefont {Weller}},
  \bibinfo {author} {\bibfnamefont {D.}~\bibnamefont {Li}}, \bibinfo {author}
  {\bibfnamefont {P.}~\bibnamefont {Bourg}}, \bibinfo {author} {\bibfnamefont
  {Y.}~\bibnamefont {Chen}},\ and\ \bibinfo {author} {\bibfnamefont
  {H.}~\bibnamefont {Yang}},\ }\bibfield  {title} {\bibinfo {title} {{Evolving
  extreme mass-ratio inspirals in a perturbed Schwarzschild spacetime}},\
  }\href {https://doi.org/10.1103/dy1z-36w5} {\bibfield  {journal} {\bibinfo
  {journal} {Phys. Rev. D}\ }\textbf {\bibinfo {volume} {113}},\ \bibinfo
  {pages} {024069} (\bibinfo {year} {2026})},\ \Eprint
  {https://arxiv.org/abs/2510.16102} {arXiv:2510.16102 [gr-qc]} \BibitemShut
  {NoStop}%
\bibitem [{\citenamefont {Zou}\ \emph {et~al.}(2025)\citenamefont {Zou},
  \citenamefont {Zhong}, \citenamefont {Han},\ and\ \citenamefont
  {Mohanty}}]{Zou:2025fsg}%
  \BibitemOpen
  \bibfield  {author} {\bibinfo {author} {\bibfnamefont {X.}~\bibnamefont
  {Zou}}, \bibinfo {author} {\bibfnamefont {X.}~\bibnamefont {Zhong}}, \bibinfo
  {author} {\bibfnamefont {W.-B.}\ \bibnamefont {Han}},\ and\ \bibinfo {author}
  {\bibfnamefont {S.~D.}\ \bibnamefont {Mohanty}},\ }\bibfield  {title}
  {\bibinfo {title} {{Constraining deviations from the Kerr metric via a bumpy
  parametrization and particle swarm optimization in extreme mass-ratio
  inspirals}},\ }\href {https://doi.org/10.1103/ng1f-ml7m} {\bibfield
  {journal} {\bibinfo  {journal} {Phys. Rev. D}\ }\textbf {\bibinfo {volume}
  {112}},\ \bibinfo {pages} {084075} (\bibinfo {year} {2025})},\ \Eprint
  {https://arxiv.org/abs/2506.21955} {arXiv:2506.21955 [gr-qc]} \BibitemShut
  {NoStop}%
\bibitem [{\citenamefont {Faraji}\ and\ \citenamefont
  {Afshordi}(2026)}]{Faraji:2025duh}%
  \BibitemOpen
  \bibfield  {author} {\bibinfo {author} {\bibfnamefont {S.}~\bibnamefont
  {Faraji}}\ and\ \bibinfo {author} {\bibfnamefont {N.}~\bibnamefont
  {Afshordi}},\ }\bibfield  {title} {\bibinfo {title} {{Self-regularized
  entropy: what does black hole entropy predict for tests of Kerr no-hair
  theorem?}},\ }\href {https://doi.org/10.1088/1361-6382/ae7793} {\bibfield
  {journal} {\bibinfo  {journal} {Classical Quantum Gravity}\ }\textbf {\bibinfo
  {volume} {43}},\ \bibinfo {pages} {135027} (\bibinfo {year} {2026})},\
  \Eprint {https://arxiv.org/abs/2508.20421} {arXiv:2508.20421 [gr-qc]}
  \BibitemShut {NoStop}%
\bibitem [{\citenamefont {Peters}\ and\ \citenamefont
  {Mathews}(1963)}]{Peters:1963pm}%
  \BibitemOpen
  \bibfield  {author} {\bibinfo {author} {\bibfnamefont {P.~C.}\ \bibnamefont
  {Peters}}\ and\ \bibinfo {author} {\bibfnamefont {J.}~\bibnamefont
  {Mathews}},\ }\bibfield  {title} {\bibinfo {title} {{Gravitational radiation
  from point masses in a Keplerian orbit}},\ }\href@noop {} {\bibfield
  {journal} {\bibinfo  {journal} {{\PR}}\ }\textbf {\bibinfo {volume} {131}},\
  \bibinfo {pages} {435} (\bibinfo {year} {1963})}\BibitemShut {NoStop}%
\bibitem [{\citenamefont {{Peters}}(1964)}]{Peters:1964PhRv..136.1224P}%
  \BibitemOpen
  \bibfield  {author} {\bibinfo {author} {\bibfnamefont {P.~C.}\ \bibnamefont
  {{Peters}}},\ }\bibfield  {title} {\bibinfo {title} {{Gravitational radiation
  and the motion of two point masses}},\ }\href
  {https://doi.org/10.1103/PhysRev.136.B1224} {\bibfield  {journal} {\bibinfo
  {journal} {{\PR}}\ }\textbf {\bibinfo {volume} {136}},\ \bibinfo {pages}
  {1224} (\bibinfo {year} {1964})}\BibitemShut {NoStop}%
\bibitem [{\citenamefont {Barack}\ and\ \citenamefont
  {Cutler}(2004)}]{Barack:2003fp}%
  \BibitemOpen
  \bibfield  {author} {\bibinfo {author} {\bibfnamefont {L.}~\bibnamefont
  {Barack}}\ and\ \bibinfo {author} {\bibfnamefont {C.}~\bibnamefont
  {Cutler}},\ }\bibfield  {title} {\bibinfo {title} {{LISA} capture sources:
  Approximate waveforms, signal-to-noise ratios, and parameter estimation
  accuracy},\ }\href@noop {} {\bibfield  {journal} {\bibinfo  {journal}
  {{\PRD}}\ }\textbf {\bibinfo {volume} {69}},\ \bibinfo {pages} {082005}
  (\bibinfo {year} {2004})},\ \Eprint {https://arxiv.org/abs/gr-qc/0310125}
  {gr-qc/0310125} \BibitemShut {NoStop}%
\bibitem [{\citenamefont {Chua}\ \emph {et~al.}(2017)\citenamefont {Chua},
  \citenamefont {Moore},\ and\ \citenamefont {Gair}}]{Chua:2017ujo}%
  \BibitemOpen
  \bibfield  {author} {\bibinfo {author} {\bibfnamefont {A.~J.~K.}\
  \bibnamefont {Chua}}, \bibinfo {author} {\bibfnamefont {C.~J.}\ \bibnamefont
  {Moore}},\ and\ \bibinfo {author} {\bibfnamefont {J.~R.}\ \bibnamefont
  {Gair}},\ }\bibfield  {title} {\bibinfo {title} {{Augmented kludge waveforms
  for detecting extreme-mass-ratio inspirals}},\ }\href
  {https://doi.org/10.1103/PhysRevD.96.044005} {\bibfield  {journal} {\bibinfo
  {journal} {{\PRD}}\ }\textbf {\bibinfo {volume} {96}},\ \bibinfo {pages}
  {044005} (\bibinfo {year} {2017})},\ \Eprint
  {https://arxiv.org/abs/1705.04259} {arXiv:1705.04259 [gr-qc]} \BibitemShut
  {NoStop}%
\bibitem [{\citenamefont {Barack}\ and\ \citenamefont
  {Cutler}(2007)}]{Barack:2006pq}%
  \BibitemOpen
  \bibfield  {author} {\bibinfo {author} {\bibfnamefont {L.}~\bibnamefont
  {Barack}}\ and\ \bibinfo {author} {\bibfnamefont {C.}~\bibnamefont
  {Cutler}},\ }\bibfield  {title} {\bibinfo {title} {{Using LISA EMRI sources
  to test off-Kerr deviations in the geometry of massive black holes}},\ }\href
  {https://doi.org/10.1103/PhysRevD.75.042003} {\bibfield  {journal} {\bibinfo
  {journal} {{\PRD}}\ }\textbf {\bibinfo {volume} {75}},\ \bibinfo {pages}
  {042003} (\bibinfo {year} {2007})},\ \Eprint
  {https://arxiv.org/abs/gr-qc/0612029} {arXiv:gr-qc/0612029} \BibitemShut
  {NoStop}%
\bibitem [{\citenamefont {Cardoso}\ and\ \citenamefont
  {Gualtieri}(2016)}]{Cardoso_2016}%
  \BibitemOpen
  \bibfield  {author} {\bibinfo {author} {\bibfnamefont {V.}~\bibnamefont
  {Cardoso}}\ and\ \bibinfo {author} {\bibfnamefont {L.}~\bibnamefont
  {Gualtieri}},\ }\bibfield  {title} {\bibinfo {title} {Testing the black hole
  ‘no-hair’ hypothesis},\ }\href
  {https://doi.org/10.1088/0264-9381/33/17/174001} {\bibfield  {journal}
  {\bibinfo  {journal} {Classical and Quantum Gravity}\ }\textbf {\bibinfo
  {volume} {33}},\ \bibinfo {pages} {174001} (\bibinfo {year}
  {2016})}\BibitemShut {NoStop}%
\bibitem [{\citenamefont {Fransen}\ and\ \citenamefont
  {Mayerson}(2022)}]{Fransen:2022jtw}%
  \BibitemOpen
  \bibfield  {author} {\bibinfo {author} {\bibfnamefont {K.}~\bibnamefont
  {Fransen}}\ and\ \bibinfo {author} {\bibfnamefont {D.~R.}\ \bibnamefont
  {Mayerson}},\ }\bibfield  {title} {\bibinfo {title} {{Detecting equatorial
  symmetry breaking with LISA}},\ }\href
  {https://doi.org/10.1103/PhysRevD.106.064035} {\bibfield  {journal} {\bibinfo
   {journal} {{\PRD}}\ }\textbf {\bibinfo {volume} {106}},\ \bibinfo {pages}
  {064035} (\bibinfo {year} {2022})},\ \Eprint
  {https://arxiv.org/abs/2201.03569} {arXiv:2201.03569 [gr-qc]} \BibitemShut
  {NoStop}%
\bibitem [{\citenamefont {Loutrel}\ \emph {et~al.}(2022)\citenamefont
  {Loutrel}, \citenamefont {Brito}, \citenamefont {Maselli},\ and\
  \citenamefont {Pani}}]{Loutrel:2022ant}%
  \BibitemOpen
  \bibfield  {author} {\bibinfo {author} {\bibfnamefont {N.}~\bibnamefont
  {Loutrel}}, \bibinfo {author} {\bibfnamefont {R.}~\bibnamefont {Brito}},
  \bibinfo {author} {\bibfnamefont {A.}~\bibnamefont {Maselli}},\ and\ \bibinfo
  {author} {\bibfnamefont {P.}~\bibnamefont {Pani}},\ }\bibfield  {title}
  {\bibinfo {title} {{Inspiraling compact objects with generic deformations}},\
  }\href {https://doi.org/10.1103/PhysRevD.105.124050} {\bibfield  {journal}
  {\bibinfo  {journal} {{\PRD}}\ }\textbf {\bibinfo {volume} {105}},\ \bibinfo
  {pages} {124050} (\bibinfo {year} {2022})},\ \Eprint
  {https://arxiv.org/abs/2203.01725} {arXiv:2203.01725 [gr-qc]} \BibitemShut
  {NoStop}%
\bibitem [{\citenamefont {Loutrel}\ \emph {et~al.}(2024)\citenamefont
  {Loutrel}, \citenamefont {Brito}, \citenamefont {Maselli},\ and\
  \citenamefont {Pani}}]{Loutrel_2024}%
  \BibitemOpen
  \bibfield  {author} {\bibinfo {author} {\bibfnamefont {N.}~\bibnamefont
  {Loutrel}}, \bibinfo {author} {\bibfnamefont {R.}~\bibnamefont {Brito}},
  \bibinfo {author} {\bibfnamefont {A.}~\bibnamefont {Maselli}},\ and\ \bibinfo
  {author} {\bibfnamefont {P.}~\bibnamefont {Pani}},\ }\bibfield  {title}
  {\bibinfo {title} {Relevance of precession for tests of the black hole no
  hair theorems},\ }\bibfield  {journal} {\bibinfo  {journal} {Physical Review
  D}\ }\textbf {\bibinfo {volume} {110}},\ \href
  {https://doi.org/10.1103/physrevd.110.044003} {10.1103/physrevd.110.044003}
  (\bibinfo {year} {2024})\BibitemShut {NoStop}%
\bibitem [{\citenamefont {Mathur}(2005)}]{Mathur:2005zp}%
  \BibitemOpen
  \bibfield  {author} {\bibinfo {author} {\bibfnamefont {S.~D.}\ \bibnamefont
  {Mathur}},\ }\bibfield  {title} {\bibinfo {title} {{The Fuzzball proposal for
  black holes: An Elementary review}},\ }\href
  {https://doi.org/10.1002/prop.200410203} {\bibfield  {journal} {\bibinfo
  {journal} {Fortsch. Phys.}\ }\textbf {\bibinfo {volume} {53}},\ \bibinfo
  {pages} {793} (\bibinfo {year} {2005})},\ \Eprint
  {https://arxiv.org/abs/hep-th/0502050} {arXiv:hep-th/0502050} \BibitemShut
  {NoStop}%
\bibitem [{\citenamefont {Skenderis}\ and\ \citenamefont
  {Taylor}(2007)}]{Skenderis:2006ah}%
  \BibitemOpen
  \bibfield  {author} {\bibinfo {author} {\bibfnamefont {K.}~\bibnamefont
  {Skenderis}}\ and\ \bibinfo {author} {\bibfnamefont {M.}~\bibnamefont
  {Taylor}},\ }\bibfield  {title} {\bibinfo {title} {{Fuzzball solutions and
  D1-D5 microstates}},\ }\href {https://doi.org/10.1103/PhysRevLett.98.071601}
  {\bibfield  {journal} {\bibinfo  {journal} {Phys. Rev. Lett.}\ }\textbf
  {\bibinfo {volume} {98}},\ \bibinfo {pages} {071601} (\bibinfo {year}
  {2007})},\ \Eprint {https://arxiv.org/abs/hep-th/0609154}
  {arXiv:hep-th/0609154} \BibitemShut {NoStop}%
\bibitem [{\citenamefont {Chowdhury}\ and\ \citenamefont
  {Mathur}(2008{\natexlab{a}})}]{Chowdhury:2007jx}%
  \BibitemOpen
  \bibfield  {author} {\bibinfo {author} {\bibfnamefont {B.~D.}\ \bibnamefont
  {Chowdhury}}\ and\ \bibinfo {author} {\bibfnamefont {S.~D.}\ \bibnamefont
  {Mathur}},\ }\bibfield  {title} {\bibinfo {title} {{Radiation from the
  non-extremal fuzzball}},\ }\href
  {https://doi.org/10.1088/0264-9381/25/13/135005} {\bibfield  {journal}
  {\bibinfo  {journal} {Classical Quantum Gravity}\ }\textbf {\bibinfo {volume}
  {25}},\ \bibinfo {pages} {135005} (\bibinfo {year} {2008}{\natexlab{a}})},\
  \Eprint {https://arxiv.org/abs/0711.4817} {arXiv:0711.4817 [hep-th]}
  \BibitemShut {NoStop}%
\bibitem [{\citenamefont {Mathur}(2010)}]{Mathur:2008kg}%
  \BibitemOpen
  \bibfield  {author} {\bibinfo {author} {\bibfnamefont {S.~D.}\ \bibnamefont
  {Mathur}},\ }\bibfield  {title} {\bibinfo {title} {{Tunneling into fuzzball
  states}},\ }\href {https://doi.org/10.1007/s10714-009-0837-3} {\bibfield
  {journal} {\bibinfo  {journal} {Gen. Rel. Grav.}\ }\textbf {\bibinfo {volume}
  {42}},\ \bibinfo {pages} {113} (\bibinfo {year} {2010})},\ \Eprint
  {https://arxiv.org/abs/0805.3716} {arXiv:0805.3716 [hep-th]} \BibitemShut
  {NoStop}%
\bibitem [{\citenamefont {Chowdhury}\ and\ \citenamefont
  {Mathur}(2008{\natexlab{b}})}]{Chowdhury:2008bd}%
  \BibitemOpen
  \bibfield  {author} {\bibinfo {author} {\bibfnamefont {B.~D.}\ \bibnamefont
  {Chowdhury}}\ and\ \bibinfo {author} {\bibfnamefont {S.~D.}\ \bibnamefont
  {Mathur}},\ }\bibfield  {title} {\bibinfo {title} {{Pair creation in
  non-extremal fuzzball geometries}},\ }\href
  {https://doi.org/10.1088/0264-9381/25/22/225021} {\bibfield  {journal}
  {\bibinfo  {journal} {Classical Quantum Gravity}\ }\textbf {\bibinfo {volume}
  {25}},\ \bibinfo {pages} {225021} (\bibinfo {year} {2008}{\natexlab{b}})},\
  \Eprint {https://arxiv.org/abs/0806.2309} {arXiv:0806.2309 [hep-th]}
  \BibitemShut {NoStop}%
\bibitem [{\citenamefont {Skenderis}\ and\ \citenamefont
  {Taylor}(2008)}]{Skenderis:2008qn}%
  \BibitemOpen
  \bibfield  {author} {\bibinfo {author} {\bibfnamefont {K.}~\bibnamefont
  {Skenderis}}\ and\ \bibinfo {author} {\bibfnamefont {M.}~\bibnamefont
  {Taylor}},\ }\bibfield  {title} {\bibinfo {title} {{The fuzzball proposal for
  black holes}},\ }\href {https://doi.org/10.1016/j.physrep.2008.08.001}
  {\bibfield  {journal} {\bibinfo  {journal} {Phys. Rept.}\ }\textbf {\bibinfo
  {volume} {467}},\ \bibinfo {pages} {117} (\bibinfo {year} {2008})},\ \Eprint
  {https://arxiv.org/abs/0804.0552} {arXiv:0804.0552 [hep-th]} \BibitemShut
  {NoStop}%
\bibitem [{\citenamefont {Bacchini}\ \emph {et~al.}(2021)\citenamefont
  {Bacchini}, \citenamefont {Mayerson}, \citenamefont {Ripperda}, \citenamefont
  {Davelaar}, \citenamefont {Olivares}, \citenamefont {Hertog},\ and\
  \citenamefont {Vercnocke}}]{Bacchini:2021fig}%
  \BibitemOpen
  \bibfield  {author} {\bibinfo {author} {\bibfnamefont {F.}~\bibnamefont
  {Bacchini}}, \bibinfo {author} {\bibfnamefont {D.~R.}\ \bibnamefont
  {Mayerson}}, \bibinfo {author} {\bibfnamefont {B.}~\bibnamefont {Ripperda}},
  \bibinfo {author} {\bibfnamefont {J.}~\bibnamefont {Davelaar}}, \bibinfo
  {author} {\bibfnamefont {H.}~\bibnamefont {Olivares}}, \bibinfo {author}
  {\bibfnamefont {T.}~\bibnamefont {Hertog}},\ and\ \bibinfo {author}
  {\bibfnamefont {B.}~\bibnamefont {Vercnocke}},\ }\bibfield  {title} {\bibinfo
  {title} {{Fuzzball shadows: Emergent horizons from microstructure}},\ }\href
  {https://doi.org/10.1103/PhysRevLett.127.171601} {\bibfield  {journal}
  {\bibinfo  {journal} {Phys. Rev. Lett.}\ }\textbf {\bibinfo {volume} {127}},\
  \bibinfo {pages} {171601} (\bibinfo {year} {2021})},\ \Eprint
  {https://arxiv.org/abs/2103.12075} {arXiv:2103.12075 [hep-th]} \BibitemShut
  {NoStop}%
\bibitem [{\citenamefont {Bianchi}\ \emph {et~al.}(2020)\citenamefont
  {Bianchi}, \citenamefont {Consoli}, \citenamefont {Grillo}, \citenamefont
  {Morales}, \citenamefont {Pani},\ and\ \citenamefont
  {Raposo}}]{Bianchi:2020bxa}%
  \BibitemOpen
  \bibfield  {author} {\bibinfo {author} {\bibfnamefont {M.}~\bibnamefont
  {Bianchi}}, \bibinfo {author} {\bibfnamefont {D.}~\bibnamefont {Consoli}},
  \bibinfo {author} {\bibfnamefont {A.}~\bibnamefont {Grillo}}, \bibinfo
  {author} {\bibfnamefont {J.~F.}\ \bibnamefont {Morales}}, \bibinfo {author}
  {\bibfnamefont {P.}~\bibnamefont {Pani}},\ and\ \bibinfo {author}
  {\bibfnamefont {G.}~\bibnamefont {Raposo}},\ }\bibfield  {title} {\bibinfo
  {title} {{Distinguishing fuzzballs from black holes through their multipolar
  structure}},\ }\href {https://doi.org/10.1103/PhysRevLett.125.221601}
  {\bibfield  {journal} {\bibinfo  {journal} {{\PRL}}\ }\textbf {\bibinfo
  {volume} {125}},\ \bibinfo {pages} {221601} (\bibinfo {year} {2020})},\
  \Eprint {https://arxiv.org/abs/2007.01743} {arXiv:2007.01743 [hep-th]}
  \BibitemShut {NoStop}%
\bibitem [{\citenamefont {Melis}\ \emph {et~al.}(2025)\citenamefont {Melis},
  \citenamefont {Brito},\ and\ \citenamefont {Pani}}]{Melis_2025}%
  \BibitemOpen
  \bibfield  {author} {\bibinfo {author} {\bibfnamefont {M.}~\bibnamefont
  {Melis}}, \bibinfo {author} {\bibfnamefont {R.}~\bibnamefont {Brito}},\ and\
  \bibinfo {author} {\bibfnamefont {P.}~\bibnamefont {Pani}},\ }\bibfield
  {title} {\bibinfo {title} {Extreme mass ratio inspirals around topological
  stars},\ }\bibfield  {journal} {\bibinfo  {journal} {Physical Review D}\
  }\textbf {\bibinfo {volume} {111}},\ \href
  {https://doi.org/10.1103/zng8-9qrn} {10.1103/zng8-9qrn} (\bibinfo {year}
  {2025})\BibitemShut {NoStop}%
\bibitem [{\citenamefont {Muguruza}\ and\ \citenamefont
  {Sopuerta}(2026)}]{Muguruza:2026zcs}%
  \BibitemOpen
  \bibfield  {author} {\bibinfo {author} {\bibfnamefont {P.~F.}\ \bibnamefont
  {Muguruza}}\ and\ \bibinfo {author} {\bibfnamefont {C.~F.}\ \bibnamefont
  {Sopuerta}},\ }\bibfield  {title} {\bibinfo {title} {{Probing fuzzballs and
  beyond: Model-independent tests of Kerr symmetry breaking with LISA}},\
  }\href@noop {} {\  (\bibinfo {year} {2026})},\ \Eprint
  {https://arxiv.org/abs/2604.06009} {arXiv:2604.06009 [hep-th]} \BibitemShut
  {NoStop}%
\bibitem [{\citenamefont {Ghosh}\ \emph {et~al.}(2025)\citenamefont {Ghosh},
  \citenamefont {Mishra},\ and\ \citenamefont {Sarkar}}]{Ghosh:2024het}%
  \BibitemOpen
  \bibfield  {author} {\bibinfo {author} {\bibfnamefont {R.}~\bibnamefont
  {Ghosh}}, \bibinfo {author} {\bibfnamefont {A.~K.}\ \bibnamefont {Mishra}},\
  and\ \bibinfo {author} {\bibfnamefont {S.}~\bibnamefont {Sarkar}},\
  }\bibfield  {title} {\bibinfo {title} {{Probing spacetime symmetries using
  gravitational wave ringdown}},\ }\href {https://doi.org/10.1103/myn2-ttrt}
  {\bibfield  {journal} {\bibinfo  {journal} {Phys. Rev. D}\ }\textbf {\bibinfo
  {volume} {112}},\ \bibinfo {pages} {064092} (\bibinfo {year} {2025})},\
  \Eprint {https://arxiv.org/abs/2412.08942} {arXiv:2412.08942 [gr-qc]}
  \BibitemShut {NoStop}%
\bibitem [{\citenamefont {Poisson}\ \emph {et~al.}(2011)\citenamefont
  {Poisson}, \citenamefont {Pound},\ and\ \citenamefont
  {Vega}}]{Poisson:2011nh}%
  \BibitemOpen
  \bibfield  {author} {\bibinfo {author} {\bibfnamefont {E.}~\bibnamefont
  {Poisson}}, \bibinfo {author} {\bibfnamefont {A.}~\bibnamefont {Pound}},\
  and\ \bibinfo {author} {\bibfnamefont {I.}~\bibnamefont {Vega}},\ }\bibfield
  {title} {\bibinfo {title} {{The motion of point particles in curved
  spacetime}},\ }\href {https://doi.org/10.12942/lrr-2011-7} {\bibfield
  {journal} {\bibinfo  {journal} {{\LRR}}\ }\textbf {\bibinfo {volume} {14}},\
  \bibinfo {pages} {7} (\bibinfo {year} {2011})},\ \Eprint
  {https://arxiv.org/abs/1102.0529} {arXiv:1102.0529 [gr-qc]} \BibitemShut
  {NoStop}%
\bibitem [{\citenamefont {Barack}\ and\ \citenamefont
  {Pound}(2019)}]{Barack:2018yvs}%
  \BibitemOpen
  \bibfield  {author} {\bibinfo {author} {\bibfnamefont {L.}~\bibnamefont
  {Barack}}\ and\ \bibinfo {author} {\bibfnamefont {A.}~\bibnamefont {Pound}},\
  }\bibfield  {title} {\bibinfo {title} {{Self-force and radiation reaction in
  general relativity}},\ }\href {https://doi.org/10.1088/1361-6633/aae552}
  {\bibfield  {journal} {\bibinfo  {journal} {{\RPP}}\ }\textbf {\bibinfo
  {volume} {82}},\ \bibinfo {pages} {016904} (\bibinfo {year} {2019})},\
  \Eprint {https://arxiv.org/abs/1805.10385} {arXiv:1805.10385 [gr-qc]}
  \BibitemShut {NoStop}%
\bibitem [{\citenamefont {van~de Meent}(2018)}]{vandeMeent:2017bcc}%
  \BibitemOpen
  \bibfield  {author} {\bibinfo {author} {\bibfnamefont {M.}~\bibnamefont
  {van~de Meent}},\ }\bibfield  {title} {\bibinfo {title} {{Gravitational
  self-force on generic bound geodesics in Kerr spacetime}},\ }\href
  {https://doi.org/10.1103/PhysRevD.97.104033} {\bibfield  {journal} {\bibinfo
  {journal} {{\PRD}}\ }\textbf {\bibinfo {volume} {97}},\ \bibinfo {pages}
  {104033} (\bibinfo {year} {2018})},\ \Eprint
  {https://arxiv.org/abs/1711.09607} {arXiv:1711.09607 [gr-qc]} \BibitemShut
  {NoStop}%
\bibitem [{\citenamefont {Drasco}\ and\ \citenamefont
  {Hughes}(2006)}]{Drasco:2005kz}%
  \BibitemOpen
  \bibfield  {author} {\bibinfo {author} {\bibfnamefont {S.}~\bibnamefont
  {Drasco}}\ and\ \bibinfo {author} {\bibfnamefont {S.~A.}\ \bibnamefont
  {Hughes}},\ }\bibfield  {title} {\bibinfo {title} {{Gravitational wave
  snapshots of generic extreme mass ratio inspirals}},\ }\href
  {https://doi.org/10.1103/PhysRevD.73.024027} {\bibfield  {journal} {\bibinfo
  {journal} {{\PRD}}\ }\textbf {\bibinfo {volume} {73}},\ \bibinfo {pages}
  {024027} (\bibinfo {year} {2006})},\ \Eprint
  {https://arxiv.org/abs/gr-qc/0509101} {arXiv:gr-qc/0509101} \BibitemShut
  {NoStop}%
\bibitem [{\citenamefont {Hinderer}\ and\ \citenamefont
  {Flanagan}(2008)}]{Hinderer:2008dm}%
  \BibitemOpen
  \bibfield  {author} {\bibinfo {author} {\bibfnamefont {T.}~\bibnamefont
  {Hinderer}}\ and\ \bibinfo {author} {\bibfnamefont {E.~E.}\ \bibnamefont
  {Flanagan}},\ }\bibfield  {title} {\bibinfo {title} {{Two timescale analysis
  of extreme mass ratio inspirals in Kerr. I. Orbital motion}},\ }\href
  {https://doi.org/10.1103/PhysRevD.78.064028} {\bibfield  {journal} {\bibinfo
  {journal} {{\PRD}}\ }\textbf {\bibinfo {volume} {78}},\ \bibinfo {pages}
  {064028} (\bibinfo {year} {2008})},\ \Eprint
  {https://arxiv.org/abs/0805.3337} {arXiv:0805.3337 [gr-qc]} \BibitemShut
  {NoStop}%
\bibitem [{\citenamefont {Barack}\ and\ \citenamefont
  {Ori}(2002)}]{Barack:2002mha}%
  \BibitemOpen
  \bibfield  {author} {\bibinfo {author} {\bibfnamefont {L.}~\bibnamefont
  {Barack}}\ and\ \bibinfo {author} {\bibfnamefont {A.}~\bibnamefont {Ori}},\
  }\bibfield  {title} {\bibinfo {title} {{Regularization parameters for the
  self force in Schwarzschild spacetime. I: Scalar case}},\ }\href
  {https://doi.org/10.1103/PhysRevD.66.084022} {\bibfield  {journal} {\bibinfo
  {journal} {{\PRD}}\ }\textbf {\bibinfo {volume} {66}},\ \bibinfo {pages}
  {084022} (\bibinfo {year} {2002})},\ \Eprint
  {https://arxiv.org/abs/gr-qc/0204093} {arXiv:gr-qc/0204093} \BibitemShut
  {NoStop}%
\bibitem [{\citenamefont {Goldstein}(1980)}]{goldstein:mechanics}%
  \BibitemOpen
  \bibfield  {author} {\bibinfo {author} {\bibfnamefont {H.}~\bibnamefont
  {Goldstein}},\ }\href@noop {} {\emph {\bibinfo {title} {Classical
  Mechanics}}}\ (\bibinfo  {publisher} {Addison-Wesley},\ \bibinfo {year}
  {1980})\BibitemShut {NoStop}%
\bibitem [{\citenamefont {Poisson}\ and\ \citenamefont
  {Will}(2014)}]{Poisson:2014pw}%
  \BibitemOpen
  \bibfield  {author} {\bibinfo {author} {\bibfnamefont {E.}~\bibnamefont
  {Poisson}}\ and\ \bibinfo {author} {\bibfnamefont {C.~M.}\ \bibnamefont
  {Will}},\ }\href@noop {} {\emph {\bibinfo {title} {Gravity: Newtonian,
  Post-Newtonian, Relativistic}}}\ (\bibinfo  {publisher} {Cambridge University
  Press},\ \bibinfo {address} {Cambridge (UK)},\ \bibinfo {year}
  {2014})\BibitemShut {NoStop}%
\bibitem [{\citenamefont {Afshordi}\ \emph {et~al.}(2025)\citenamefont
  {Afshordi} \emph {et~al.}}]{LISAConsortiumWaveformWorkingGroup:2023arg}%
  \BibitemOpen
  \bibfield  {author} {\bibinfo {author} {\bibfnamefont {N.}~\bibnamefont
  {Afshordi}} \emph {et~al.} (\bibinfo {Collaboration} {LISA Consortium
  Waveform Working Group}),\ }\bibfield  {title} {\bibinfo {title} {{Waveform
  modelling for the Laser Interferometer Space Antenna}},\ }\href
  {https://doi.org/10.1007/s41114-025-00056-1} {\bibfield  {journal} {\bibinfo
  {journal} {Living Rev. Rel.}\ }\textbf {\bibinfo {volume} {28}},\ \bibinfo
  {pages} {9} (\bibinfo {year} {2025})},\ \Eprint
  {https://arxiv.org/abs/2311.01300} {arXiv:2311.01300 [gr-qc]} \BibitemShut
  {NoStop}%
\bibitem [{\citenamefont {Landau}\ and\ \citenamefont
  {Lifschits}(1975)}]{Landau:1975pou}%
  \BibitemOpen
  \bibfield  {author} {\bibinfo {author} {\bibfnamefont {L.~D.}\ \bibnamefont
  {Landau}}\ and\ \bibinfo {author} {\bibfnamefont {E.~M.}\ \bibnamefont
  {Lifschits}},\ }\href@noop {} {\emph {\bibinfo {title} {{The Classical Theory
  of Fields}}}},\ \bibinfo {series} {Course of Theoretical Physics},
  Vol.~\bibinfo {volume} {2}\ (\bibinfo  {publisher} {Pergamon Press},\
  \bibinfo {address} {Oxford},\ \bibinfo {year} {1975})\BibitemShut {NoStop}%
\bibitem [{\citenamefont {Jos\'e}\ and\ \citenamefont
  {Saletan}(1998)}]{Jose:1998bt}%
  \BibitemOpen
  \bibfield  {author} {\bibinfo {author} {\bibfnamefont {J.~V.}\ \bibnamefont
  {Jos\'e}}\ and\ \bibinfo {author} {\bibfnamefont {E.~J.}\ \bibnamefont
  {Saletan}},\ }\href@noop {} {\emph {\bibinfo {title} {Classical Dynamics: A
  Contemporary Approach}}}\ (\bibinfo  {publisher} {Cambridge University
  Press},\ \bibinfo {address} {Cambridge},\ \bibinfo {year} {1998})\BibitemShut
  {NoStop}%
\bibitem [{\citenamefont {Hinderer}\ and\ \citenamefont
  {Flanagan}(2010)}]{Hinderer:2010}%
  \BibitemOpen
  \bibfield  {author} {\bibinfo {author} {\bibfnamefont {T.}~\bibnamefont
  {Hinderer}}\ and\ \bibinfo {author} {\bibfnamefont {E.~E.}\ \bibnamefont
  {Flanagan}},\ }\bibfield  {title} {\bibinfo {title} {{Evolution of the Carter
  constant for inspirals into a black hole: effect of the black hole
  quadrupole}},\ }\href {https://doi.org/10.1103/PhysRevD.82.129903} {\bibfield
   {journal} {\bibinfo  {journal} {Phys. Rev. D}\ }\textbf {\bibinfo {volume}
  {82}},\ \bibinfo {pages} {129903} (\bibinfo {year} {2010})},\ \Eprint
  {https://arxiv.org/abs/0704.0389} {arXiv:0704.0389 [gr-qc]} \BibitemShut
  {NoStop}%
\bibitem [{\citenamefont {Misner}\ \emph {et~al.}(1973)\citenamefont {Misner},
  \citenamefont {Thorne},\ and\ \citenamefont {Wheeler}}]{Misner:1973cw}%
  \BibitemOpen
  \bibfield  {author} {\bibinfo {author} {\bibfnamefont {C.~W.}\ \bibnamefont
  {Misner}}, \bibinfo {author} {\bibfnamefont {K.}~\bibnamefont {Thorne}},\
  and\ \bibinfo {author} {\bibfnamefont {J.~A.}\ \bibnamefont {Wheeler}},\
  }\href@noop {} {\emph {\bibinfo {title} {{Gravitation}}}}\ (\bibinfo
  {publisher} {W. H. Freeman \& Co.},\ \bibinfo {address} {San Francisco},\
  \bibinfo {year} {1973})\BibitemShut {NoStop}%
\bibitem [{\citenamefont {Maggiore}(2007)}]{Maggiore:2007mm}%
  \BibitemOpen
  \bibfield  {author} {\bibinfo {author} {\bibfnamefont {M.}~\bibnamefont
  {Maggiore}},\ }\href@noop {} {\emph {\bibinfo {title} {{Gravitational Waves:
  Volume 1: Theory and Experiments}}}}\ (\bibinfo  {publisher} {Oxford
  University Press},\ \bibinfo {address} {Oxford},\ \bibinfo {year}
  {2007})\BibitemShut {NoStop}%
\bibitem [{\citenamefont {Zi}\ \emph {et~al.}(2021)\citenamefont {Zi},
  \citenamefont {Zhang}, \citenamefont {Fan}, \citenamefont {Zhang},
  \citenamefont {Hu}, \citenamefont {Shi},\ and\ \citenamefont
  {Mei}}]{Zi:2021pdp}%
  \BibitemOpen
  \bibfield  {author} {\bibinfo {author} {\bibfnamefont {T.-G.}\ \bibnamefont
  {Zi}}, \bibinfo {author} {\bibfnamefont {J.-D.}\ \bibnamefont {Zhang}},
  \bibinfo {author} {\bibfnamefont {H.-M.}\ \bibnamefont {Fan}}, \bibinfo
  {author} {\bibfnamefont {X.-T.}\ \bibnamefont {Zhang}}, \bibinfo {author}
  {\bibfnamefont {Y.-M.}\ \bibnamefont {Hu}}, \bibinfo {author} {\bibfnamefont
  {C.}~\bibnamefont {Shi}},\ and\ \bibinfo {author} {\bibfnamefont
  {J.}~\bibnamefont {Mei}},\ }\bibfield  {title} {\bibinfo {title} {{Science
  with the TianQin Observatory: Preliminary results on testing the no-hair
  theorem with extreme mass ratio inspirals}},\ }\href
  {https://doi.org/10.1103/PhysRevD.104.064008} {\bibfield  {journal} {\bibinfo
   {journal} {{\PRD}}\ }\textbf {\bibinfo {volume} {104}},\ \bibinfo {pages}
  {064008} (\bibinfo {year} {2021})},\ \Eprint
  {https://arxiv.org/abs/2104.06047} {arXiv:2104.06047 [gr-qc]} \BibitemShut
  {NoStop}%
\bibitem [{\citenamefont {Amaro-Seoane}\ \emph
  {et~al.}(2013{\natexlab{b}})\citenamefont {Amaro-Seoane}, \citenamefont
  {Sopuerta},\ and\ \citenamefont {Freitag}}]{Amaro-Seoane:2012jcd}%
  \BibitemOpen
  \bibfield  {author} {\bibinfo {author} {\bibfnamefont {P.}~\bibnamefont
  {Amaro-Seoane}}, \bibinfo {author} {\bibfnamefont {C.~F.}\ \bibnamefont
  {Sopuerta}},\ and\ \bibinfo {author} {\bibfnamefont {M.~D.}\ \bibnamefont
  {Freitag}},\ }\bibfield  {title} {\bibinfo {title} {{The role of the
  supermassive black hole spin in the estimation of the EMRI event rate}},\
  }\href {https://doi.org/10.1093/mnras/sts572} {\bibfield  {journal} {\bibinfo
   {journal} {{\MNRAS}}\ }\textbf {\bibinfo {volume} {429}},\ \bibinfo {pages}
  {3155} (\bibinfo {year} {2013}{\natexlab{b}})},\ \Eprint
  {https://arxiv.org/abs/1205.4713} {arXiv:1205.4713 [astro-ph.CO]}
  \BibitemShut {NoStop}%
\bibitem [{\citenamefont {Cornish}\ and\ \citenamefont
  {Rubbo}(2003)}]{Cornish:2003d}%
  \BibitemOpen
  \bibfield  {author} {\bibinfo {author} {\bibfnamefont {N.~J.}\ \bibnamefont
  {Cornish}}\ and\ \bibinfo {author} {\bibfnamefont {L.~J.}\ \bibnamefont
  {Rubbo}},\ }\bibfield  {title} {\bibinfo {title} {{The LISA response
  function}},\ }\href {https://doi.org/10.48550/arXiv.gr-qc/0209011} {\bibfield
   {journal} {\bibinfo  {journal} {{\PRD}}\ }\textbf {\bibinfo {volume} {67}},\
  \bibinfo {pages} {6} (\bibinfo {year} {2003})}\BibitemShut {NoStop}%
\bibitem [{\citenamefont {Marsat}\ and\ \citenamefont
  {Baker}(2018)}]{Marsat:2018oam}%
  \BibitemOpen
  \bibfield  {author} {\bibinfo {author} {\bibfnamefont {S.}~\bibnamefont
  {Marsat}}\ and\ \bibinfo {author} {\bibfnamefont {J.~G.}\ \bibnamefont
  {Baker}},\ }\bibfield  {title} {\bibinfo {title} {{Fourier-domain modulations
  and delays of gravitational-wave signals}},\ }\href@noop {} {\  (\bibinfo
  {year} {2018})},\ \Eprint {https://arxiv.org/abs/1806.10734}
  {arXiv:1806.10734 [gr-qc]} \BibitemShut {NoStop}%
\bibitem [{\citenamefont {Marsat}\ \emph {et~al.}(2021)\citenamefont {Marsat},
  \citenamefont {Baker},\ and\ \citenamefont {Dal~Canton}}]{Marsat:2020rtl}%
  \BibitemOpen
  \bibfield  {author} {\bibinfo {author} {\bibfnamefont {S.}~\bibnamefont
  {Marsat}}, \bibinfo {author} {\bibfnamefont {J.~G.}\ \bibnamefont {Baker}},\
  and\ \bibinfo {author} {\bibfnamefont {T.}~\bibnamefont {Dal~Canton}},\
  }\bibfield  {title} {\bibinfo {title} {{Exploring the Bayesian parameter
  estimation of binary black holes with LISA}},\ }\href
  {https://doi.org/10.1103/PhysRevD.103.083011} {\bibfield  {journal} {\bibinfo
   {journal} {Phys. Rev. D}\ }\textbf {\bibinfo {volume} {103}},\ \bibinfo
  {pages} {083011} (\bibinfo {year} {2021})},\ \Eprint
  {https://arxiv.org/abs/2003.00357} {arXiv:2003.00357 [gr-qc]} \BibitemShut
  {NoStop}%
\bibitem [{\citenamefont {Tinto}\ \emph {et~al.}(2002)\citenamefont {Tinto},
  \citenamefont {Estabrook},\ and\ \citenamefont {Armstrong}}]{Tinto:2002de}%
  \BibitemOpen
  \bibfield  {author} {\bibinfo {author} {\bibfnamefont {M.}~\bibnamefont
  {Tinto}}, \bibinfo {author} {\bibfnamefont {F.~B.}\ \bibnamefont
  {Estabrook}},\ and\ \bibinfo {author} {\bibfnamefont {J.~W.}\ \bibnamefont
  {Armstrong}},\ }\bibfield  {title} {\bibinfo {title} {{Time delay
  interferometry for LISA}},\ }\href
  {https://doi.org/10.1103/PhysRevD.65.082003} {\bibfield  {journal} {\bibinfo
  {journal} {Phys. Rev. D}\ }\textbf {\bibinfo {volume} {65}},\ \bibinfo
  {pages} {082003} (\bibinfo {year} {2002})}\BibitemShut {NoStop}%
\bibitem [{\citenamefont {Shaddock}\ \emph {et~al.}(2003)\citenamefont
  {Shaddock}, \citenamefont {Tinto}, \citenamefont {Estabrook},\ and\
  \citenamefont {Armstrong}}]{Shaddock:2003dj}%
  \BibitemOpen
  \bibfield  {author} {\bibinfo {author} {\bibfnamefont {D.~A.}\ \bibnamefont
  {Shaddock}}, \bibinfo {author} {\bibfnamefont {M.}~\bibnamefont {Tinto}},
  \bibinfo {author} {\bibfnamefont {F.~B.}\ \bibnamefont {Estabrook}},\ and\
  \bibinfo {author} {\bibfnamefont {J.~W.}\ \bibnamefont {Armstrong}},\
  }\bibfield  {title} {\bibinfo {title} {{Data combinations accounting for LISA
  spacecraft motion}},\ }\href {https://doi.org/10.1103/PhysRevD.68.061303}
  {\bibfield  {journal} {\bibinfo  {journal} {Phys. Rev. D}\ }\textbf {\bibinfo
  {volume} {68}},\ \bibinfo {pages} {061303} (\bibinfo {year} {2003})},\
  \Eprint {https://arxiv.org/abs/gr-qc/0307080} {arXiv:gr-qc/0307080}
  \BibitemShut {NoStop}%
\bibitem [{\citenamefont {Tinto}\ and\ \citenamefont
  {Dhurandhar}(2021)}]{Tinto:2020fcc}%
  \BibitemOpen
  \bibfield  {author} {\bibinfo {author} {\bibfnamefont {M.}~\bibnamefont
  {Tinto}}\ and\ \bibinfo {author} {\bibfnamefont {S.~V.}\ \bibnamefont
  {Dhurandhar}},\ }\bibfield  {title} {\bibinfo {title} {{Time-delay
  interferometry}},\ }\href {https://doi.org/10.1007/s41114-020-00029-6}
  {\bibfield  {journal} {\bibinfo  {journal} {{\LRR}}\ }\textbf {\bibinfo
  {volume} {24}},\ \bibinfo {pages} {1} (\bibinfo {year} {2021})}\BibitemShut
  {NoStop}%
\bibitem [{\citenamefont {Vallisneri}(2008)}]{Vallisneri:2007ev}%
  \BibitemOpen
  \bibfield  {author} {\bibinfo {author} {\bibfnamefont {M.}~\bibnamefont
  {Vallisneri}},\ }\bibfield  {title} {\bibinfo {title} {{Use and abuse of the
  Fisher information matrix in the assessment of gravitational-wave
  parameter-estimation prospects}},\ }\href
  {https://doi.org/10.1103/PhysRevD.77.042001} {\bibfield  {journal} {\bibinfo
  {journal} {{\PRD}}\ }\textbf {\bibinfo {volume} {77}},\ \bibinfo {pages}
  {042001} (\bibinfo {year} {2008})},\ \Eprint
  {https://arxiv.org/abs/gr-qc/0703086} {arXiv:gr-qc/0703086} \BibitemShut
  {NoStop}%
\bibitem [{\citenamefont {Finn}(1992)}]{Finn:1992wt}%
  \BibitemOpen
  \bibfield  {author} {\bibinfo {author} {\bibfnamefont {L.~S.}\ \bibnamefont
  {Finn}},\ }\bibfield  {title} {\bibinfo {title} {{Detection, measurement and
  gravitational radiation}},\ }\href {https://doi.org/10.1103/PhysRevD.46.5236}
  {\bibfield  {journal} {\bibinfo  {journal} {Phys. Rev. D}\ }\textbf {\bibinfo
  {volume} {46}},\ \bibinfo {pages} {5236} (\bibinfo {year} {1992})},\ \Eprint
  {https://arxiv.org/abs/gr-qc/9209010} {arXiv:gr-qc/9209010} \BibitemShut
  {NoStop}%
\bibitem [{\citenamefont {Cutler}\ and\ \citenamefont
  {Flanagan}(1994)}]{Cutler:1994ys}%
  \BibitemOpen
  \bibfield  {author} {\bibinfo {author} {\bibfnamefont {C.}~\bibnamefont
  {Cutler}}\ and\ \bibinfo {author} {\bibfnamefont {E.~E.}\ \bibnamefont
  {Flanagan}},\ }\bibfield  {title} {\bibinfo {title} {{Gravitational waves
  from merging compact binaries: How accurately can one extract the binary's
  parameters from the inspiral wave form?}},\ }\href
  {https://doi.org/10.1103/PhysRevD.49.2658} {\bibfield  {journal} {\bibinfo
  {journal} {{\PRD}}\ }\textbf {\bibinfo {volume} {49}},\ \bibinfo {pages}
  {2658} (\bibinfo {year} {1994})},\ \Eprint
  {https://arxiv.org/abs/gr-qc/9402014} {arXiv:gr-qc/9402014} \BibitemShut
  {NoStop}%
\bibitem [{\citenamefont {Kay}(1993)}]{Kay:1993estimation}%
  \BibitemOpen
  \bibfield  {author} {\bibinfo {author} {\bibfnamefont {S.~M.}\ \bibnamefont
  {Kay}},\ }\href@noop {} {\emph {\bibinfo {title} {{Fundamentals of
  Statistical Signal Processing - Volume I: Estimation Theory}}}},\
  {Prentice-Hall Signal Processing Series}\ (\bibinfo  {publisher} {Prentice
  Hall},\ \bibinfo {address} {New Jersey},\ \bibinfo {year} {1993})\BibitemShut
  {NoStop}%
\bibitem [{\citenamefont {Khalvati}\ \emph {et~al.}(2025)\citenamefont
  {Khalvati}, \citenamefont {Santini}, \citenamefont {Duque}, \citenamefont
  {Speri}, \citenamefont {Gair}, \citenamefont {Yang},\ and\ \citenamefont
  {Brito}}]{Khalvati:2024tzz}%
  \BibitemOpen
  \bibfield  {author} {\bibinfo {author} {\bibfnamefont {H.}~\bibnamefont
  {Khalvati}}, \bibinfo {author} {\bibfnamefont {A.}~\bibnamefont {Santini}},
  \bibinfo {author} {\bibfnamefont {F.}~\bibnamefont {Duque}}, \bibinfo
  {author} {\bibfnamefont {L.}~\bibnamefont {Speri}}, \bibinfo {author}
  {\bibfnamefont {J.}~\bibnamefont {Gair}}, \bibinfo {author} {\bibfnamefont
  {H.}~\bibnamefont {Yang}},\ and\ \bibinfo {author} {\bibfnamefont
  {R.}~\bibnamefont {Brito}},\ }\bibfield  {title} {\bibinfo {title} {{Impact
  of relativistic waveforms in LISA{\textquoteright}s science objectives with
  extreme-mass-ratio inspirals}},\ }\href
  {https://doi.org/10.1103/PhysRevD.111.082010} {\bibfield  {journal} {\bibinfo
   {journal} {Phys. Rev. D}\ }\textbf {\bibinfo {volume} {111}},\ \bibinfo
  {pages} {082010} (\bibinfo {year} {2025})},\ \Eprint
  {https://arxiv.org/abs/2410.17310} {arXiv:2410.17310 [gr-qc]} \BibitemShut
  {NoStop}%
\bibitem [{\citenamefont {Speri}\ \emph
  {et~al.}(2026{\natexlab{a}})\citenamefont {Speri}, \citenamefont {Duque},
  \citenamefont {Barsanti}, \citenamefont {Santini}, \citenamefont {Kejriwal},
  \citenamefont {Burke},\ and\ \citenamefont {Chapman-Bird}}]{Speri:2026ade}%
  \BibitemOpen
  \bibfield  {author} {\bibinfo {author} {\bibfnamefont {L.}~\bibnamefont
  {Speri}}, \bibinfo {author} {\bibfnamefont {F.}~\bibnamefont {Duque}},
  \bibinfo {author} {\bibfnamefont {S.}~\bibnamefont {Barsanti}}, \bibinfo
  {author} {\bibfnamefont {A.}~\bibnamefont {Santini}}, \bibinfo {author}
  {\bibfnamefont {S.}~\bibnamefont {Kejriwal}}, \bibinfo {author}
  {\bibfnamefont {O.}~\bibnamefont {Burke}},\ and\ \bibinfo {author}
  {\bibfnamefont {C.~E.~A.}\ \bibnamefont {Chapman-Bird}},\ }\bibfield  {title}
  {\bibinfo {title} {{Quantifying the scientific potential of intermediate and
  extreme mass ratio inspirals with the Laser Interferometer Space Antenna}},\
  }\href@noop {} {\  (\bibinfo {year} {2026}{\natexlab{a}})},\ \Eprint
  {https://arxiv.org/abs/2603.17072} {arXiv:2603.17072 [astro-ph.IM]}
  \BibitemShut {NoStop}%
\bibitem [{\citenamefont {Khalvati}\ \emph {et~al.}(2026)\citenamefont
  {Khalvati}, \citenamefont {Lynch}, \citenamefont {Burke}, \citenamefont
  {Speri}, \citenamefont {van~de Meent},\ and\ \citenamefont
  {Nasipak}}]{Khalvati:2025znb}%
  \BibitemOpen
  \bibfield  {author} {\bibinfo {author} {\bibfnamefont {H.}~\bibnamefont
  {Khalvati}}, \bibinfo {author} {\bibfnamefont {P.}~\bibnamefont {Lynch}},
  \bibinfo {author} {\bibfnamefont {O.}~\bibnamefont {Burke}}, \bibinfo
  {author} {\bibfnamefont {L.}~\bibnamefont {Speri}}, \bibinfo {author}
  {\bibfnamefont {M.}~\bibnamefont {van~de Meent}},\ and\ \bibinfo {author}
  {\bibfnamefont {Z.}~\bibnamefont {Nasipak}},\ }\bibfield  {title} {\bibinfo
  {title} {{Systematic errors in fast relativistic waveforms for extreme mass
  ratio inspirals}},\ }\href {https://doi.org/10.1103/4ly7-zn15} {\bibfield
  {journal} {\bibinfo  {journal} {Phys. Rev. D}\ }\textbf {\bibinfo {volume}
  {113}},\ \bibinfo {pages} {084042} (\bibinfo {year} {2026})},\ \Eprint
  {https://arxiv.org/abs/2509.08875} {arXiv:2509.08875 [gr-qc]} \BibitemShut
  {NoStop}%
\bibitem [{\citenamefont {Strub}\ \emph {et~al.}(2026)\citenamefont {Strub},
  \citenamefont {Speri},\ and\ \citenamefont {Giardini}}]{Strub:2025dfs}%
  \BibitemOpen
  \bibfield  {author} {\bibinfo {author} {\bibfnamefont {S.~H.}\ \bibnamefont
  {Strub}}, \bibinfo {author} {\bibfnamefont {L.}~\bibnamefont {Speri}},\ and\
  \bibinfo {author} {\bibfnamefont {D.}~\bibnamefont {Giardini}},\ }\bibfield
  {title} {\bibinfo {title} {{Searching for extreme-mass-ratio inspirals in
  LISA: From identification to parameter estimation}},\ }\href
  {https://doi.org/10.1103/yp6d-8vqv} {\bibfield  {journal} {\bibinfo
  {journal} {Phys. Rev. D}\ }\textbf {\bibinfo {volume} {113}},\ \bibinfo
  {pages} {084048} (\bibinfo {year} {2026})},\ \Eprint
  {https://arxiv.org/abs/2505.17814} {arXiv:2505.17814 [gr-qc]} \BibitemShut
  {NoStop}%
\bibitem [{\citenamefont {Cole}\ \emph {et~al.}(2026)\citenamefont {Cole},
  \citenamefont {Alvey}, \citenamefont {Speri}, \citenamefont {Weniger},
  \citenamefont {Bhardwaj}, \citenamefont {Gerosa},\ and\ \citenamefont
  {Bertone}}]{Cole:2025sqo}%
  \BibitemOpen
  \bibfield  {author} {\bibinfo {author} {\bibfnamefont {P.~S.}\ \bibnamefont
  {Cole}}, \bibinfo {author} {\bibfnamefont {J.}~\bibnamefont {Alvey}},
  \bibinfo {author} {\bibfnamefont {L.}~\bibnamefont {Speri}}, \bibinfo
  {author} {\bibfnamefont {C.}~\bibnamefont {Weniger}}, \bibinfo {author}
  {\bibfnamefont {U.}~\bibnamefont {Bhardwaj}}, \bibinfo {author}
  {\bibfnamefont {D.}~\bibnamefont {Gerosa}},\ and\ \bibinfo {author}
  {\bibfnamefont {G.}~\bibnamefont {Bertone}},\ }\bibfield  {title} {\bibinfo
  {title} {{Sequential simulation-based inference for extreme mass ratio
  inspirals}},\ }\href {https://doi.org/10.1103/4cd3-wfjr} {\bibfield
  {journal} {\bibinfo  {journal} {Phys. Rev. D}\ }\textbf {\bibinfo {volume}
  {113}},\ \bibinfo {pages} {063030} (\bibinfo {year} {2026})},\ \Eprint
  {https://arxiv.org/abs/2505.16795} {arXiv:2505.16795 [gr-qc]} \BibitemShut
  {NoStop}%
\bibitem [{\citenamefont {Babak}\ \emph {et~al.}(2007)\citenamefont {Babak},
  \citenamefont {Fang}, \citenamefont {Gair}, \citenamefont {Glampedakis},\
  and\ \citenamefont {Hughes}}]{Babak:2006uv}%
  \BibitemOpen
  \bibfield  {author} {\bibinfo {author} {\bibfnamefont {S.}~\bibnamefont
  {Babak}}, \bibinfo {author} {\bibfnamefont {H.}~\bibnamefont {Fang}},
  \bibinfo {author} {\bibfnamefont {J.~R.}\ \bibnamefont {Gair}}, \bibinfo
  {author} {\bibfnamefont {K.}~\bibnamefont {Glampedakis}},\ and\ \bibinfo
  {author} {\bibfnamefont {S.~A.}\ \bibnamefont {Hughes}},\ }\bibfield  {title}
  {\bibinfo {title} {{`Kludge' gravitational waveforms for a test-body orbiting
  a Kerr black hole}},\ }\href {https://doi.org/10.1103/PhysRevD.75.024005}
  {\bibfield  {journal} {\bibinfo  {journal} {{\PRD}}\ }\textbf {\bibinfo
  {volume} {75}},\ \bibinfo {pages} {024005} (\bibinfo {year} {2007})},\
  \bibinfo {note} {[Erratum: Phys.Rev.D 77, 04990 (2008)]},\ \Eprint
  {https://arxiv.org/abs/gr-qc/0607007} {arXiv:gr-qc/0607007} \BibitemShut
  {NoStop}%
\bibitem [{\citenamefont {Sopuerta}\ and\ \citenamefont
  {Yunes}(2011)}]{Sopuerta:2011te}%
  \BibitemOpen
  \bibfield  {author} {\bibinfo {author} {\bibfnamefont {C.~F.}\ \bibnamefont
  {Sopuerta}}\ and\ \bibinfo {author} {\bibfnamefont {N.}~\bibnamefont
  {Yunes}},\ }\bibfield  {title} {\bibinfo {title} {{New kludge scheme for the
  construction of approximate waveforms for extreme-mass-ratio inspirals}},\
  }\href {https://doi.org/10.1103/PhysRevD.84.124060} {\bibfield  {journal}
  {\bibinfo  {journal} {{\PRD}}\ }\textbf {\bibinfo {volume} {84}},\ \bibinfo
  {pages} {124060} (\bibinfo {year} {2011})},\ \Eprint
  {https://arxiv.org/abs/1109.0572} {arXiv:1109.0572 [gr-qc]} \BibitemShut
  {NoStop}%
\bibitem [{\citenamefont {Sopuerta}\ and\ \citenamefont
  {Yunes}(2012)}]{Sopuerta:2012de}%
  \BibitemOpen
  \bibfield  {author} {\bibinfo {author} {\bibfnamefont {C.~F.}\ \bibnamefont
  {Sopuerta}}\ and\ \bibinfo {author} {\bibfnamefont {N.}~\bibnamefont
  {Yunes}},\ }\bibfield  {title} {\bibinfo {title} {{Approximate waveforms for
  extreme-mass-ratio inspirals: The Chimera Scheme}},\ }\href
  {https://doi.org/10.1088/1742-6596/363/1/012021} {\bibfield  {journal}
  {\bibinfo  {journal} {{\JPCS}}\ }\textbf {\bibinfo {volume} {363}},\ \bibinfo
  {pages} {012021} (\bibinfo {year} {2012})},\ \Eprint
  {https://arxiv.org/abs/1201.5715} {arXiv:1201.5715 [gr-qc]} \BibitemShut
  {NoStop}%
\bibitem [{\citenamefont {Chua}\ \emph {et~al.}(2021)\citenamefont {Chua},
  \citenamefont {Katz}, \citenamefont {Warburton},\ and\ \citenamefont
  {Hughes}}]{Chua:2020stf}%
  \BibitemOpen
  \bibfield  {author} {\bibinfo {author} {\bibfnamefont {A.~J.~K.}\
  \bibnamefont {Chua}}, \bibinfo {author} {\bibfnamefont {M.~L.}\ \bibnamefont
  {Katz}}, \bibinfo {author} {\bibfnamefont {N.}~\bibnamefont {Warburton}},\
  and\ \bibinfo {author} {\bibfnamefont {S.~A.}\ \bibnamefont {Hughes}},\
  }\bibfield  {title} {\bibinfo {title} {{Rapid generation of fully
  relativistic extreme-mass-ratio-inspiral waveform templates for LISA data
  analysis}},\ }\href {https://doi.org/10.1103/PhysRevLett.126.051102}
  {\bibfield  {journal} {\bibinfo  {journal} {{\PRL}}\ }\textbf {\bibinfo
  {volume} {126}},\ \bibinfo {pages} {051102} (\bibinfo {year} {2021})},\
  \Eprint {https://arxiv.org/abs/2008.06071} {arXiv:2008.06071 [gr-qc]}
  \BibitemShut {NoStop}%
\bibitem [{\citenamefont {Katz}\ \emph {et~al.}(2021)\citenamefont {Katz},
  \citenamefont {Chua}, \citenamefont {Speri}, \citenamefont {Warburton},\ and\
  \citenamefont {Hughes}}]{Katz:2021yft}%
  \BibitemOpen
  \bibfield  {author} {\bibinfo {author} {\bibfnamefont {M.~L.}\ \bibnamefont
  {Katz}}, \bibinfo {author} {\bibfnamefont {A.~J.~K.}\ \bibnamefont {Chua}},
  \bibinfo {author} {\bibfnamefont {L.}~\bibnamefont {Speri}}, \bibinfo
  {author} {\bibfnamefont {N.}~\bibnamefont {Warburton}},\ and\ \bibinfo
  {author} {\bibfnamefont {S.~A.}\ \bibnamefont {Hughes}},\ }\bibfield  {title}
  {\bibinfo {title} {{Fast extreme-mass-ratio-inspiral waveforms: New tools for
  millihertz gravitational-wave data analysis}},\ }\href
  {https://doi.org/10.1103/PhysRevD.104.064047} {\bibfield  {journal} {\bibinfo
   {journal} {{\PRD}}\ }\textbf {\bibinfo {volume} {104}},\ \bibinfo {pages}
  {064047} (\bibinfo {year} {2021})},\ \Eprint
  {https://arxiv.org/abs/2104.04582} {arXiv:2104.04582 [gr-qc]} \BibitemShut
  {NoStop}%
\bibitem [{\citenamefont {Speri}\ \emph {et~al.}(2024)\citenamefont {Speri},
  \citenamefont {Katz}, \citenamefont {Chua}, \citenamefont {Hughes},
  \citenamefont {Warburton}, \citenamefont {Thompson}, \citenamefont
  {Chapman-Bird},\ and\ \citenamefont {Gair}}]{Speri:2023jte}%
  \BibitemOpen
  \bibfield  {author} {\bibinfo {author} {\bibfnamefont {L.}~\bibnamefont
  {Speri}}, \bibinfo {author} {\bibfnamefont {M.~L.}\ \bibnamefont {Katz}},
  \bibinfo {author} {\bibfnamefont {A.~J.~K.}\ \bibnamefont {Chua}}, \bibinfo
  {author} {\bibfnamefont {S.~A.}\ \bibnamefont {Hughes}}, \bibinfo {author}
  {\bibfnamefont {N.}~\bibnamefont {Warburton}}, \bibinfo {author}
  {\bibfnamefont {J.~E.}\ \bibnamefont {Thompson}}, \bibinfo {author}
  {\bibfnamefont {C.~E.~A.}\ \bibnamefont {Chapman-Bird}},\ and\ \bibinfo
  {author} {\bibfnamefont {J.~R.}\ \bibnamefont {Gair}},\ }\bibfield  {title}
  {\bibinfo {title} {{Fast and Fourier: Extreme mass ratio inspiral waveforms
  in the frequency domain}},\ }\bibfield  {journal} {\bibinfo  {journal}
  {Front. Appl. Math. Stat.}\ }\textbf {\bibinfo {volume} {9}},\ \href
  {https://doi.org/10.3389/fams.2023.1266739} {10.3389/fams.2023.1266739}
  (\bibinfo {year} {2024}),\ \Eprint {https://arxiv.org/abs/2307.12585}
  {arXiv:2307.12585 [gr-qc]} \BibitemShut {NoStop}%
\bibitem [{\citenamefont {Chapman-Bird}\ \emph {et~al.}(2025)\citenamefont
  {Chapman-Bird} \emph {et~al.}}]{Chapman-Bird:2025xtd}%
  \BibitemOpen
  \bibfield  {author} {\bibinfo {author} {\bibfnamefont {C.~E.~A.}\
  \bibnamefont {Chapman-Bird}} \emph {et~al.},\ }\bibfield  {title} {\bibinfo
  {title} {{Efficient waveforms for asymmetric-mass eccentric equatorial
  inspirals into rapidly spinning black holes}},\ }\href
  {https://doi.org/10.1103/scbp-75pf} {\bibfield  {journal} {\bibinfo
  {journal} {Phys. Rev. D}\ }\textbf {\bibinfo {volume} {112}},\ \bibinfo
  {pages} {104023} (\bibinfo {year} {2025})},\ \Eprint
  {https://arxiv.org/abs/2506.09470} {arXiv:2506.09470 [gr-qc]} \BibitemShut
  {NoStop}%
\bibitem [{\citenamefont {Gair}\ and\ \citenamefont
  {Yunes}(2011)}]{Gair:2011ym}%
  \BibitemOpen
  \bibfield  {author} {\bibinfo {author} {\bibfnamefont {J.}~\bibnamefont
  {Gair}}\ and\ \bibinfo {author} {\bibfnamefont {N.}~\bibnamefont {Yunes}},\
  }\bibfield  {title} {\bibinfo {title} {{Approximate waveforms for
  extreme-mass-ratio inspirals in modified gravity spacetimes}},\ }\href
  {https://doi.org/10.1103/PhysRevD.84.064016} {\bibfield  {journal} {\bibinfo
  {journal} {{\PRD}}\ }\textbf {\bibinfo {volume} {84}},\ \bibinfo {pages}
  {064016} (\bibinfo {year} {2011})},\ \Eprint
  {https://arxiv.org/abs/1106.6313} {arXiv:1106.6313 [gr-qc]} \BibitemShut
  {NoStop}%
\bibitem [{\citenamefont {Kumar}\ \emph {et~al.}(2024)\citenamefont {Kumar},
  \citenamefont {Singh}, \citenamefont {Chowdhuri},\ and\ \citenamefont
  {Bhattacharyya}}]{Kumar:2024utz}%
  \BibitemOpen
  \bibfield  {author} {\bibinfo {author} {\bibfnamefont {S.}~\bibnamefont
  {Kumar}}, \bibinfo {author} {\bibfnamefont {R.~K.}\ \bibnamefont {Singh}},
  \bibinfo {author} {\bibfnamefont {A.}~\bibnamefont {Chowdhuri}},\ and\
  \bibinfo {author} {\bibfnamefont {A.}~\bibnamefont {Bhattacharyya}},\
  }\bibfield  {title} {\bibinfo {title} {{Exploring waveforms with non-GR
  deviations for extreme mass-ratio inspirals}},\ }\href
  {https://doi.org/10.1088/1475-7516/2024/10/047} {\bibfield  {journal}
  {\bibinfo  {journal} {{JCAP}}\ }\textbf {\bibinfo {volume} {10}},\ \bibinfo
  {pages} {047} (\bibinfo {year} {2024})},\ \Eprint
  {https://arxiv.org/abs/2405.18508} {arXiv:2405.18508 [gr-qc]} \BibitemShut
  {NoStop}%
\bibitem [{\citenamefont {Maselli}\ \emph {et~al.}(2020)\citenamefont
  {Maselli}, \citenamefont {Franchini}, \citenamefont {Gualtieri},\ and\
  \citenamefont {Sotiriou}}]{Maselli:2020zgv}%
  \BibitemOpen
  \bibfield  {author} {\bibinfo {author} {\bibfnamefont {A.}~\bibnamefont
  {Maselli}}, \bibinfo {author} {\bibfnamefont {N.}~\bibnamefont {Franchini}},
  \bibinfo {author} {\bibfnamefont {L.}~\bibnamefont {Gualtieri}},\ and\
  \bibinfo {author} {\bibfnamefont {T.~P.}\ \bibnamefont {Sotiriou}},\
  }\bibfield  {title} {\bibinfo {title} {{Detecting scalar fields with extreme
  mass ratio inspirals}},\ }\href
  {https://doi.org/10.1103/PhysRevLett.125.141101} {\bibfield  {journal}
  {\bibinfo  {journal} {{\PRL}}\ }\textbf {\bibinfo {volume} {125}},\ \bibinfo
  {pages} {141101} (\bibinfo {year} {2020})},\ \Eprint
  {https://arxiv.org/abs/2004.11895} {arXiv:2004.11895 [gr-qc]} \BibitemShut
  {NoStop}%
\bibitem [{\citenamefont {Maselli}\ \emph {et~al.}(2022)\citenamefont
  {Maselli}, \citenamefont {Franchini}, \citenamefont {Gualtieri},
  \citenamefont {Sotiriou}, \citenamefont {Barsanti},\ and\ \citenamefont
  {Pani}}]{Maselli:2021men}%
  \BibitemOpen
  \bibfield  {author} {\bibinfo {author} {\bibfnamefont {A.}~\bibnamefont
  {Maselli}}, \bibinfo {author} {\bibfnamefont {N.}~\bibnamefont {Franchini}},
  \bibinfo {author} {\bibfnamefont {L.}~\bibnamefont {Gualtieri}}, \bibinfo
  {author} {\bibfnamefont {T.~P.}\ \bibnamefont {Sotiriou}}, \bibinfo {author}
  {\bibfnamefont {S.}~\bibnamefont {Barsanti}},\ and\ \bibinfo {author}
  {\bibfnamefont {P.}~\bibnamefont {Pani}},\ }\bibfield  {title} {\bibinfo
  {title} {{Detecting fundamental fields with LISA observations of
  gravitational waves from extreme mass-ratio inspirals}},\ }\href
  {https://doi.org/10.1038/s41550-021-01589-5} {\bibfield  {journal} {\bibinfo
  {journal} {{\NA}}\ }\textbf {\bibinfo {volume} {6}},\ \bibinfo {pages} {464}
  (\bibinfo {year} {2022})},\ \Eprint {https://arxiv.org/abs/2106.11325}
  {arXiv:2106.11325 [gr-qc]} \BibitemShut {NoStop}%
\bibitem [{\citenamefont {Della~Rocca}\ \emph {et~al.}(2024)\citenamefont
  {Della~Rocca}, \citenamefont {Barsanti}, \citenamefont {Gualtieri},\ and\
  \citenamefont {Maselli}}]{DellaRocca:2024pnm}%
  \BibitemOpen
  \bibfield  {author} {\bibinfo {author} {\bibfnamefont {M.}~\bibnamefont
  {Della~Rocca}}, \bibinfo {author} {\bibfnamefont {S.}~\bibnamefont
  {Barsanti}}, \bibinfo {author} {\bibfnamefont {L.}~\bibnamefont
  {Gualtieri}},\ and\ \bibinfo {author} {\bibfnamefont {A.}~\bibnamefont
  {Maselli}},\ }\bibfield  {title} {\bibinfo {title} {{Extreme mass-ratio
  inspirals as probes of scalar fields: Inclined circular orbits around Kerr
  black holes}},\ }\href {https://doi.org/10.1103/PhysRevD.109.104079}
  {\bibfield  {journal} {\bibinfo  {journal} {Phys. Rev. D}\ }\textbf {\bibinfo
  {volume} {109}},\ \bibinfo {pages} {104079} (\bibinfo {year} {2024})},\
  \Eprint {https://arxiv.org/abs/2401.09542} {arXiv:2401.09542 [gr-qc]}
  \BibitemShut {NoStop}%
\bibitem [{\citenamefont {Speri}\ \emph
  {et~al.}(2026{\natexlab{b}})\citenamefont {Speri}, \citenamefont {Barsanti},
  \citenamefont {Maselli}, \citenamefont {Sotiriou}, \citenamefont {Warburton},
  \citenamefont {van~de Meent}, \citenamefont {Chua}, \citenamefont {Burke},\
  and\ \citenamefont {Gair}}]{Speri:2024qak}%
  \BibitemOpen
  \bibfield  {author} {\bibinfo {author} {\bibfnamefont {L.}~\bibnamefont
  {Speri}}, \bibinfo {author} {\bibfnamefont {S.}~\bibnamefont {Barsanti}},
  \bibinfo {author} {\bibfnamefont {A.}~\bibnamefont {Maselli}}, \bibinfo
  {author} {\bibfnamefont {T.~P.}\ \bibnamefont {Sotiriou}}, \bibinfo {author}
  {\bibfnamefont {N.}~\bibnamefont {Warburton}}, \bibinfo {author}
  {\bibfnamefont {M.}~\bibnamefont {van~de Meent}}, \bibinfo {author}
  {\bibfnamefont {A.~J.~K.}\ \bibnamefont {Chua}}, \bibinfo {author}
  {\bibfnamefont {O.}~\bibnamefont {Burke}},\ and\ \bibinfo {author}
  {\bibfnamefont {J.}~\bibnamefont {Gair}},\ }\bibfield  {title} {\bibinfo
  {title} {{Probing fundamental physics with extreme mass ratio inspirals: Full
  Bayesian inference for scalar charge}},\ }\href
  {https://doi.org/10.1103/cnhz-6zlk} {\bibfield  {journal} {\bibinfo
  {journal} {Phys. Rev. D}\ }\textbf {\bibinfo {volume} {113}},\ \bibinfo
  {pages} {023036} (\bibinfo {year} {2026}{\natexlab{b}})},\ \Eprint
  {https://arxiv.org/abs/2406.07607} {arXiv:2406.07607 [gr-qc]} \BibitemShut
  {NoStop}%
\bibitem [{\citenamefont {Zi}\ and\ \citenamefont {Kumar}(2026)}]{Zi:2025lio}%
  \BibitemOpen
  \bibfield  {author} {\bibinfo {author} {\bibfnamefont {T.}~\bibnamefont
  {Zi}}\ and\ \bibinfo {author} {\bibfnamefont {S.}~\bibnamefont {Kumar}},\
  }\bibfield  {title} {\bibinfo {title} {{Probing scalar field with generic
  extreme mass-ratio inspirals around Kerr black holes}},\ }\href
  {https://doi.org/10.1088/1475-7516/2026/04/074} {\bibfield  {journal}
  {\bibinfo  {journal} {{JCAP}}\ }\textbf {\bibinfo {volume} {04}},\ \bibinfo
  {pages} {074} (\bibinfo {year} {2026})},\ \Eprint
  {https://arxiv.org/abs/2508.00516} {arXiv:2508.00516 [gr-qc]} \BibitemShut
  {NoStop}%
\bibitem [{\citenamefont {Zi}\ and\ \citenamefont {Shu}(2025)}]{Zi:2025qos}%
  \BibitemOpen
  \bibfield  {author} {\bibinfo {author} {\bibfnamefont {T.}~\bibnamefont
  {Zi}}\ and\ \bibinfo {author} {\bibfnamefont {F.-W.}\ \bibnamefont {Shu}},\
  }\bibfield  {title} {\bibinfo {title} {{Eccentric extreme-mass-ratio
  inspirals: A new window into ultra-light vector fields}},\ }\href
  {https://doi.org/10.1140/epjc/s10052-025-14990-5} {\bibfield  {journal}
  {\bibinfo  {journal} {Eur. Phys. J. C}\ }\textbf {\bibinfo {volume} {85}},\
  \bibinfo {pages} {1251} (\bibinfo {year} {2025})},\ \Eprint
  {https://arxiv.org/abs/2510.22275} {arXiv:2510.22275 [gr-qc]} \BibitemShut
  {NoStop}%
\bibitem [{\citenamefont {Zi}\ \emph {et~al.}(2026)\citenamefont {Zi},
  \citenamefont {Rahman},\ and\ \citenamefont {Kumar}}]{Zi:2026zpw}%
  \BibitemOpen
  \bibfield  {author} {\bibinfo {author} {\bibfnamefont {T.}~\bibnamefont
  {Zi}}, \bibinfo {author} {\bibfnamefont {M.}~\bibnamefont {Rahman}},\ and\
  \bibinfo {author} {\bibfnamefont {S.}~\bibnamefont {Kumar}},\ }\bibfield
  {title} {\bibinfo {title} {{Probing beyond-vacuum general relativistic
  effects with extreme mass-ratio inspirals}},\ }\href
  {https://doi.org/10.1103/ckd6-tgn6} {\bibfield  {journal} {\bibinfo
  {journal} {Phys. Rev. D}\ }\textbf {\bibinfo {volume} {114}},\ \bibinfo
  {pages} {024022} (\bibinfo {year} {2026})},\ \Eprint
  {https://arxiv.org/abs/2601.03374} {arXiv:2601.03374 [gr-qc]} \BibitemShut
  {NoStop}%
\bibitem [{\citenamefont {Sopuerta}\ and\ \citenamefont
  {Yunes}(2009)}]{Sopuerta:2009iy}%
  \BibitemOpen
  \bibfield  {author} {\bibinfo {author} {\bibfnamefont {C.~F.}\ \bibnamefont
  {Sopuerta}}\ and\ \bibinfo {author} {\bibfnamefont {N.}~\bibnamefont
  {Yunes}},\ }\bibfield  {title} {\bibinfo {title} {{Extreme and
  intermediate-mass ratio inspirals in dynamical Chern-Simons modified
  gravity}},\ }\href {https://doi.org/10.1103/PhysRevD.80.064006} {\bibfield
  {journal} {\bibinfo  {journal} {Phys. Rev. D}\ }\textbf {\bibinfo {volume}
  {80}},\ \bibinfo {pages} {064006} (\bibinfo {year} {2009})},\ \Eprint
  {https://arxiv.org/abs/0904.4501} {arXiv:0904.4501 [gr-qc]} \BibitemShut
  {NoStop}%
\bibitem [{\citenamefont {Pani}\ \emph {et~al.}(2011)\citenamefont {Pani},
  \citenamefont {Cardoso},\ and\ \citenamefont {Gualtieri}}]{Pani:2011xj}%
  \BibitemOpen
  \bibfield  {author} {\bibinfo {author} {\bibfnamefont {P.}~\bibnamefont
  {Pani}}, \bibinfo {author} {\bibfnamefont {V.}~\bibnamefont {Cardoso}},\ and\
  \bibinfo {author} {\bibfnamefont {L.}~\bibnamefont {Gualtieri}},\ }\bibfield
  {title} {\bibinfo {title} {{Gravitational waves from extreme mass-ratio
  inspirals in dynamical Chern-Simons gravity}},\ }\href
  {https://doi.org/10.1103/PhysRevD.83.104048} {\bibfield  {journal} {\bibinfo
  {journal} {{\PRD}}\ }\textbf {\bibinfo {volume} {83}},\ \bibinfo {pages}
  {104048} (\bibinfo {year} {2011})},\ \Eprint
  {https://arxiv.org/abs/1104.1183} {arXiv:1104.1183 [gr-qc]} \BibitemShut
  {NoStop}%
\bibitem [{\citenamefont {Canizares}\ \emph {et~al.}(2013)\citenamefont
  {Canizares}, \citenamefont {Gair},\ and\ \citenamefont
  {Sopuerta}}]{Canizares:2012he}%
  \BibitemOpen
  \bibfield  {author} {\bibinfo {author} {\bibfnamefont {P.}~\bibnamefont
  {Canizares}}, \bibinfo {author} {\bibfnamefont {J.~R.}\ \bibnamefont
  {Gair}},\ and\ \bibinfo {author} {\bibfnamefont {C.~F.}\ \bibnamefont
  {Sopuerta}},\ }\bibfield  {title} {\bibinfo {title} {{Constraining gravity
  with LISA detections of binaries}},\ }\href@noop {} {\bibfield  {journal}
  {\bibinfo  {journal} {{\ACS}}\ }\textbf {\bibinfo {volume} {467}},\ \bibinfo
  {pages} {319} (\bibinfo {year} {2013})},\ \Eprint
  {https://arxiv.org/abs/1209.2534} {arXiv:1209.2534 [gr-qc]} \BibitemShut
  {NoStop}%
\bibitem [{\citenamefont {Yunes}\ \emph {et~al.}(2012)\citenamefont {Yunes},
  \citenamefont {Pani},\ and\ \citenamefont {Cardoso}}]{Yunes:2011aa}%
  \BibitemOpen
  \bibfield  {author} {\bibinfo {author} {\bibfnamefont {N.}~\bibnamefont
  {Yunes}}, \bibinfo {author} {\bibfnamefont {P.}~\bibnamefont {Pani}},\ and\
  \bibinfo {author} {\bibfnamefont {V.}~\bibnamefont {Cardoso}},\ }\bibfield
  {title} {\bibinfo {title} {{Gravitational waves from quasicircular extreme
  mass-ratio inspirals as probes of scalar-tensor theories}},\ }\href
  {https://doi.org/10.1103/PhysRevD.85.102003} {\bibfield  {journal} {\bibinfo
  {journal} {{\PRD}}\ }\textbf {\bibinfo {volume} {85}},\ \bibinfo {pages}
  {102003} (\bibinfo {year} {2012})},\ \Eprint
  {https://arxiv.org/abs/1112.3351} {arXiv:1112.3351 [gr-qc]} \BibitemShut
  {NoStop}%
\bibitem [{\citenamefont {Daniel}\ \emph {et~al.}(2024)\citenamefont {Daniel},
  \citenamefont {Jenks},\ and\ \citenamefont {Alexander}}]{Daniel:2024lev}%
  \BibitemOpen
  \bibfield  {author} {\bibinfo {author} {\bibfnamefont {T.}~\bibnamefont
  {Daniel}}, \bibinfo {author} {\bibfnamefont {L.}~\bibnamefont {Jenks}},\ and\
  \bibinfo {author} {\bibfnamefont {S.}~\bibnamefont {Alexander}},\ }\bibfield
  {title} {\bibinfo {title} {{Gravitational waves in Chern-Simons-Gauss-Bonnet
  gravity}},\ }\href {https://doi.org/10.1103/PhysRevD.109.124012} {\bibfield
  {journal} {\bibinfo  {journal} {Phys. Rev. D}\ }\textbf {\bibinfo {volume}
  {109}},\ \bibinfo {pages} {124012} (\bibinfo {year} {2024})},\ \Eprint
  {https://arxiv.org/abs/2403.09373} {arXiv:2403.09373 [gr-qc]} \BibitemShut
  {NoStop}%
\bibitem [{\citenamefont {Rahman}\ \emph {et~al.}(2023)\citenamefont {Rahman},
  \citenamefont {Kumar},\ and\ \citenamefont {Bhattacharyya}}]{Rahman:2022fay}%
  \BibitemOpen
  \bibfield  {author} {\bibinfo {author} {\bibfnamefont {M.}~\bibnamefont
  {Rahman}}, \bibinfo {author} {\bibfnamefont {S.}~\bibnamefont {Kumar}},\ and\
  \bibinfo {author} {\bibfnamefont {A.}~\bibnamefont {Bhattacharyya}},\
  }\bibfield  {title} {\bibinfo {title} {{Gravitational wave from extreme
  mass-ratio inspirals as a probe of extra dimensions}},\ }\href
  {https://doi.org/10.1088/1475-7516/2023/01/046} {\bibfield  {journal}
  {\bibinfo  {journal} {{JCAP}}\ }\textbf {\bibinfo {volume} {01}},\ \bibinfo
  {pages} {046} (\bibinfo {year} {2023})},\ \Eprint
  {https://arxiv.org/abs/2212.01404} {arXiv:2212.01404 [gr-qc]} \BibitemShut
  {NoStop}%
\bibitem [{\citenamefont {Zi}(2024)}]{Zi:2024dpi}%
  \BibitemOpen
  \bibfield  {author} {\bibinfo {author} {\bibfnamefont {T.}~\bibnamefont
  {Zi}},\ }\bibfield  {title} {\bibinfo {title} {{Extreme mass-ratio inspiral
  as a probe of extra dimensions: The case of spinning massive object}},\
  }\href {https://doi.org/10.1016/j.physletb.2024.138538} {\bibfield  {journal}
  {\bibinfo  {journal} {Phys. Lett. B}\ }\textbf {\bibinfo {volume} {850}},\
  \bibinfo {pages} {138538} (\bibinfo {year} {2024})}\BibitemShut {NoStop}%
\bibitem [{\citenamefont {Kumar}\ \emph {et~al.}(2026)\citenamefont {Kumar},
  \citenamefont {Zi},\ and\ \citenamefont {Bhattacharyya}}]{Kumar:2025jsi}%
  \BibitemOpen
  \bibfield  {author} {\bibinfo {author} {\bibfnamefont {S.}~\bibnamefont
  {Kumar}}, \bibinfo {author} {\bibfnamefont {T.}~\bibnamefont {Zi}},\ and\
  \bibinfo {author} {\bibfnamefont {A.}~\bibnamefont {Bhattacharyya}},\
  }\bibfield  {title} {\bibinfo {title} {{Extreme mass-ratio inspirals and
  extra dimensions: Insights from modified Teukolsky framework}},\ }\href
  {https://doi.org/10.1103/pnh5-9yqn} {\bibfield  {journal} {\bibinfo
  {journal} {Phys. Rev. D}\ }\textbf {\bibinfo {volume} {113}},\ \bibinfo
  {pages} {124079} (\bibinfo {year} {2026})},\ \Eprint
  {https://arxiv.org/abs/2507.03380} {arXiv:2507.03380 [gr-qc]} \BibitemShut
  {NoStop}%
\bibitem [{\citenamefont {Barausse}\ \emph {et~al.}(2014)\citenamefont
  {Barausse}, \citenamefont {Cardoso},\ and\ \citenamefont
  {Pani}}]{Barausse:2014tra}%
  \BibitemOpen
  \bibfield  {author} {\bibinfo {author} {\bibfnamefont {E.}~\bibnamefont
  {Barausse}}, \bibinfo {author} {\bibfnamefont {V.}~\bibnamefont {Cardoso}},\
  and\ \bibinfo {author} {\bibfnamefont {P.}~\bibnamefont {Pani}},\ }\bibfield
  {title} {\bibinfo {title} {{Can environmental effects spoil precision
  gravitational-wave astrophysics?}},\ }\href
  {https://doi.org/10.1103/PhysRevD.89.104059} {\bibfield  {journal} {\bibinfo
  {journal} {{\PRD}}\ }\textbf {\bibinfo {volume} {89}},\ \bibinfo {pages}
  {104059} (\bibinfo {year} {2014})},\ \Eprint
  {https://arxiv.org/abs/1404.7149} {arXiv:1404.7149 [gr-qc]} \BibitemShut
  {NoStop}%
\bibitem [{\citenamefont {Zwick}\ \emph {et~al.}(2023)\citenamefont {Zwick},
  \citenamefont {Capelo},\ and\ \citenamefont {Mayer}}]{Zwick:2022dih}%
  \BibitemOpen
  \bibfield  {author} {\bibinfo {author} {\bibfnamefont {L.}~\bibnamefont
  {Zwick}}, \bibinfo {author} {\bibfnamefont {P.~R.}\ \bibnamefont {Capelo}},\
  and\ \bibinfo {author} {\bibfnamefont {L.}~\bibnamefont {Mayer}},\ }\bibfield
   {title} {\bibinfo {title} {{Priorities in gravitational waveforms for future
  space-borne detectors: vacuum accuracy or environment?}},\ }\href
  {https://doi.org/10.1093/mnras/stad707} {\bibfield  {journal} {\bibinfo
  {journal} {{\MNRAS}}\ }\textbf {\bibinfo {volume} {521}},\ \bibinfo {pages}
  {4645} (\bibinfo {year} {2023})},\ \Eprint {https://arxiv.org/abs/2209.04060}
  {arXiv:2209.04060 [gr-qc]} \BibitemShut {NoStop}%
\bibitem [{\citenamefont {Cole}\ \emph {et~al.}(2023)\citenamefont {Cole},
  \citenamefont {Bertone}, \citenamefont {Coogan}, \citenamefont {Gaggero},
  \citenamefont {Karydas}, \citenamefont {Kavanagh}, \citenamefont {Spieksma},\
  and\ \citenamefont {Tomaselli}}]{Cole:2022yzw}%
  \BibitemOpen
  \bibfield  {author} {\bibinfo {author} {\bibfnamefont {P.~S.}\ \bibnamefont
  {Cole}}, \bibinfo {author} {\bibfnamefont {G.}~\bibnamefont {Bertone}},
  \bibinfo {author} {\bibfnamefont {A.}~\bibnamefont {Coogan}}, \bibinfo
  {author} {\bibfnamefont {D.}~\bibnamefont {Gaggero}}, \bibinfo {author}
  {\bibfnamefont {T.}~\bibnamefont {Karydas}}, \bibinfo {author} {\bibfnamefont
  {B.~J.}\ \bibnamefont {Kavanagh}}, \bibinfo {author} {\bibfnamefont
  {T.~F.~M.}\ \bibnamefont {Spieksma}},\ and\ \bibinfo {author} {\bibfnamefont
  {G.~M.}\ \bibnamefont {Tomaselli}},\ }\bibfield  {title} {\bibinfo {title}
  {{Distinguishing environmental effects on binary black hole gravitational
  waveforms}},\ }\href {https://doi.org/10.1038/s41550-023-01990-2} {\bibfield
  {journal} {\bibinfo  {journal} {Nat. Astron.}\ }\textbf {\bibinfo {volume}
  {7}},\ \bibinfo {pages} {943} (\bibinfo {year} {2023})},\ \Eprint
  {https://arxiv.org/abs/2211.01362} {arXiv:2211.01362 [gr-qc]} \BibitemShut
  {NoStop}%
\bibitem [{\citenamefont {Shen}\ \emph {et~al.}(2025)\citenamefont {Shen},
  \citenamefont {Cui},\ and\ \citenamefont {Han}}]{Shen:2025svs}%
  \BibitemOpen
  \bibfield  {author} {\bibinfo {author} {\bibfnamefont {P.}~\bibnamefont
  {Shen}}, \bibinfo {author} {\bibfnamefont {Q.}~\bibnamefont {Cui}},\ and\
  \bibinfo {author} {\bibfnamefont {W.-B.}\ \bibnamefont {Han}},\ }\bibfield
  {title} {\bibinfo {title} {{Assessing the systematic errors of
  extreme-mass-ratio inspirals waveforms for testing general relativity}},\
  }\href {https://doi.org/10.1103/PhysRevD.111.024004} {\bibfield  {journal}
  {\bibinfo  {journal} {Phys. Rev. D}\ }\textbf {\bibinfo {volume} {111}},\
  \bibinfo {pages} {024004} (\bibinfo {year} {2025})},\ \Eprint
  {https://arxiv.org/abs/2501.07264} {arXiv:2501.07264 [gr-qc]} \BibitemShut
  {NoStop}%
\bibitem [{\citenamefont {Lyu}\ \emph {et~al.}(2026)\citenamefont {Lyu},
  \citenamefont {Pan}, \citenamefont {Mao}, \citenamefont {Jiang},\ and\
  \citenamefont {Yang}}]{Lyu:2024gnk}%
  \BibitemOpen
  \bibfield  {author} {\bibinfo {author} {\bibfnamefont {Z.}~\bibnamefont
  {Lyu}}, \bibinfo {author} {\bibfnamefont {Z.}~\bibnamefont {Pan}}, \bibinfo
  {author} {\bibfnamefont {J.}~\bibnamefont {Mao}}, \bibinfo {author}
  {\bibfnamefont {N.}~\bibnamefont {Jiang}},\ and\ \bibinfo {author}
  {\bibfnamefont {H.}~\bibnamefont {Yang}},\ }\bibfield  {title} {\bibinfo
  {title} {{Science opportunities of wet extreme mass-ratio inspirals}},\
  }\href {https://doi.org/10.1103/337c-g6x1} {\bibfield  {journal} {\bibinfo
  {journal} {Phys. Rev. D}\ }\textbf {\bibinfo {volume} {113}},\ \bibinfo
  {pages} {043002} (\bibinfo {year} {2026})},\ \Eprint
  {https://arxiv.org/abs/2501.03252} {arXiv:2501.03252 [astro-ph.HE]}
  \BibitemShut {NoStop}%
\bibitem [{\citenamefont {Li}\ \emph {et~al.}(2025{\natexlab{a}})\citenamefont
  {Li}, \citenamefont {Yang},\ and\ \citenamefont {Pan}}]{Li:2025zgo}%
  \BibitemOpen
  \bibfield  {author} {\bibinfo {author} {\bibfnamefont {Y.-P.}\ \bibnamefont
  {Li}}, \bibinfo {author} {\bibfnamefont {H.}~\bibnamefont {Yang}},\ and\
  \bibinfo {author} {\bibfnamefont {Z.}~\bibnamefont {Pan}},\ }\bibfield
  {title} {\bibinfo {title} {{Extreme mass-ratio inspirals in active galactic
  nucleus disks: The role of circumsingle disks}},\ }\href
  {https://doi.org/10.1103/PhysRevD.111.063074} {\bibfield  {journal} {\bibinfo
   {journal} {Phys. Rev. D}\ }\textbf {\bibinfo {volume} {111}},\ \bibinfo
  {pages} {063074} (\bibinfo {year} {2025}{\natexlab{a}})},\ \Eprint
  {https://arxiv.org/abs/2503.04042} {arXiv:2503.04042 [astro-ph.HE]}
  \BibitemShut {NoStop}%
\bibitem [{\citenamefont {Speri}\ \emph {et~al.}(2023)\citenamefont {Speri},
  \citenamefont {Antonelli}, \citenamefont {Sberna}, \citenamefont {Babak},
  \citenamefont {Barausse}, \citenamefont {Gair},\ and\ \citenamefont
  {Katz}}]{Speri:2022upm}%
  \BibitemOpen
  \bibfield  {author} {\bibinfo {author} {\bibfnamefont {L.}~\bibnamefont
  {Speri}}, \bibinfo {author} {\bibfnamefont {A.}~\bibnamefont {Antonelli}},
  \bibinfo {author} {\bibfnamefont {L.}~\bibnamefont {Sberna}}, \bibinfo
  {author} {\bibfnamefont {S.}~\bibnamefont {Babak}}, \bibinfo {author}
  {\bibfnamefont {E.}~\bibnamefont {Barausse}}, \bibinfo {author}
  {\bibfnamefont {J.~R.}\ \bibnamefont {Gair}},\ and\ \bibinfo {author}
  {\bibfnamefont {M.~L.}\ \bibnamefont {Katz}},\ }\bibfield  {title} {\bibinfo
  {title} {{Probing accretion physics with gravitational waves}},\ }\href
  {https://doi.org/10.1103/PhysRevX.13.021035} {\bibfield  {journal} {\bibinfo
  {journal} {{\PRX}}\ }\textbf {\bibinfo {volume} {13}},\ \bibinfo {pages}
  {021035} (\bibinfo {year} {2023})},\ \Eprint
  {https://arxiv.org/abs/2207.10086} {arXiv:2207.10086 [gr-qc]} \BibitemShut
  {NoStop}%
\bibitem [{\citenamefont {Duque}\ \emph {et~al.}(2025)\citenamefont {Duque},
  \citenamefont {Kejriwal}, \citenamefont {Sberna}, \citenamefont {Speri},\
  and\ \citenamefont {Gair}}]{Duque:2024mfw}%
  \BibitemOpen
  \bibfield  {author} {\bibinfo {author} {\bibfnamefont {F.}~\bibnamefont
  {Duque}}, \bibinfo {author} {\bibfnamefont {S.}~\bibnamefont {Kejriwal}},
  \bibinfo {author} {\bibfnamefont {L.}~\bibnamefont {Sberna}}, \bibinfo
  {author} {\bibfnamefont {L.}~\bibnamefont {Speri}},\ and\ \bibinfo {author}
  {\bibfnamefont {J.}~\bibnamefont {Gair}},\ }\bibfield  {title} {\bibinfo
  {title} {{Constraining accretion physics with gravitational waves from
  eccentric extreme-mass-ratio inspirals}},\ }\href
  {https://doi.org/10.1103/PhysRevD.111.084006} {\bibfield  {journal} {\bibinfo
   {journal} {Phys. Rev. D}\ }\textbf {\bibinfo {volume} {111}},\ \bibinfo
  {pages} {084006} (\bibinfo {year} {2025})},\ \Eprint
  {https://arxiv.org/abs/2411.03436} {arXiv:2411.03436 [gr-qc]} \BibitemShut
  {NoStop}%
\bibitem [{\citenamefont {Hegade K.~R.}\ \emph
  {et~al.}(2025{\natexlab{a}})\citenamefont {Hegade K.~R.}, \citenamefont
  {Gammie},\ and\ \citenamefont {Yunes}}]{HegadeKR:2025dur}%
  \BibitemOpen
  \bibfield  {author} {\bibinfo {author} {\bibfnamefont {A.}~\bibnamefont
  {Hegade K.~R.}}, \bibinfo {author} {\bibfnamefont {C.~F.}\ \bibnamefont
  {Gammie}},\ and\ \bibinfo {author} {\bibfnamefont {N.}~\bibnamefont
  {Yunes}},\ }\bibfield  {title} {\bibinfo {title} {{Relativistic treatment of
  accretion disk torques on extreme mass-ratio inspirals around nonspinning
  black holes}},\ }\href {https://doi.org/10.1103/9src-p7sp} {\bibfield
  {journal} {\bibinfo  {journal} {Phys. Rev. D}\ }\textbf {\bibinfo {volume}
  {112}},\ \bibinfo {pages} {124012} (\bibinfo {year} {2025}{\natexlab{a}})},\
  \Eprint {https://arxiv.org/abs/2509.20457} {arXiv:2509.20457 [gr-qc]}
  \BibitemShut {NoStop}%
\bibitem [{\citenamefont {Hegade K.~R.}\ \emph
  {et~al.}(2025{\natexlab{b}})\citenamefont {Hegade K.~R.}, \citenamefont
  {Gammie},\ and\ \citenamefont {Yunes}}]{HegadeKR:2025rpr}%
  \BibitemOpen
  \bibfield  {author} {\bibinfo {author} {\bibfnamefont {A.}~\bibnamefont
  {Hegade K.~R.}}, \bibinfo {author} {\bibfnamefont {C.~F.}\ \bibnamefont
  {Gammie}},\ and\ \bibinfo {author} {\bibfnamefont {N.}~\bibnamefont
  {Yunes}},\ }\bibfield  {title} {\bibinfo {title} {{Relativistic treatment of
  accretion disk torques on extreme mass ratio inspirals around spinning black
  holes}},\ }\href {https://doi.org/10.1103/g83s-jdld} {\bibfield  {journal}
  {\bibinfo  {journal} {Phys. Rev. D}\ }\textbf {\bibinfo {volume} {112}},\
  \bibinfo {pages} {124068} (\bibinfo {year} {2025}{\natexlab{b}})},\ \Eprint
  {https://arxiv.org/abs/2510.03564} {arXiv:2510.03564 [gr-qc]} \BibitemShut
  {NoStop}%
\bibitem [{\citenamefont {Vicente}\ \emph {et~al.}(2025)\citenamefont
  {Vicente}, \citenamefont {Karydas},\ and\ \citenamefont
  {Bertone}}]{Vicente:2025gsg}%
  \BibitemOpen
  \bibfield  {author} {\bibinfo {author} {\bibfnamefont {R.}~\bibnamefont
  {Vicente}}, \bibinfo {author} {\bibfnamefont {T.~K.}\ \bibnamefont
  {Karydas}},\ and\ \bibinfo {author} {\bibfnamefont {G.}~\bibnamefont
  {Bertone}},\ }\bibfield  {title} {\bibinfo {title} {{Fully relativistic
  treatment of extreme mass-ratio inspirals in collisionless environments}},\
  }\href {https://doi.org/10.1103/s4wh-x6c4} {\bibfield  {journal} {\bibinfo
  {journal} {Phys. Rev. Lett.}\ }\textbf {\bibinfo {volume} {135}},\ \bibinfo
  {pages} {211401} (\bibinfo {year} {2025})},\ \Eprint
  {https://arxiv.org/abs/2505.09715} {arXiv:2505.09715 [gr-qc]} \BibitemShut
  {NoStop}%
\bibitem [{\citenamefont {Duque}\ \emph {et~al.}(2026)\citenamefont {Duque},
  \citenamefont {Sberna}, \citenamefont {Spiers},\ and\ \citenamefont
  {Vicente}}]{Duque:2025yfm}%
  \BibitemOpen
  \bibfield  {author} {\bibinfo {author} {\bibfnamefont {F.}~\bibnamefont
  {Duque}}, \bibinfo {author} {\bibfnamefont {L.}~\bibnamefont {Sberna}},
  \bibinfo {author} {\bibfnamefont {A.}~\bibnamefont {Spiers}},\ and\ \bibinfo
  {author} {\bibfnamefont {R.}~\bibnamefont {Vicente}},\ }\bibfield  {title}
  {\bibinfo {title} {{Extreme-mass-ratio inspirals in relativistic accretion
  discs}},\ }\href {https://doi.org/10.1103/jcv5-ssfd} {\bibfield  {journal}
  {\bibinfo  {journal} {Phys. Rev. D}\ }\textbf {\bibinfo {volume} {113}},\
  \bibinfo {pages} {084028} (\bibinfo {year} {2026})},\ \Eprint
  {https://arxiv.org/abs/2510.02433} {arXiv:2510.02433 [gr-qc]} \BibitemShut
  {NoStop}%
\bibitem [{\citenamefont {Straten{\'y}}\ \emph {et~al.}(2026)\citenamefont
  {Straten{\'y}}, \citenamefont {Lukes-Gerakopoulos},\ and\ \citenamefont
  {Zelenka}}]{Strateny:2025zkd}%
  \BibitemOpen
  \bibfield  {author} {\bibinfo {author} {\bibfnamefont {M.}~\bibnamefont
  {Straten{\'y}}}, \bibinfo {author} {\bibfnamefont {G.}~\bibnamefont
  {Lukes-Gerakopoulos}},\ and\ \bibinfo {author} {\bibfnamefont
  {O.}~\bibnamefont {Zelenka}},\ }\bibfield  {title} {\bibinfo {title}
  {{Extreme mass ratio inspirals into black holes surrounded by matter:
  Resonance crossings}},\ }\href {https://doi.org/10.1103/qxx6-vvvk} {\bibfield
   {journal} {\bibinfo  {journal} {Phys. Rev. D}\ }\textbf {\bibinfo {volume}
  {113}},\ \bibinfo {pages} {044054} (\bibinfo {year} {2026})},\ \Eprint
  {https://arxiv.org/abs/2511.08057} {arXiv:2511.08057 [gr-qc]} \BibitemShut
  {NoStop}%
\bibitem [{\citenamefont {Zeng}\ and\ \citenamefont
  {Pan}(2026)}]{Zeng:2026ydj}%
  \BibitemOpen
  \bibfield  {author} {\bibinfo {author} {\bibfnamefont {Y.}~\bibnamefont
  {Zeng}}\ and\ \bibinfo {author} {\bibfnamefont {Z.}~\bibnamefont {Pan}},\
  }\bibfield  {title} {\bibinfo {title} {{Relativistic treatment of extreme
  mass ratio inspirals crossing accretion disks}},\ }\href
  {https://doi.org/10.1103/fxnd-r4bs} {\bibfield  {journal} {\bibinfo
  {journal} {Phys. Rev. D}\ }\textbf {\bibinfo {volume} {113}},\ \bibinfo
  {pages} {123002} (\bibinfo {year} {2026})},\ \Eprint
  {https://arxiv.org/abs/2601.11925} {arXiv:2601.11925 [astro-ph.HE]}
  \BibitemShut {NoStop}%
\bibitem [{\citenamefont {Brito}\ and\ \citenamefont
  {Shah}(2023)}]{Brito:2023pyl}%
  \BibitemOpen
  \bibfield  {author} {\bibinfo {author} {\bibfnamefont {R.}~\bibnamefont
  {Brito}}\ and\ \bibinfo {author} {\bibfnamefont {S.}~\bibnamefont {Shah}},\
  }\bibfield  {title} {\bibinfo {title} {{Extreme mass-ratio inspirals into
  black holes surrounded by scalar clouds}},\ }\href
  {https://doi.org/10.1103/PhysRevD.108.084019} {\bibfield  {journal} {\bibinfo
   {journal} {{\PRD}}\ }\textbf {\bibinfo {volume} {108}},\ \bibinfo {pages}
  {084019} (\bibinfo {year} {2023})},\ \Eprint
  {https://arxiv.org/abs/2307.16093} {arXiv:2307.16093 [gr-qc]} \BibitemShut
  {NoStop}%
\bibitem [{\citenamefont {Zhang}\ \emph {et~al.}(2024)\citenamefont {Zhang},
  \citenamefont {Fu},\ and\ \citenamefont {Dai}}]{Zhang:2024ugv}%
  \BibitemOpen
  \bibfield  {author} {\bibinfo {author} {\bibfnamefont {C.}~\bibnamefont
  {Zhang}}, \bibinfo {author} {\bibfnamefont {G.}~\bibnamefont {Fu}},\ and\
  \bibinfo {author} {\bibfnamefont {N.}~\bibnamefont {Dai}},\ }\bibfield
  {title} {\bibinfo {title} {{Detecting dark matter halos with extreme
  mass-ratio inspirals}},\ }\href
  {https://doi.org/10.1088/1475-7516/2024/04/088} {\bibfield  {journal}
  {\bibinfo  {journal} {{JCAP}}\ }\textbf {\bibinfo {volume} {2024}},\ \bibinfo
  {pages} {088} (\bibinfo {year} {2024})},\ \Eprint
  {https://arxiv.org/abs/2401.04467} {arXiv:2401.04467 [gr-qc]} \BibitemShut
  {NoStop}%
\bibitem [{\citenamefont {Gliorio}\ \emph {et~al.}(2025)\citenamefont
  {Gliorio}, \citenamefont {Berti}, \citenamefont {Maselli},\ and\
  \citenamefont {Speeney}}]{Gliorio:2025cbh}%
  \BibitemOpen
  \bibfield  {author} {\bibinfo {author} {\bibfnamefont {S.}~\bibnamefont
  {Gliorio}}, \bibinfo {author} {\bibfnamefont {E.}~\bibnamefont {Berti}},
  \bibinfo {author} {\bibfnamefont {A.}~\bibnamefont {Maselli}},\ and\ \bibinfo
  {author} {\bibfnamefont {N.}~\bibnamefont {Speeney}},\ }\bibfield  {title}
  {\bibinfo {title} {{Extreme mass ratio inspirals in dark matter halos:
  Dynamics and distinguishability of halo models}},\ }\href
  {https://doi.org/10.1103/dw6c-14pt} {\bibfield  {journal} {\bibinfo
  {journal} {Phys. Rev. D}\ }\textbf {\bibinfo {volume} {112}},\ \bibinfo
  {pages} {124050} (\bibinfo {year} {2025})},\ \Eprint
  {https://arxiv.org/abs/2503.16649} {arXiv:2503.16649 [gr-qc]} \BibitemShut
  {NoStop}%
\bibitem [{\citenamefont {Mitra}\ \emph {et~al.}(2025)\citenamefont {Mitra},
  \citenamefont {Speeney}, \citenamefont {Chakraborty},\ and\ \citenamefont
  {Berti}}]{Mitra:2025tag}%
  \BibitemOpen
  \bibfield  {author} {\bibinfo {author} {\bibfnamefont {S.}~\bibnamefont
  {Mitra}}, \bibinfo {author} {\bibfnamefont {N.}~\bibnamefont {Speeney}},
  \bibinfo {author} {\bibfnamefont {S.}~\bibnamefont {Chakraborty}},\ and\
  \bibinfo {author} {\bibfnamefont {E.}~\bibnamefont {Berti}},\ }\bibfield
  {title} {\bibinfo {title} {{Extreme mass ratio inspirals in rotating dark
  matter spikes}},\ }\href {https://doi.org/10.1103/ycl1-kx7d} {\bibfield
  {journal} {\bibinfo  {journal} {Phys. Rev. D}\ }\textbf {\bibinfo {volume}
  {112}},\ \bibinfo {pages} {044030} (\bibinfo {year} {2025})},\ \Eprint
  {https://arxiv.org/abs/2505.04697} {arXiv:2505.04697 [gr-qc]} \BibitemShut
  {NoStop}%
\bibitem [{\citenamefont {Li}\ \emph {et~al.}(2025{\natexlab{b}})\citenamefont
  {Li}, \citenamefont {Weller}, \citenamefont {Bourg}, \citenamefont {LaHaye},
  \citenamefont {Yunes},\ and\ \citenamefont {Yang}}]{Li:2025ffh}%
  \BibitemOpen
  \bibfield  {author} {\bibinfo {author} {\bibfnamefont {D.}~\bibnamefont
  {Li}}, \bibinfo {author} {\bibfnamefont {C.}~\bibnamefont {Weller}}, \bibinfo
  {author} {\bibfnamefont {P.}~\bibnamefont {Bourg}}, \bibinfo {author}
  {\bibfnamefont {M.}~\bibnamefont {LaHaye}}, \bibinfo {author} {\bibfnamefont
  {N.}~\bibnamefont {Yunes}},\ and\ \bibinfo {author} {\bibfnamefont
  {H.}~\bibnamefont {Yang}},\ }\bibfield  {title} {\bibinfo {title} {{Extreme
  mass-ratio inspiral within an ultralight scalar cloud: Scalar radiation}},\
  }\href {https://doi.org/10.1103/7l9s-g21j} {\bibfield  {journal} {\bibinfo
  {journal} {Phys. Rev. D}\ }\textbf {\bibinfo {volume} {112}},\ \bibinfo
  {pages} {084057} (\bibinfo {year} {2025}{\natexlab{b}})},\ \Eprint
  {https://arxiv.org/abs/2507.02045} {arXiv:2507.02045 [gr-qc]} \BibitemShut
  {NoStop}%
\bibitem [{\citenamefont {Dyson}\ \emph {et~al.}(2025)\citenamefont {Dyson},
  \citenamefont {Spieksma}, \citenamefont {Brito}, \citenamefont {van~de
  Meent},\ and\ \citenamefont {Dolan}}]{Dyson:2025dlj}%
  \BibitemOpen
  \bibfield  {author} {\bibinfo {author} {\bibfnamefont {C.}~\bibnamefont
  {Dyson}}, \bibinfo {author} {\bibfnamefont {T.~F.~M.}\ \bibnamefont
  {Spieksma}}, \bibinfo {author} {\bibfnamefont {R.}~\bibnamefont {Brito}},
  \bibinfo {author} {\bibfnamefont {M.}~\bibnamefont {van~de Meent}},\ and\
  \bibinfo {author} {\bibfnamefont {S.}~\bibnamefont {Dolan}},\ }\bibfield
  {title} {\bibinfo {title} {{Environmental effects in extreme-mass-ratio
  inspirals: Perturbations to the environment in Kerr spacetimes}},\ }\href
  {https://doi.org/10.1103/PhysRevLett.134.211403} {\bibfield  {journal}
  {\bibinfo  {journal} {Phys. Rev. Lett.}\ }\textbf {\bibinfo {volume} {134}},\
  \bibinfo {pages} {211403} (\bibinfo {year} {2025})},\ \Eprint
  {https://arxiv.org/abs/2501.09806} {arXiv:2501.09806 [gr-qc]} \BibitemShut
  {NoStop}%
\bibitem [{\citenamefont {Kakehi}\ \emph {et~al.}(2025)\citenamefont {Kakehi},
  \citenamefont {Omiya}, \citenamefont {Takahashi},\ and\ \citenamefont
  {Tanaka}}]{Kakehi:2025peb}%
  \BibitemOpen
  \bibfield  {author} {\bibinfo {author} {\bibfnamefont {T.}~\bibnamefont
  {Kakehi}}, \bibinfo {author} {\bibfnamefont {H.}~\bibnamefont {Omiya}},
  \bibinfo {author} {\bibfnamefont {T.}~\bibnamefont {Takahashi}},\ and\
  \bibinfo {author} {\bibfnamefont {T.}~\bibnamefont {Tanaka}},\ }\bibfield
  {title} {\bibinfo {title} {{Resonant DM scattering in the galactic Center
  under the influence of extreme mass ratio inspirals}},\ }\href
  {https://doi.org/10.1103/l9dn-5s25} {\bibfield  {journal} {\bibinfo
  {journal} {Phys. Rev. D}\ }\textbf {\bibinfo {volume} {112}},\ \bibinfo
  {pages} {104061} (\bibinfo {year} {2025})},\ \Eprint
  {https://arxiv.org/abs/2505.10036} {arXiv:2505.10036 [gr-qc]} \BibitemShut
  {NoStop}%
\bibitem [{\citenamefont {Das}\ \emph {et~al.}(2025)\citenamefont {Das},
  \citenamefont {Dalui}, \citenamefont {Lee},\ and\ \citenamefont
  {Cai}}]{Das:2025eiv}%
  \BibitemOpen
  \bibfield  {author} {\bibinfo {author} {\bibfnamefont {S.}~\bibnamefont
  {Das}}, \bibinfo {author} {\bibfnamefont {S.}~\bibnamefont {Dalui}}, \bibinfo
  {author} {\bibfnamefont {B.-H.}\ \bibnamefont {Lee}},\ and\ \bibinfo {author}
  {\bibfnamefont {Y.-F.}\ \bibnamefont {Cai}},\ }\bibfield  {title} {\bibinfo
  {title} {{Extreme-mass-ratio inspirals embedded in dark matter halo II:
  Chaotic Imprints in gravitational waves}},\ }\href@noop {} {\  (\bibinfo
  {year} {2025})},\ \Eprint {https://arxiv.org/abs/2512.04848}
  {arXiv:2512.04848 [gr-qc]} \BibitemShut {NoStop}%
\bibitem [{\citenamefont {Feng}\ \emph {et~al.}(2026)\citenamefont {Feng},
  \citenamefont {Tang},\ and\ \citenamefont {Wu}}]{Feng:2025fkc}%
  \BibitemOpen
  \bibfield  {author} {\bibinfo {author} {\bibfnamefont {C.}~\bibnamefont
  {Feng}}, \bibinfo {author} {\bibfnamefont {Y.}~\bibnamefont {Tang}},\ and\
  \bibinfo {author} {\bibfnamefont {Y.-L.}\ \bibnamefont {Wu}},\ }\bibfield
  {title} {\bibinfo {title} {{Probing the dark matter spike with gravitational
  waves from early extreme mass-ratio inspirals in the Milky~Way Center}},\
  }\href {https://doi.org/10.1103/3qmb-lgvt} {\bibfield  {journal} {\bibinfo
  {journal} {Phys. Rev. D}\ }\textbf {\bibinfo {volume} {113}},\ \bibinfo
  {pages} {023016} (\bibinfo {year} {2026})},\ \Eprint
  {https://arxiv.org/abs/2506.02937} {arXiv:2506.02937 [astro-ph.GA]}
  \BibitemShut {NoStop}%
\bibitem [{\citenamefont {Zhao}\ and\ \citenamefont
  {Gong}(2026)}]{Zhao:2026yis}%
  \BibitemOpen
  \bibfield  {author} {\bibinfo {author} {\bibfnamefont {Y.}~\bibnamefont
  {Zhao}}\ and\ \bibinfo {author} {\bibfnamefont {Y.}~\bibnamefont {Gong}},\
  }\bibfield  {title} {\bibinfo {title} {{Dark matter distributions around
  extreme mass ratio inspirals: effects of radial pressure and relativistic
  treatment}},\ }\href@noop {} {\  (\bibinfo {year} {2026})},\ \Eprint
  {https://arxiv.org/abs/2602.12022} {arXiv:2602.12022 [gr-qc]} \BibitemShut
  {NoStop}%
\bibitem [{\citenamefont {Li}\ \emph {et~al.}(2026)\citenamefont {Li},
  \citenamefont {Qiao},\ and\ \citenamefont {Tao}}]{Li:2026uva}%
  \BibitemOpen
  \bibfield  {author} {\bibinfo {author} {\bibfnamefont {G.-H.}\ \bibnamefont
  {Li}}, \bibinfo {author} {\bibfnamefont {C.-K.}\ \bibnamefont {Qiao}},\ and\
  \bibinfo {author} {\bibfnamefont {J.}~\bibnamefont {Tao}},\ }\bibfield
  {title} {\bibinfo {title} {{Orbital dynamics and gravitational wave
  signatures of extreme mass ratio inspirals in galactic dark matter halos}},\
  }\href@noop {} {\  (\bibinfo {year} {2026})},\ \Eprint
  {https://arxiv.org/abs/2603.02414} {arXiv:2603.02414 [gr-qc]} \BibitemShut
  {NoStop}%
\bibitem [{\citenamefont {Kejriwal}\ \emph {et~al.}(2024)\citenamefont
  {Kejriwal}, \citenamefont {Speri},\ and\ \citenamefont
  {Chua}}]{Kejriwal:2023djc}%
  \BibitemOpen
  \bibfield  {author} {\bibinfo {author} {\bibfnamefont {S.}~\bibnamefont
  {Kejriwal}}, \bibinfo {author} {\bibfnamefont {L.}~\bibnamefont {Speri}},\
  and\ \bibinfo {author} {\bibfnamefont {A.~J.~K.}\ \bibnamefont {Chua}},\
  }\bibfield  {title} {\bibinfo {title} {{Impact of correlations on the
  modeling and inference of beyond vacuum{\textendash}general relativistic
  effects in extreme-mass-ratio inspirals}},\ }\href
  {https://doi.org/10.1103/PhysRevD.110.084060} {\bibfield  {journal} {\bibinfo
   {journal} {Phys. Rev. D}\ }\textbf {\bibinfo {volume} {110}},\ \bibinfo
  {pages} {084060} (\bibinfo {year} {2024})},\ \Eprint
  {https://arxiv.org/abs/2312.13028} {arXiv:2312.13028 [gr-qc]} \BibitemShut
  {NoStop}%
\bibitem [{\citenamefont {Mancieri}\ \emph {et~al.}(2026)\citenamefont
  {Mancieri}, \citenamefont {Broggi}, \citenamefont {Vinciguerra},
  \citenamefont {Sesana},\ and\ \citenamefont {Bonetti}}]{Mancieri:2025cmx}%
  \BibitemOpen
  \bibfield  {author} {\bibinfo {author} {\bibfnamefont {D.}~\bibnamefont
  {Mancieri}}, \bibinfo {author} {\bibfnamefont {L.}~\bibnamefont {Broggi}},
  \bibinfo {author} {\bibfnamefont {M.}~\bibnamefont {Vinciguerra}}, \bibinfo
  {author} {\bibfnamefont {A.}~\bibnamefont {Sesana}},\ and\ \bibinfo {author}
  {\bibfnamefont {M.}~\bibnamefont {Bonetti}},\ }\bibfield  {title} {\bibinfo
  {title} {{Eccentricity distribution of extreme mass ratio inspirals}},\
  }\href {https://doi.org/10.1103/vq2j-kbhs} {\bibfield  {journal} {\bibinfo
  {journal} {Phys. Rev. D}\ }\textbf {\bibinfo {volume} {113}},\ \bibinfo
  {pages} {043062} (\bibinfo {year} {2026})},\ \Eprint
  {https://arxiv.org/abs/2509.02394} {arXiv:2509.02394 [astro-ph.HE]}
  \BibitemShut {NoStop}%
\bibitem [{\citenamefont {Zhang}\ and\ \citenamefont
  {Seoane}(2026)}]{Zhang:2025jmm}%
  \BibitemOpen
  \bibfield  {author} {\bibinfo {author} {\bibfnamefont {F.}~\bibnamefont
  {Zhang}}\ and\ \bibinfo {author} {\bibfnamefont {P.~A.}\ \bibnamefont
  {Seoane}},\ }\bibfield  {title} {\bibinfo {title} {{Coevolution of nuclear
  star clusters and massive black holes: Extreme-mass-ratio inspirals}},\
  }\href {https://doi.org/10.3847/1538-4357/ae3ca3} {\bibfield  {journal}
  {\bibinfo  {journal} {Astrophys. J.}\ }\textbf {\bibinfo {volume} {999}},\
  \bibinfo {pages} {224} (\bibinfo {year} {2026})},\ \Eprint
  {https://arxiv.org/abs/2510.10821} {arXiv:2510.10821 [astro-ph.GA]}
  \BibitemShut {NoStop}%
\bibitem [{\citenamefont {Bayle}\ \emph {et~al.}(2022)\citenamefont {Bayle},
  \citenamefont {Hartwig}, \citenamefont {Petiteau},\ and\ \citenamefont
  {Lilley}}]{Bayle2022LISANode}%
  \BibitemOpen
  \bibfield  {author} {\bibinfo {author} {\bibfnamefont {J.-B.}\ \bibnamefont
  {Bayle}}, \bibinfo {author} {\bibfnamefont {O.}~\bibnamefont {Hartwig}},
  \bibinfo {author} {\bibfnamefont {A.}~\bibnamefont {Petiteau}},\ and\
  \bibinfo {author} {\bibfnamefont {M.}~\bibnamefont {Lilley}},\ }\href@noop {}
  {\bibinfo {title} {{LISANode}}},\ \bibinfo {howpublished} {Zenodo} (\bibinfo
  {year} {2022}),\ \bibinfo {note} {10.5281/zenodo.6461078}\BibitemShut
  {NoStop}%
\bibitem [{\citenamefont {Babak}\ \emph {et~al.}(2021)\citenamefont {Babak},
  \citenamefont {Petiteau},\ and\ \citenamefont {Hewitson}}]{Babak:2021mhe}%
  \BibitemOpen
  \bibfield  {author} {\bibinfo {author} {\bibfnamefont {S.}~\bibnamefont
  {Babak}}, \bibinfo {author} {\bibfnamefont {A.}~\bibnamefont {Petiteau}},\
  and\ \bibinfo {author} {\bibfnamefont {M.}~\bibnamefont {Hewitson}},\
  }\href@noop {} {\emph {\bibinfo {title} {{LISA sensitivity and SNR
  calculations}}}},\ \bibinfo {type} {Tech. Rep.}\ \bibinfo {number}
  {LISA-LCST-SGS-TN-001}\ (\bibinfo  {institution} {{APC} and {AEI}},\ \bibinfo
  {year} {2021})\ \Eprint {https://arxiv.org/abs/2108.01167} {arXiv:2108.01167
  [astro-ph.IM]} \BibitemShut {NoStop}%
\bibitem [{\citenamefont {{LISA Data Challenge working
  group}}(2022)}]{LISADataWG}%
  \BibitemOpen
  \bibfield  {author} {\bibinfo {author} {\bibnamefont {{LISA Data Challenge
  Working Group}}},\ }\href@noop {} {\bibinfo {title} {{LISA Data Challenge
  software}}},\ \bibinfo {howpublished} {Zenodo} (\bibinfo {year} {2022}),\
  \bibinfo {note} {10.5281/zenodo.7332221}\BibitemShut {NoStop}%
\end{thebibliography}

%

\end{document}